\documentclass{aa}  
\usepackage{graphicx}
\usepackage{xcolor}
\usepackage{cancel}
\usepackage{deluxetable}
\usepackage{comment}
\usepackage{hyperref}
\usepackage{orcidlink}
\usepackage[switch]{lineno}

\usepackage{txfonts}
\newcommand{\swift}{\textit{Swift}} 

\newcommand{\program}{\textit{}} 

\begin{document}

   \title{SN\,2021lwz: Another Exotic Luminous and Fast Evolving Optical Stripped Envelope Supernova ?}


   \author{F. Poidevin\orcidlink{0000-0002-5391-5568} \inst{1,2}\thanks{E-mail: fpoidevin@iac.es}
	\and  
	S. L. West\orcidlink{0000-0002-4575-6157}$^{3}$
	\and
  	C. M. B. Omand\orcidlink{0000-0002-9646-8710}$^{4}$
    \and
    R. K\"onyves-T\'oth\orcidlink{0000-0002-8770-6764}\inst{5,6,7} 
	\and
  	S. Schulze\orcidlink{0000-0001-6797-1889}$^{8}$ 
	\and
    L. Yan\orcidlink{0000-0003-1710-9339} \inst{9}\fnmsep
    \and
    T. Kangas\orcidlink{0000-0002-5477-0217}$^{10,11}$
    \and
  	I. P\'{e}rez-Fournon\orcidlink{0000-0002-2807-6459}$^{1,2}$
	\and
    S. Geier\orcidlink{0000-0003-3154-2120}$^{1,12}$
	\and
    J. Sollerman\orcidlink{0000-0003-1546-6615}$^{3}$
    \and
    P.~J. Pessi\orcidlink{0000-0002-8041-8559},$^{3}$
    \and
    C. M. Guti\'{e}rrez\orcidlink{0000-0001-7854-783X}$^{1,2}$
	\and
    T.-W. Chen\orcidlink{0000-0002-1066-6098}$^{9}$
    \and
    K-Ryan Hinds\orcidlink{0000-0002-0129-806X}$^{4}$
    \and
  	R. Marques-Chaves\orcidlink{0000-0001-8442-1846}$^{13}$
    \and
  	R. Shirley\orcidlink{0000-0002-1114-0135}$^{14}$
	\and
  	C. Jimenez Angel\orcidlink{0000-0003-3100-7718}$^{1,2,9}$
	\and
    R. Lunnan\orcidlink{0000-0001-9454-4639}$^{3}$
    \and
    D. A. Perley\orcidlink{0000-0001-8472-1996}$^{4}$
    \and
    N. Sarin\orcidlink{0000-0003-2700-1030}$^{15,16}$
	\and
    Y. Yao\orcidlink{0000-0001-6747-8509}$^{17,18}$
    \and
    R. Dekany\orcidlink{0000-0002-5884-7867}$^{19}$
    \and
    J. Purdum\orcidlink{0000-0003-1227-3738}$^{19}$
    \and
    A. Wold\orcidlink{0000-0002-9998-6732}$^{20}$
    \and
    R.~R. Laher\orcidlink{0000-0003-2451-5482}$^{20}$
    \and
    M.~J. Graham\orcidlink{0000-0002-3168-0139}$^{21}$
    \and
    M.~M. Kasliwal\orcidlink{0000-0002-5619-4938}$^{21}$
    \and
    T. Jegou Du Laz\orcidlink{0009-0003-6181-4526}$^{22}$
}
   \institute{Instituto de Astrof\'{\i}sica de Canarias, V\'{\i}a   L\'actea, 38205 La Laguna, Tenerife, Spain
         \and
    Universidad de La Laguna, Departamento de Astrof\'{\i}sica,  38206 La Laguna, Tenerife, Spain
         \and
	The Oskar Klein Centre, Department of Astronomy, Stockholm University, AlbaNova, SE-10691 Stockholm, Sweden
         \and
    Astrophysics Research Institute, Liverpool John Moores University, Liverpool Science Park IC2, 146 Brownlow Hill, Liverpool, UK, L3 5RF 
        \and
	Konkoly Observatory,  CSFK, Konkoly-Thege M. \'ut 15-17, Budapest, 1121, Hungary
        \and
	Department of Optics \& Quantum Electronics, University of Szeged, D\'om t\'er 9, Szeged, 6720, Hungary
        \and
	ELTE Eötvös Loránd University, Gothard Astrophysical Observatory, 9700 Szombathely, Szent Imre h. u. 112, Hungary
         \and
    Center for Interdisciplinary Exploration and Research in Astrophysics (CIERA), Northwestern University, 1800 Sherman Ave, Evanston, IL 60201, USA
        \and
    Graduate Institute of Astronomy, National Central University, 300 Jhongda Road, 32001 Jhongli, Taiwan
    \and        
    Finnish Centre for Astronomy with ESO (FINCA), FI-20014 University of Turku, Finland 
       \and
    Department of Physics and Astronomy, FI-20014 University of Turku, Finland
        \and
    GRANTECAN: Cuesta de San Jos\'{e} s/n, 38712 Bre{\~n}a Baja, La Palma, Spain 
       \and
	Geneva Observatory, University of Geneva, Chemin Pegasi 51, CH-1290 Versoix, Switzerland
       \and
	Max-Planck-Institut für extraterrestrische Physik, Giessenbachstr. 1, 85748 Garching, Germany
    \and
    Kavli Institute for Cosmology, University of Cambridge, Madingley Road, CB3 0HA, UK
    \and
    Institute of Astronomy, University of Cambridge, Madingley Road, CB3 0HA, UK
    \and
    Miller Institute for Basic Research in Science, 206B Stanley Hall, Berkeley, CA 94720, USA
    \and
    Department of Astronomy, University of California, Berkeley, CA 94720-3411, USA
    \and
    Caltech Optical Observatories, California Institute of Technology, Pasadena, CA 91125, USA
    \and
    IPAC, California Institute of Technology, 1200 E. California, Blvd, Pasadena, CA 91125, USA
    \and
    Division of Physics, Mathematics, and Astronomy, California Institute of Technology, Pasadena, CA 91125, USA
    \and
    Cahill Center for Astrophysics, California Institute of Technology, MC 249-17, 1200 E California Boulevard, Pasadena, CA, 91125, USA
    }

   \date{}
 
  \abstract
   {Current large-scale, high-cadence surveys, such as the Zwicky Transient Facility (ZTF), provide detections of new and rare types of transients and supernovae whose physical origins are not well understood.} 
   {We investigate the nature of SN\,2021lwz at a redshift $z=0.065$, an overluminous supernova (SN) of absolute magnitude, $M_{g} \sim -20.1$ AB, falling in the lower range of superluminous supernovae (SLSNe) luminosities, and discovered in a faint dwarf galaxy with an absolute magnitude of $M_{g} \simeq -14.5$ AB.}
   {SN\,2021lwz is studied using optical spectroscopy, photometry and imaging linear polarimetry obtained during several follow-up campaigns. All the data are used to analyse and model the evolution of the explosion. Comparisons with other SNe of well known or rarer types are investigated.}
   {SN\,2021lwz belongs to the rare class of rapidly evolving transients. The bolometric light curve rises in about $7$ days to a peak luminosity of about $5 \times 10^{43}$ erg/s, at a rate of 0.2 mag day$^{-1}$ close to the peak. Spectroscopy modelling reveals more similarities with a normal Type Ic-like SN than with a SLSN before peak, showing slightly broadened lines after peak. Light curve modelling shows that the Arnett model of the bolometric light curve using a radioactive source ($^{56}$ Ni) is not able to reasonably explain the light curve evolution. A magnetar model seems more appropriate, suggesting that the explosion of low ejecta mass ($M_{\rm ej} \sim 0.24 ~M_\odot$) took place in a low mass ($M \sim 10^{6.66}~M_\odot$) dwarf galaxy of specific star-formation rate about ten times larger than typical star-forming galaxies.}
   {In conclusion SN\,2021lwz is an uncommon transient showing many similarities with several classes of transients, and with rare transients. It may be an interesting example pointing on how differences in ejecta mass and engine parameters could produce a wide range of engine-driven SESNe.}
   \keywords{supernova: general -- supernova: Individual (SN\,2021lwz, iPTF\,16asu, SN\,1998bw, SN\,2018gep, SN\,2018bgv, SN\,2014ft, AT\,2020xnd) -- techniques: spectroscopy, photometry, polarimetry}

   \maketitle

\section{Introduction} \label{intro}

The advent of large-scale high-cadence surveys like Catalina Real-Time Transient Survey \citep[CRTS;][]{2009ApJ...696..870D}, the Panoramic Survey Telescope and Rapid Response System \citep[PanSTARRS;][]{2010SPIE.7733E..0EK}, the Palomar Transient Factory (PTF) \citep[see][]{2009PASP..121.1334R}, the Zwicky Transient Facility \citep[ZTF;][]{bellm2019, Graham2019a} and the Asteroid Terrestrial- impact Last Alert System \citep[ATLAS;][]{2018ApJ...867..105T}, developed over the last two decades, have opened the possibility to detect new types of transients. They have been useful to start populating the parameter space of the rising time, $T_{\rm rise}$, defined as the time between the explosion and peak magnitude, versus absolute magnitude at peak, $M_{\rm abs}$, with new populations of objects. Among those, superluminous supernovae (SLSNe) are explosions expected to originate from core collapsing stars. Such explosions may be forming a newborn rotating neutron star \citep[][]{og71}, or a magnetar \citep[][]{kb10, woo10}. 
Interactions of the explosion products with the circumstellar medium (CSM) coming from previous progenitor mass loss is an additional mechanism often considered to interpret the diversity of light curves of H-poor SLSNe \citep[e.g.][]{2022ApJ...933...14H}. 
A SLSN has often been considered as a type of stellar explosion that exhibits a peak luminosity ten or more times higher than that of a typical SN. This definition is quite general. In practice H-poor SLSNe are detected with absolute r-band magnitudes lying between $\sim -22.5$ and $\sim -19.2$ magnitudes \citep[][]{Gomez2024}.
The light curves of the largest sample from ZTF (78 SLSNe) evolve slowly with a mean rest-frame rise time of 41.9 $\pm$ 17.8 days \citep[see][]{chen2023a, chen2023b}, while \cite{Gomez2024} derive a mean rise time of 27$^{+25}_{-13}$ days on a larger sample. In both distributions low luminosity ($\sim 20$ mag) extremely fast rising ($<$ 10 days) transients are rare. From the ZTF sample, with a rising time of $\sim$ 10 days, and an absolute magnitude peaking close to $M_{g} = -21$, SN\,2018bgv was the fastest SLSN observed by ZTF during the first months of the survey \citep[][]{2020ApJ...901...61L}.

\begin{figure}
\begin{center}
\vspace*{2mm}
\centering
\hspace*{0.cm}
\includegraphics[width=90mm,angle=0]{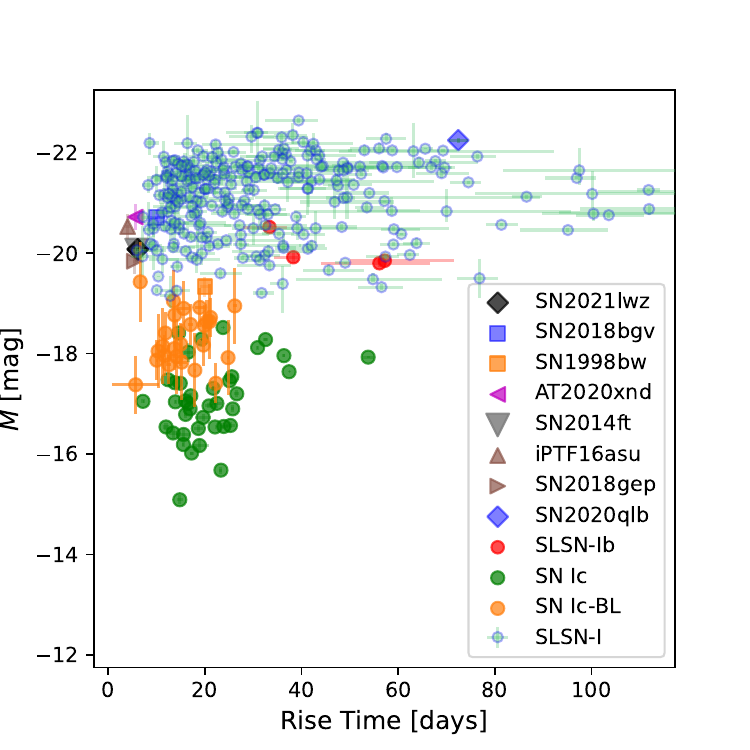}
\vspace*{0.cm}
\caption{Rise times $t_{rise,10\%}$ versus $g$-band absolute magnitudes $M_{g}$ for SLSNe-I \citep[][]{Gomez2024}, normal SNe Ic \citep[from][]{2021A&A...651A..81B} and SNe Ic-BL \citep[from][]{2019A&A...621A..71T}. As in \cite{chen2023a} the $M_{g}$ of the Sn Ic and SN Ic-BL is computed from the r-band magnitudes using a colour correction of $\sim 0.36$ mag. The rise times of these two populations are measured from the explosion dates in the r-band, which are slightly longer than $t_{rise,10\%}$. The data used for the labelled individual sources are from our compilation discussed in Section\,\ref{RandT} and Section\,\ref{discussion}.}
\label{fig:mrpeak_vs_trise}
\end{center}
\end{figure}

The formation of stars as single objects, or in binary, and even triple stars systems, which is followed by their evolution up to a final stage (e.g. core collapse or disruptive events) suggest the existence of a complex diversity of star evolution channels. Stars evolving in binary systems are subject to strong interactions with their environment (e.g. binary star companion, black hole proximity, winds, shock waves) potentially leading to new types of transients. 
As such, a variety of new types of transients has recently been discovered among which Luminous Fast Blue Optical Transients \citep[LFBOTs like AT\,2018cow (the Cow) and AT\,2020xnd (the Camel); see for example][and references therein]{2021MNRAS.508.5138P, 2023ApJ...949..120H}, Ultra-Stripped Supernovae \citep[USSNe; like SN\,2014ft (or iPTF\,14gqr) and SN\,2019ehk, see for example][and references therein]{2018Sci...362..201D, 2021ApJ...912...30N}. A more general terminology encompassing these events is Rapidly Evolving Transients \citep[RETs, e.g.][]{2014ApJ...794...23D, 2016ApJ...819...35A, 2016ApJ...819....5T, 2018MNRAS.481..894P}. Ideally, each of these types would be filling a given area in the phase space ($T_{\rm rise}$ , $M_{\rm abs}$) but, in practice distinct populations sometimes fill similar areas leading to a degeneracy in this parameter space, and more data-driven information is required. This happens for example with well studied populations like Type Ia and Type II SNe, and with rarer populations like Type Ib, Ic and broad line Ic (or Ic-BL) SNe \citep[see e.g. Figure 1 in][]{2017ApJ...851..107W}. Such a plot is presented in Figure~\ref{fig:mrpeak_vs_trise} where the position of SN\,2021lwz (see Section\ref{Light_Curves_analysis}) and several other objects of various types are displayed.

The empirical method used to study such new populations of transients, and to try to understand the nature of their progenitors and their environments, is therefore to gather as much data as possible. This includes getting multi-band, multi-messenger, high cadence photometry allowing to estimate the full energy budget spatially deployed by the explosion over time \citep[e.g.][]{2016MNRAS.458.2973P,2023MNRAS.526.1822K}. This requires to get high cadence spectroscopy to study the chemical evolution of the photosphere \citep[e.g.][]{quimby2011,ktr22,ktr23}. When possible this means to get polarimetry data to study the time-evolution of the geometry of the systems closely connected to the explosion \citep[e.g. inner core, jets, see][]{2024A&A...687L..17N, 2024OJAp....7E..31S} or products resulting from the explosion \citep[e.g. photosphere evolution, ejecta, see][]{inserra2016,leloudas2017}. Obviously gathering such data is a complex task even more complex for fast-rising events, and the lack of data obtained on some specific objects often prevents us from getting the full picture in one go. Nevertheless, gathering data and increasing the sample on several members of the same class of objects, retrospectively helps to better understand the said class. Such a progress is done in parallel to theoretical and numerical modelling developments \citep[e.g.][]{2024ApJ...972..140K, Omand2023}, which are crucial to understand and interpret the data, and ultimately to put constraints on, or to identify the channel formation of such new or peculiar transients or SNe. In this context of luminous transients SN\,2021lwz is a particularly intriguing object deserving a dedicated in depth study.

SN\,2021lwz (also known as ZTF21abaiono, PS21fcj, ATLAS21oii and Gaia21cvf), at R.A.$=$09h44mn47.390s, Dec$= +34 ^{\circ} 42 \arcmin 44.21 \arcsec$, was discovered by
\citet{2021TNSTR1564....1F} on May 10, 2021
from ZTF \footnote{Zwicky Transient Facility, {\tt
    https://www.ztf.caltech.edu}.} public alerts \citep[][]{2019PASP..131a8001P}. The discovery magnitude obtained with the ZTF-cam mounted on the Palomar 1.2 meter Oschin was of g$=$18.76 mag (AB system). The transient was classified on May 15, 2021 as a SLSN-I at a redshift z $=0.065$ by \citet{2021TNSCR1649....1P} from the analysis of a Keck1 Low Resolution Imaging Spectrometer (LRIS, \citealt{OkeEtal1995}) 
\footnote{LRIS, {\tt
    https://www2.keck.hawaii.edu/inst/lris/lrishome.html}} spectrum. 
For that reason SN\,2021lwz entered in our sample of H-poor SLSNe for follow-up. Subsequently, though, we realized that some of the pre-peak spectra are not similar to other typical H-poor SLSN spectra at similar phases, and we therefore decided to investigate this source in more detail. Indeed, SN\,2021lwz is also part of the sample of rare Luminous SNe discussed by \cite{gomez2022}. This type of transient appearing to form a continuum between type Ic SNe and SLSNe is expected to make up $\sim 0.4 \%$ of all observed Core Collapse SNe (CCSNe) \citep{gomez2022}. 
Additionally, SN\,2021lwz is a fast-rising SLSN ranging in the low absolute magnitude range, as do most of the He-rich SLSNe. 
This type is rare but a sample of seven Helium-Rich SLSNe was reported by \citet{2020ApJ...902L...8Y}. In this sample, the spectroscopically well sampled SLSN SN\,2019hge (ZTF\,19aawfbtg) was shown to be of similar spectral type as PTF\,10hgi, the only known He-rich SLSN at the time. Both were used to identify 6 more He-rich SLSNe thus increasing the sample of such known events to a total of 8 members.

In this work we present an intensive study of SN\,2021lwz. All the data obtained on SN\,2021lwz and its host galaxy are presented in section\, \ref{data}. The data analysis of SN\,2021lwz provided in section\,\ref{data_analysis} is followed by detailed modelling of the SN and its host galaxy in section\,\ref{data_modeling}. In order to better understand and contextualize the nature of SN\,2021lwz, comparisons with some SLSNe and other typical SNe and exotic transients are discussed in section\,\ref{discussion}. Our conclusions are given in section\,\ref{conclusions}.

Along this work, the photometric measurements are reported in the AB system and the uncertainties are provided at 1$\sigma$ confidence. We use the \textit{Planck} 2018 Flat $\Lambda-$CDM cosmology model ($\Omega_{m}=0.31, H_{0}=67.7$ km$/$s) \citep[][]{planck2018i}

\begin{figure}
\begin{center}
\vspace*{2mm}
\centering
\hspace*{0.cm}
\includegraphics[width=85mm,angle=0]{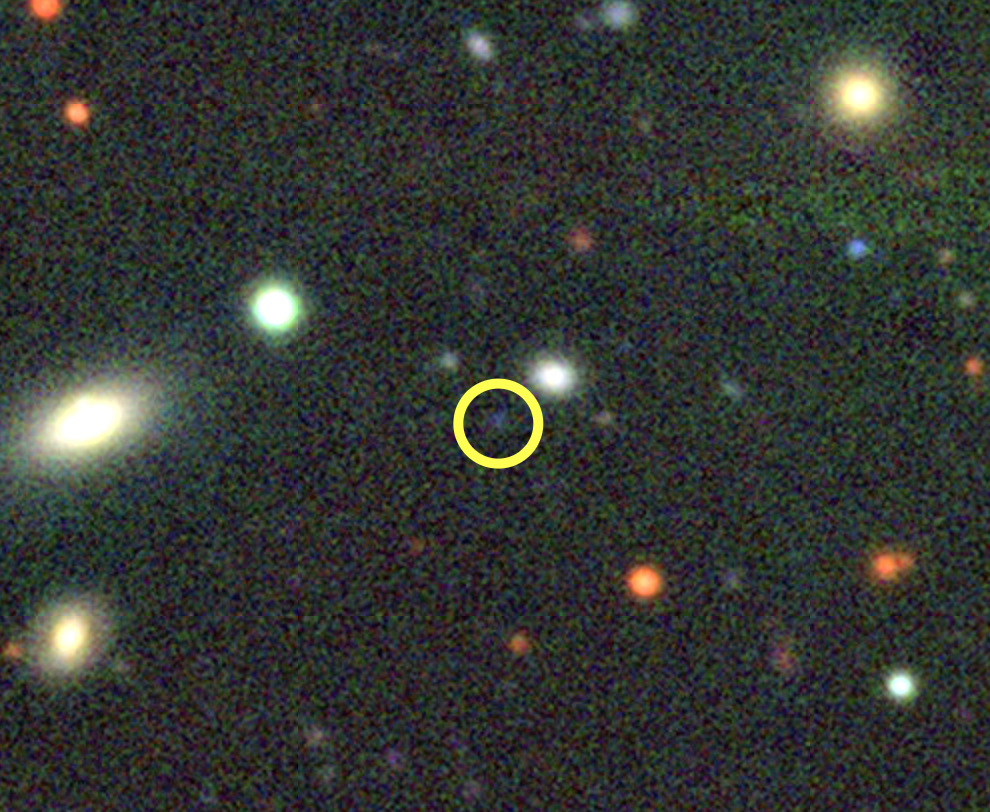}
\vspace*{0.cm}
\caption{LS DR 10 image of SN\,2021lwz's host, as shown by the faint feature centred in the yellow circle, and its environment. The field-of-view is of about 120$\arcsec$ by 90$\arcsec$ in size. North is up and East is left in the Equatorial frame.
}
\label{fig:host_desi_image}
\end{center}
\end{figure}

\section{Data} \label{data} 

In this section, we present data and information about the host galaxy. We also present the photometry, spectroscopy and polarimetry data obtained at several epochs on SN\,2021lwz. Most of these observations were coordinated through the ZTF Fritz Marshal system \citep{Walt2019a, Coughlin2023a}.
The reference epoch of maximum light is defined as MJD$_{\rm max} = 59350.9 \pm 1.0$ days with respect to g-band, as from a visual inspection. The extinction contribution from our Galaxy in the line-of-sight to SN\,2021lwz and its host is $A_{\rm{MW}, u}=0.054, A_{\rm{MW}, g}=0.042, A_{\rm{MW}, r}=0.029, A_{\rm{MW}, i}=0.021$ and $A_{\rm{MW}, z}=0.016$ \citep[][]{2011ApJ...737..103S}. 

\subsection{Host photometry} \label{data_phot_host}

The highly accurate spectroscopic redshift z $=0.065$ of SN\,2021lwz was determined from narrow H-alpha host-galaxy emission and probable [OIII] emission at a redshift consistent with the less accurate redshift obtained with the SNID analysis \citep[][]{2021TNSCR1649....1P}. 
The host galaxy at R.A.$=$09h44mn47.397s, Dec$= +34 ^{\circ} 42 \arcmin 44.63 \arcsec$ is detected in archival Sloan Digitized Sky Survey (SDSS) \footnote{{\tt http://skyserver.sdss.org/dr17/VisualTools/explore/summary}} and Legacy Survey (LS) \footnote{Legacy Survey, {\tt https://www.legacysurvey.org/}} images. It can be seen as the blue feature marked by the yellow circle in the LS image plotted in Figure~\ref{fig:host_desi_image}. 

We retrieved science-ready co-added images from the Panoramic Survey Telescope and Rapid Response System (Pan-STARRS, PS1) DR1 \citep{Chambers2016a} and the Sloan Digital Sky Survey data release 9 (SDSS DR 9; \citealt{Ahn2012a}). In addition, the field was observed with the 3.58-m Canada France Hawaii Telescope telescope equipped with MegaPrime in u and r band between 2016 and 2019. We retrieved the level-2 data from the CADC and combined them using Swarp version 2.38.0 \citep{Bertin2010a} to create a co-added image. We measure the brightness in all images using an aperture photometry tool presented in \citet{Schulze2018a} that is based on Source Extractor \citep{Bertin2010b}. The measurements in these different bands are given in Table \ref{tab:host_phot} (see Appendix\,\ref{host_appendix}). 

One survey observed that part of the sky in the near-infrared (NIR) with the Wide Field Camera (WFCAM) mounted on the United Kingdom Infrared Telescope \footnote{UKIRT, {\tt https://about.ifa.hawaii.edu/ukirt/}}. The Large Area Scale (LAS) survey footprint of the UKIDSS Data Release 9 \footnote{{\tt https://cdsarc.cds.unistra.fr/viz-bin/cat/II/319}} includes the coordinates position of SN\,2021lwz. This survey whose limit magnitudes are all lower than 20.2 mag in the Y, J, H and K bands does not show any detection of the host of SN\,2021lwz \citep[see][]{2007MNRAS.379.1599L}, in agreement with our results from the modelling of the host (see Section\,\ref{host_modeling}).

\begin{figure}
	\centering
    \includegraphics[width=100mm,angle=0]{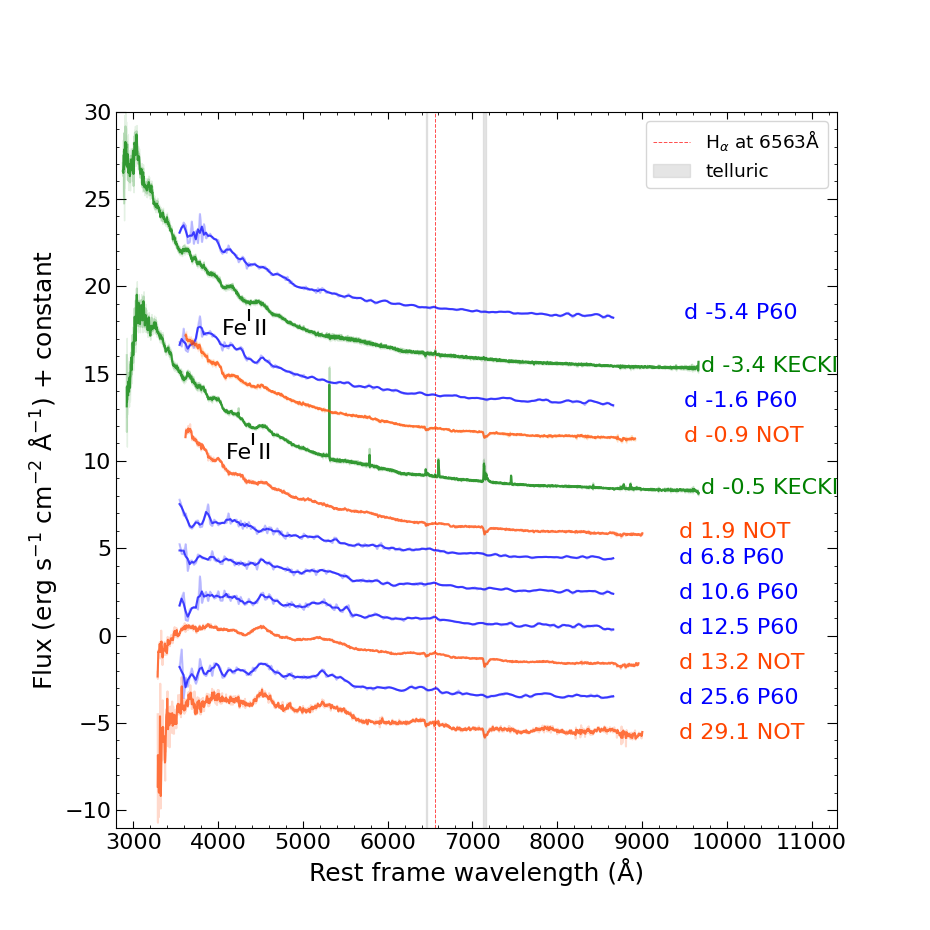}
	\caption{\label{SpectralEv}Spectral sequence of SN\,2021lwz. Spectra taken with the P60 are shown in blue; with the NOT in orange; and with the Keck I telescope in green. All the spectra were smoothed using a Savitzky-Golay low pass filter. Un-smoothed data and measurement errors are shown with a lighter shade of the same colour. Each spectrum is marked with its phase in rest frame days relative to maximum g-band flux as well as the telescope used for measurement. The Fe II P-Cygni profile minima used for velocity estimation are marked in dark red. Figure~\ref{fig:PCygni} zooms in onto the Fe II lines.}
\end{figure}

\subsection{Spectroscopy} \label{spectra}  

In addition to the Keck1 LRIS public spectrum used to classify the transient, we acquired spectra with the Spectral Energy Distribution Machine (SEDM; \citealt{BlagorodnovaEtal2018}) on the Palomar Observatory 60 inch telescope (P60), the LRIS on the 10 m Keck I telescope and the Andalucia Faint Object Spectrograph and Camera (ALFOSC)\footnote{\url{http://www.not.iac.es/instruments/alfosc/}} on the 2.56~m Nordic Optical Telescope (NOT).
SEDM spectra were reduced according to \citet{RigaultEtal2019, Kim2022a}. The LRIS spectrum was reduced using LPipe as described by \citet{Perley2019}. ALFOSC spectra were reduced using PyNOT\footnote{\url{https://github.com/jkrogager/PyNOT}}. A spectrophotometric standard star was used to calibrate each spectrum.
A summary of the spectra obtained on SN\,2021lwz is given in Table~\ref{tab:spectra}.
In Figure~\ref{SpectralEv} we show the spectral evolution of SN\,2021lwz. The spectra were taken between $-5.4$ rest frame days and $+29.1$ days relative to the g-band maximum.

\begin{figure*}
	\centering
	\includegraphics[width=140mm,angle=0]{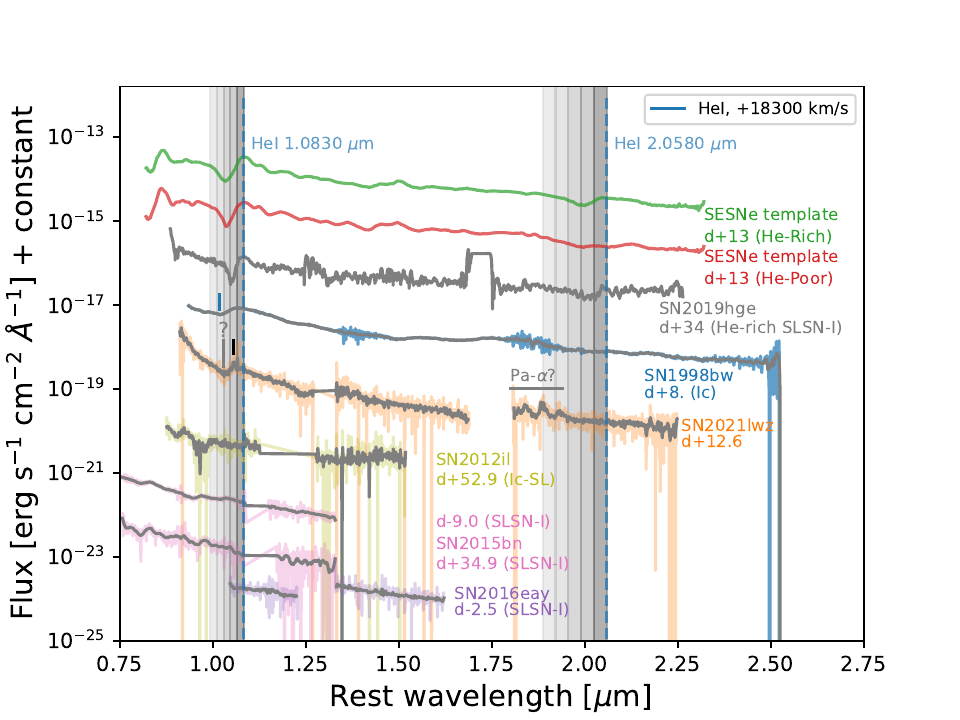}
	\caption{\label{SpecNIR} NIR spectrum of SN\,2021lwz obtained with NIRES on Keck II. Also shown are the SESNe templates spectra of Helium-Rich (green) and Helium-Poor (red) at phase $+13$ days produced by \cite{2022ApJ...925..175S}. The NIR spectra of the He-Rich SLSN-I SN\,2019hge at phase $+34$ days \citep[gray colour, smoothed,][]{2020ApJ...902L...8Y}, of the peculiar and bright Type Ic SN\,1998bw at phase $+8$ days \citep[blue colour,][]{2001ApJ...555..900P}, of Ic-SL SN\,2012il at phase $+12.6$ days \citep[][]{2013ApJ...770..128I}, of H-poor SLSN SN\,2015bn at phases -9.0 days and +34.9 days \citep[][]{2016ApJ...828L..18N}, and of SN\,2016eay (or Gaia16apd) at phase -2.5 days \citep[][]{YanEtal2017b} are also shown for comparisons. The observed spectra smoothed using Savitzky-Golay low pass filters are shown in gray. The vertical gray bands show velocity line ranges in bins of 5000\,km$/$s.}
\end{figure*}

The brightness and low-redshift of SN\,2021lwz motivated observations in the NIR. One spectrum was obtained at $+12.6$ days, rest frame, with the Near-Infrared Echellette Spectrometer \citep[NIRES][]{2004SPIE.5492.1295W}, a prism cross-dispersed near-infrared spectrograph mounted on the Keck II telescope. The spectrograph has a power resolution of R $\sim 2700$. It simultaneously covers the $YJHK$ bands over the wavelength range of 0.8 -- 2.4 $\mu$m. The spectra were reduced with \textit{spextool} \citep[][]{2004PASP..116..362C} and \textit{Xtellcor} \citep[][]{2003PASP..115..389V}, then combined together. The final spectrum is shown in Figure~\ref{SpecNIR}. 

\subsection{SN photometry} \label{data_phot}

\begin{figure*}
    \sidecaption
	\centering
	\includegraphics[width=120mm, angle=0]{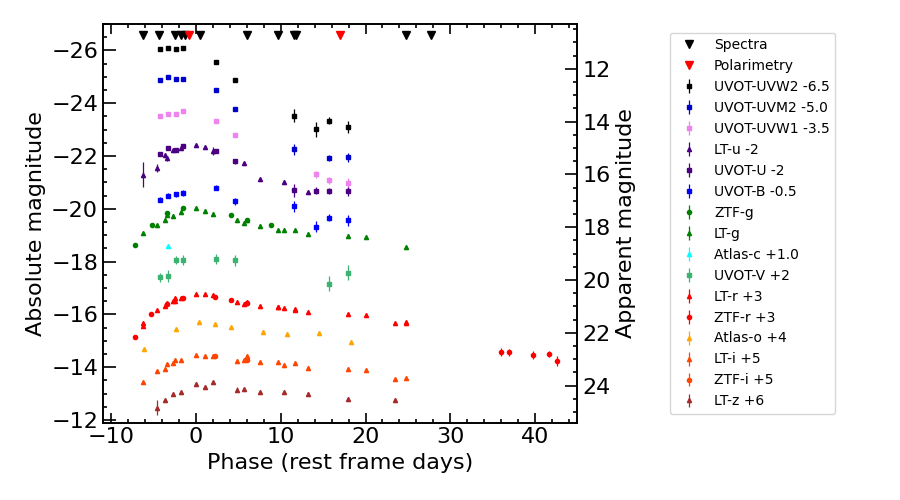}
	\caption{\label{fig:AbsMag} Absolute magnitude (rest frame) light curves (see Sec.~\ref{Light_Curves_analysis} for details) from the ATLAS and Liverpool Telescope SDSS filters as triangles, the \textit{Swift} UVOT filters as squares and the ZTF filters as circles. The phases of available spectra are shown as black triangles at the top along with the phases of the polarimetry shown as red triangles. Phase $=0$ is at peak ZTF $g$-band flux. The light curves were shifted by the indicated amounts for illustration purposes. 
    }
\end{figure*}

Photometry of SN\,2021lwz was obtained in several filters with the \swift\, satellite \citep{Gehrels2004a}, from the ZTF survey \citep[][]{bellm2019}, and with the Liverpool Telescope (LT) \citep{SteeleEtal2004}.
Public photometry was retrieved from the Asteroid Terrestrial-impact Last Alert System (ATLAS) \footnote{\href{https://atlas.fallingstar.com}{https://atlas.fallingstar.com}} \citep[][]{tonry2018} public server (as explained in Section~\ref{AtlasData} )
Details about the data are given in the following sections. 
The photometry is shown in Figure\,\ref{fig:AbsMag}. 
Table~\ref{tab:sn21lwz_all_phot} summarises all these measurements.    

\subsubsection{Swift UVOT data} 

We observed the field with the 30 cm Ultraviolet/Optical Telescope \citep[UVOT;][]{Roming2005a} aboard the The Neil Gehrels Swift Observatory, hereafter \swift\, satellite \citep{Gehrels2004a} between $-$2.6 and 20.9~days before and after maximum light (observer-frame) in 6 different filters ($w2$, $m2$, $w1$, $u$, $b$, $v$) from 2086 to 5411~\AA. We retrieved the science-ready data from the \swift\,archive \footnote{\href{https://www.swift.ac.uk/swift_portal}{https://www.swift.ac.uk/swift\_portal}}. We co-added all sky exposures for a given epoch and filter to boost the S/N using \texttt{uvotimsum} in HEAsoft\footnote{\href{https://heasarc.gsfc.nasa.gov/docs/software/heasoft/}{https://heasarc.gsfc.nasa.gov/docs/software/heasoft/}} version 6.26.1. Afterwards, we measured the brightness of SN\,2021lwz with the \swift\ tool {\tt uvotsource}. The source aperture had a radius of $5''$ while the background region had a significantly larger radius. The photometry was calibrated with the latest calibration files from September 2020 and converted to the AB system using \citet{Breeveld2011a}. 

\subsubsection{ZTF data} 

Data were acquired with the Zwicky Transient Facility (ZTF) that uses the Samuel Oschin 48-inch (1.22\,m) Schmidt telescope at Palomar Observatory on Mount Palomar (USA). It is equipped with a 47-square-degree camera \citep{Dekany2020} and monitors the entire northern hemisphere every 2--3 days in the $g$ and $r$ bands to a depth of $\sim20.7$~mag ($5\sigma$; \citealt{bellm2019, Graham2019a}) as part of the public ZTF Northern Sky Survey \citep{Bellm2019b}. 

We retrieved the host-subtracted photometry via the ZTF forced-photometry service \citep[][]{Masci2023a} hosted by NASA/IPAC Infrared Science Archive (IRSA) at IPAC.

\subsubsection{Liverpool Telescope data} \label{LiverpoolData}

Observations were taken with the IO:O optical imager on the LT 
in the Sloan Digital Sky Survey (SDSS) $u$, $g$, $r$, $i$ and $z$ filters.  Reduced images were downloaded from the LT archive and processed with custom image-subtraction and analysis software (K. Hinds and K. Taggart et al., in prep.).  Image stacking and alignment is performed using SWarp \citep[][]{2002ASPC..281..228B} where required.  Image subtraction was performed using a pre-explosion reference image in the appropriate filter from the Panoramic Survey Telescope and Rapid Response System 1 (Pan-STARRS1) or SDSS. The photometry is measured using PSF fitting methodology relative to Pan-STARRS1 or SDSS standards and is based on techniques in \cite{2016A&A...593A..68F}.

\subsubsection{ATLAS data} \label{AtlasData}

ATLAS forced photometry data were retrieved from the ATLAS public server \footnote{\url{https://fallingstar-data.com/forcedphot/}} \citep[][]{tonry2018, 2020PASP..132h5002S, 2021TNSAN...7....1S}.
They were clipped and binned to one day using the publicly available code plot$\_$atlas$\_$fp.py \footnote{\url{https://gist.github.com/thespacedoctor/86777fa5a9567b7939e8d84fd8cf6a76}}. 

\subsection{SN polarimetry} \label{data_pol}

The observations log of the imaging polarimetry obtained with ALFOSC on the Nordic Optical Telescope (NOT) is given in Table~\ref{tab:log_polarimetry}. 
Linear polarimetry was obtained at +1.0 and +19.0 days, rest-frame.
The polarimetry was obtained using a half wave plate in the FAPOL unit and a calcite plate mounted in the aperture wheel. The calcite plate provides the simultaneous measurement of the ordinary and the extraordinary components of two orthogonal polarized beams (see
Figure~\ref{fig:eo_images}). The half wave
plate can be rotated in steps of 22.5$^{\circ}$ 
from 0$^{\circ}$ to $337.5^{\circ}$. 
As a standard, we used 4 angle positions ($0^{\circ},
22.5^{\circ}, 45^{\circ}$, and $67.5^{\circ}$) to sample the Stokes parameters space. The photometry of the ordinary and extra-ordinary beams was done using aperture photometry of size $\sim$ 2 to 3 times the Full-Width at Half-Maximum (FWHM) of point sources in the images. For multiple sequences of 4 Half-Wave Plate (HWP) angles the polarization is obtained by summing-up the fluxes from the ordinary and extra-ordinary beams to minimize uncertainty propagation.

\section{Data analysis} \label{data_analysis}

In this section we derive physical parameters associated to the explosion using an analytical approach.

\subsection{Spectral analysis using template fitting}\label{snid_analysis}

In order to get a first understanding of the spectral evolution of SN\,2021lwz, we ran all the optical spectra through the SuperNova IDentifier (SNID) \citep[][and references therein]{blondin2007}. We first performed a general classification by running the code over our upgraded template database which, in addition to the standard Types (Ia, Ib, Ic, II) and their subtypes, contains SLSN-I and SLSN-II templates including those of PTF\,10hgi (19 spectra with epochs lying between -30 and +312 days with respect to peak in the SN rest-frame) and SN\,2019hge (2 spectra at epochs 0 and +48 days, respectively) which are used as He-rich SLSN-I templates in \citet{2020ApJ...902L...8Y} (see Section~\ref{intro}).

The analysis provided by SNID shows that none of the spectra of SN\,2021lwz obtained before peak (see Table~\ref{tab:spectra} and Figure~\ref{SpectralEv}) can be matched properly without decreasing some of the values of the default SNID parameters aimed to ensure a statistically robust diagnostic. Fit attempts using only He-rich SLSNe PTF\,10hgi or SN\,2019hge templates were conducted on the pre-peak spectra without producing any good match. Using the full database, a relatively poor fit is obtained on the -3.4 days Keck spectrum with the Type Ic-SL (Ic-superluminous) PTF10cwr, whose absorption features are smoothed compared to other SL templates. At maximum light the Keck spectrum at -0.5 days show similarities with the fast-evolving transient SN\,2018bgv. These fits are shown in Figure~\ref{fig:snid_fits}. Accurate fits of the ALFOSC spectra of SN\,2021wz obtained post-peak are obtained without degrading the by-default SNID parameters. They suggest that the photosphere of SN\,2021lwz evolved as a Ic-BL looking-like Type, in agreement with the first statement provided in the TNS classification report. 

At early times, the spectra look (by eye) like they could be consistent with SLSNe, but with higher velocities than we are used to seeing. This could explain why the fits with SLSNe favour lower redshifts. 
This may also happen because SNID tries to adjust the W-feature often attributed to  \ion{O}{ii}, but it is known that different lines of other elements like  \ion{Fe}{ii} are expected in this wavelength range. This will be discussed further in our analysis. The fits obtained with SLSN-I SN\,2016wi and with the Type Ic-SL SN\,LSQ12dlf, shown in Figure~\ref{fig:snid_fits}; middle, point to more accurate redshifts consistency. Using a slightly lower value of 4 for the \textit{rlap} factor, it is possible to produce lower quality fits with late time spectra of the peculiar Type Ic SN\,1998bw (bottom) whose NIR spectral properties will be discussed and compared with those of SN\,2021lwz NIR spectrum in Section\,\ref{nires}.  

\subsection{NIRES spectrum}\label{nires}

The NIR spectrum of SN\,2021lwz, obtained +12.6 days, rest frame, after peak is shown in Figure~\ref{SpecNIR}. It provides valuable information for comparison with several types of supernovae observed in the NIR. The spectrum is smooth and relatively featureless with an absorption feature centred at $\sim 1.0290$ $\mu$m (see gray vertical line with exclamation mark). If that feature was to be produced by  \ion{He}{i}, $1.0830$ $\mu$m, it's velocity would be $\sim 16200$ km s$^{-1}$, i.e. about the same order as the ejecta velocity derived a few days before maximum light, and about two times larger than the photosphere velocities at late phase ($\sim 7500$ km s$^{-1}$), as derived from the optical spectra analysis discussed in the following sections. At this late phase velocity, any  \ion{He}{i}, $1.0830$ $\mu$m absorption feature would be encountered at a wavelength position centred close to 1.0559 $\mu$m as indicated by the black vertical line. For comparison, the high velocity $18000$ km s$^{-1}$,  \ion{He}{i}, $1.0830$ $\mu$m absorption feature possibly detected in the peculiar Type Ic SN\,1998bw \citep[][]{2001ApJ...555..900P} is shown by the blue vertical line. Similar comparisons can be done with the  He-Rich and  He-Poor template spectra {at phase +13 days (d$+13$) of Stripped-Envelope SNe (SESNe) derived from a principal component analysis of Types IIb, Ib, Ic and Ic-BL by \cite{2022ApJ...925..175S}, and with the $+34$ days He-Rich spectrum of SLSN-I SN\,2019hge \citep[][]{2020ApJ...902L...8Y}. At higher wavelengths, these 3 spectra show spectral features which are not seen in the spectrum of SN\,2021lwz, whose smooth features look more similar to those seen in the spectrum of SN\,1998bw. 

Additional unidentified absorption or emission features are seen in the spectrum of SN\,2021lwz in the wavelength range $\sim 1.800-1.950$ $\mu$m, which are not seen in the SESNe template spectra. In this wavelength range, some features are seen in the spectra of SN\,2008ax and SN\,1998bw in sky-line contaminated regions of the spectra making any direct comparison difficult. At higher wavelength, the SN\,2021lwz spectrum does not show any hint of an absorption feature that could be produced by  \ion{He}{i}, $2.0580$ $\mu$m.  

To assess if SN\,2021lwz could be a Type I SLSN, we also made comparisons with spectra of SN\,2015bn at phases -9 days and +34 days \citep[][]{2016ApJ...826...39N}, of Gaia16apd at phase -2.5 days \citep[][and references therein]{YanEtal2017b} and of SN\,2012il at phase \citep[][]{2013ApJ...770..128I}. None of these spectra are at a phase similar to the NIR spectrum of SN\,2021lwz, and the wavelength coverage is sometimes limited to make meaningful comparisons, but all in all, they are all looking different from the spectrum of SN\,2021lwz. 

\subsection{P-Cygni velocity estimation}\label{PCygni}

We use the Fritz data science platform to cycle through all of the available lines and find that  \ion{Fe}{ii} is the best one that matches the broad feature near 4515\,\AA\,seen in some of the spectra at a redshift z=0.065. 

With this respect, we attribute the prominent P-Cygni-like profiles in the Keck I spectra at $-3.4$ and $-0.5$ days to be mainly due to  \ion{Fe}{ii} at 4515\,\AA\,. They are marked with black in Figure~\ref{SpectralEv}. 
We estimate the ejecta velocity according to the longitudinal relativistic Doppler shift. We use the \texttt{scipy.optimize.curve\_fit} algorithm to fit an equation consisting of a straight line component to match the continuum, and a Gaussian component to match the absorption feature to P-Cygni profile data as shown in Figure~\ref{fig:PCygni}. The resulting covariance matrices and the corresponding parameter fit values of $\lambda_{min}=4293.8$\,\AA\, and $\lambda_{min}=4379.5$\,\AA\, are then used to estimate the absorption line velocity of $15040 \pm 570$ km s$^{-1}$ at $-3.4$ days and $9130 \pm 350$ km s$^{-1}$ at $-0.5$ days, respectively.

\begin{figure}
	\centering
	\includegraphics[width=95mm,angle=0]{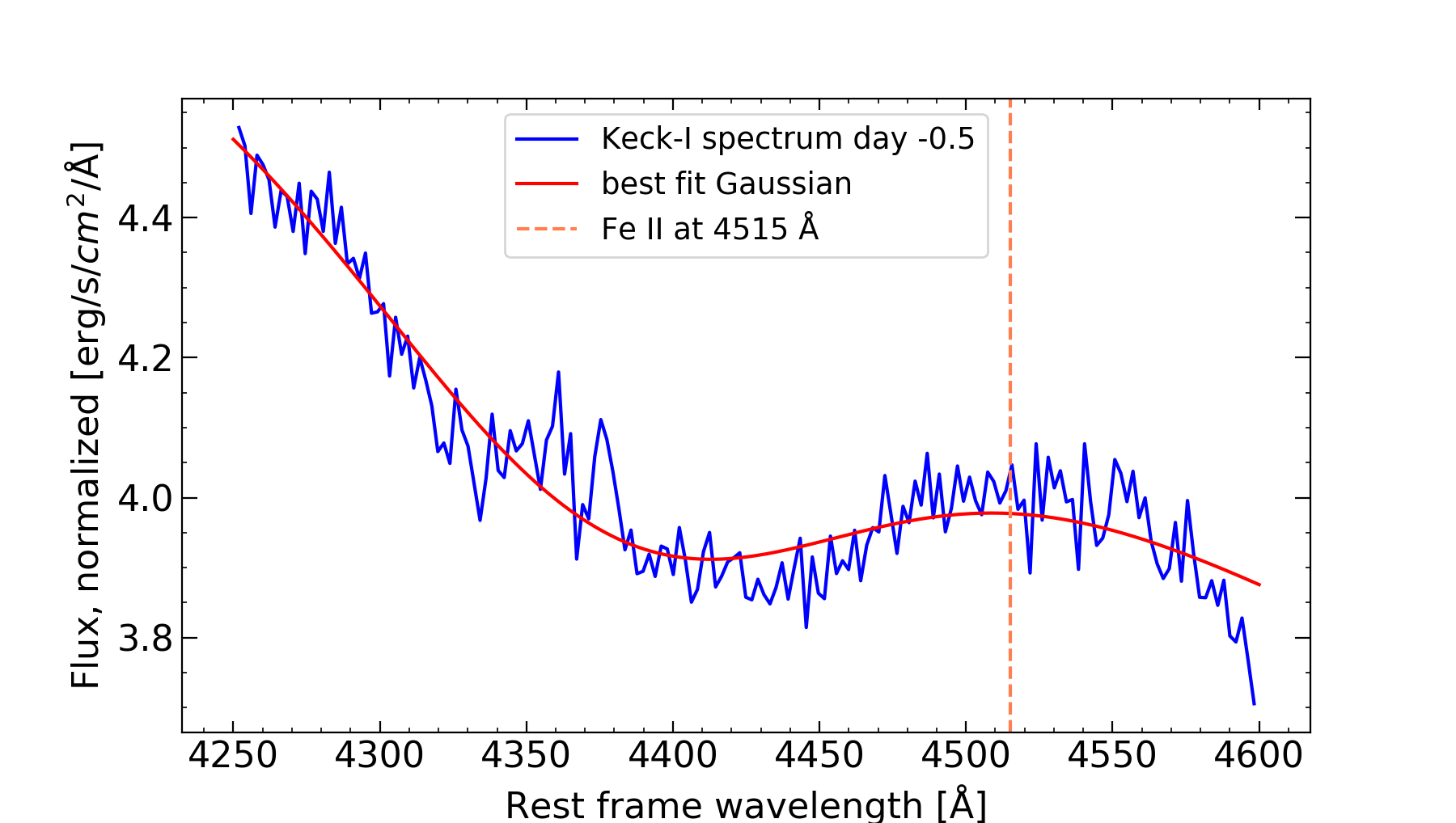}
	\caption{\label{fig:PCygni}SN\,2021lwz P-Cygni profile for Fe II at 4515\,\AA\, from the Keck I spectrum at -0.5 rest frame days. }
\end{figure}

\subsection{Light curves analysis} \label{Light_Curves_analysis} 

The absolute magnitude is calculated as
\begin{equation}
\label{eq:1}
    \mathrm{M} = \mathrm{m} - \mu - K_{\mathrm{corr}} - \mathrm{A}_{\mathrm{MW}} - \mathrm{A}_{\mathrm{host}}\,,
\end{equation}
where $\mu$ represents the distance modulus (which is equal to 37.40 for a redshift of 0.065); $K_{\text{corr}}$ is the correction factor for the difference in filter bandpass between the observer and the rest frame; A$_{\text{MW}}$ is the extinction caused by the Milky Way; and A$_{\text{host}}$ is the extinction caused by the host galaxy.

We use only the first term of the K-correction as described by \citet{HoggEtal2002} as shown here.
 \begin{equation}\label{eq:2}
    K_{\rm corr} = -2.5 \times \log_{10}(1+z) = -0.07 \quad \text{mag},
\end{equation}
The \citet{Fitzpatrick1999} extinction model is used to calculate the MW dust extinction based on the parameters R$_{V}=3.1$ and E(B-V)~$=0.013$ mag. 
We assume that the extinction from the host galaxy is negligible since no absorption lines like for example, \ion{Na}{I} Doublet absorption lines, are visible. 
The resulting absolute magnitude (rest frame) light curves from LT, UVOT, ATLAS and ZTF are plotted in Figure~\ref{fig:AbsMag}. 

The rising rate is calculated using the g-band data. At -7.1 days the g-band absolute magnitude is -18.7 mags. 
At zero days the g-band absolute magnitude is -17.2 mags. 
The delta mag on the rise close to the peak is therefore 1.5 mags / 7.1 days = 0.2 mags per day.

\subsection{Photospheric radius and temperature evolution}\label{RandT}

We estimate the photospheric temperature and radius evolution at the phases where full sets of UVOT measurements are available, and interpolate ZTF, LT and ATLAS-o light curve measurements using Gaussian Process (GP) regression that utilizes the Python package \texttt{GEORGE} \citep{AmbikasaranEtal2015} with a Matern 3/2 kernel. We then use the \texttt{scipy.optimize.minimize}\footnote{\url{https://docs.scipy.org/doc/scipy/reference/generated/scipy.optimize.minimize.html}} algorithm to fit a Planck function (blackbody) to the datasets. 

We find that the three UVOT UV filters are consistently incompatible with the best-fit blackbody. We attribute the problem to line blanketing as seen in other SLSNe \citep[see e.g.][]{YanEtal2017b}. We therefore exclude the three UV filter measurements from UVOT before rerunning the blackbody fits to create the photospheric temperature and radius evolutions.

The temperature evolution and the radius evolution of SN\,2021lwz are compared to a sample including those of SN\,2020qlb \citep{WestEtal2023} and those of the 31 SLSNe-I provided in the online version of Table A67 of  \citet{chen2023a} in Figure~\ref{fig:TempEvol} and in Figure~\ref{fig:RadiusEvol}, respectively.

\begin{figure}
	\centering
	\includegraphics[width=95mm, angle=0]
    {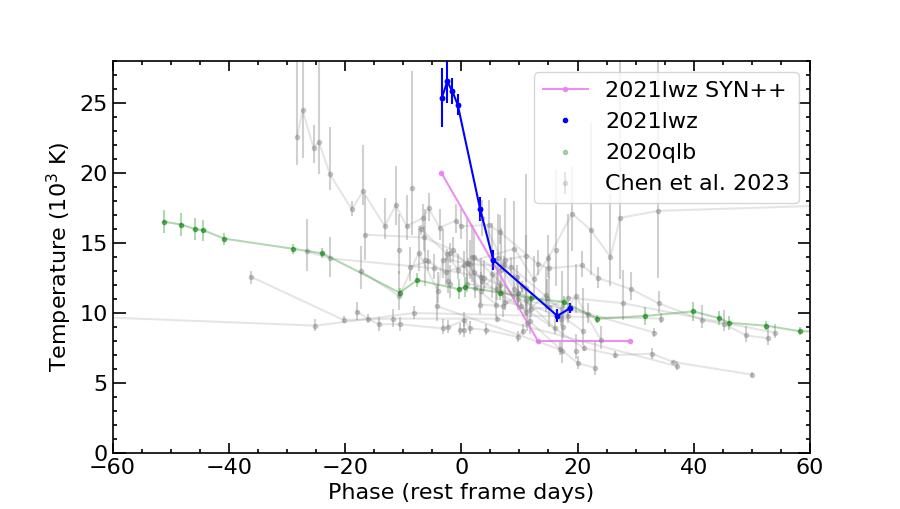}
	\caption{\label{fig:TempEvol}SN\,2021lwz temperature evolution compared to SN\,2020qlb \citep{WestEtal2023}, data from Table \ref{tab:globparams} and 31 SLSNe-I from \citet{chen2023a}.}
\end{figure}

\begin{figure}
	\centering
	\includegraphics[width=95mm, angle=0]
    {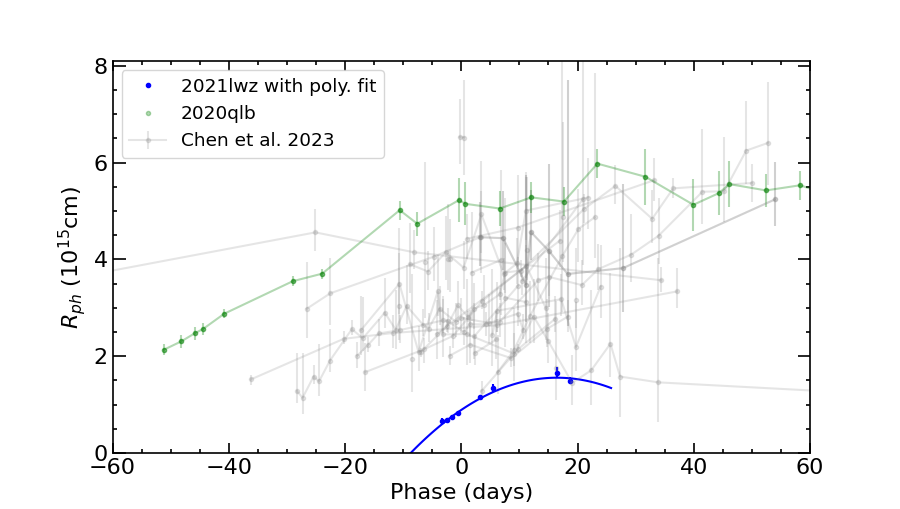}
	\caption{\label{fig:RadiusEvol}SN\,2021lwz radius evolution compared to SN\,2020qlb \citep{WestEtal2023} and to 31 SLSNe-I 
 from \citet{chen2023a}.
    A best-fit second degree polynomial is also plotted for SN\,2021lwz.}
\end{figure}

\subsection{Ejecta velocity estimation}\label{EjectaVelocity}

The time derivative of the best fit polynomial of the SN\,2021lwz radius evolution gives rough estimates of the photospheric velocity at relevant phases. At epochs just after explosion the photospheric velocity should approximate the ejecta velocity. At later epochs, P-Cygni absorption line features estimate velocities outside the photosphere. In Figure~\ref{fig:Velocity} we compare the photospheric velocity evolution to the P-Cygni velocity estimate from Sec.~\ref{PCygni}.

\begin{figure}
	\centering
	\includegraphics[width=95mm, angle=0]
    {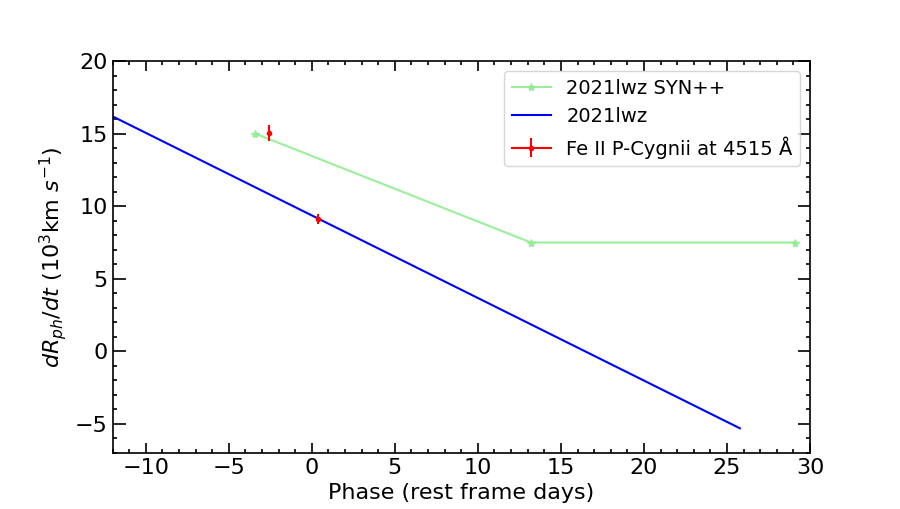}
	\caption{\label{fig:Velocity} SN\,2021lwz photosphere velocity estimates from the time derivative of the best-fit polynomial to the radius evolution of SN\,2021lwz shown in Figure~\ref{fig:RadiusEvol}, P-Cygni velocity estimates from the Keck I spectra at $-3.4$ and $-0.5$ days, and SYN++ data from Table \ref{tab:globparams}.}
\end{figure}

\subsection{Bolometric light curve estimation}\label{BoloLC}

At the eight epochs where UVOT data is available we calculate the spectral luminosity (L$_{\lambda}$) for each UVOT filter measurement and GP interpolate filter value from the other filters. Short of the lowest filter value we create L$_{\lambda}$ estimates based on the best fit blackbody to the UVOT-UV, UVOT-U and LT-u filter measurements. Long of the highest filter value we create L$_{\lambda}$ estimates based on the best fit blackbody to the optical filter measurements. These extrapolation methods are also used in \texttt{SuperBol} software \citep{Nicholl2018}. We then integrate over all wavelengths to create bolometric luminosity estimates, at each of the eight epochs, which form the bolometric light curve. These results are shown, modelled and discussed in section\,\ref{LCmodeling}.

\subsection{Photosphere geometry} \label{pol_analysis}

The polarization degrees measured on the unpolarised star, BD$+$52913, on the polarized calibration star, HD\,127769, on the field star STAR\,1 (shown in Figure~\ref{fig:eo_images}), and on SN\,2021lwz are all displayed in Table~\ref{tab:pol_results}. The instrumental polarization ($IP$) degree is of order of $0.1\%$ at each epoch. This component, which is retrieved in the parameter space of the Stokes $Q$ and $U$ parameters, is removed from the components of the other sources ($IP$ correction).
We use the polarized star HD\,127769 to calculate the zero polarization angle ($ZPA$) which is used to rotate the Stokes $Q,U$ parameters from the ALFOSC FAPOL instrument reference frame to the sky reference frame in Equatorial coordinates. The polarization angles are counted positively from North to East. We use the polarimetry available on STAR 1, in the field of view of FAPOL (see Figure~\ref{fig:eo_images}), to get an estimate of the polarized component of the Galaxy in the vicinity of SN\,2021lwz (see values shown in bold in Table~\ref{tab:pol_results}). These estimates are of the same order as their uncertainties suggesting a low level of the interstellar polarization ($ISP$) in this region of the sky. This result is consistent with the polarization degrees of the two stars HD\,82885 ($P =0.05 \pm 0.12$) and HD\,82635 ($P =0.03 \pm 0.12$) retrieved from the \citet{heiles2000} agglomeration catalog. These two stars are at distances lower than two degrees on the plane-of-sky from the central coordinate position of SN\,2021lwz.  
The levels of polarization obtained on SN\,2021lwz after $ISP$ correction using STAR\,1 Stoke $Q$ and $U$ parameters, 
are displayed in the last column of Table~\ref{tab:pol_results}.
The final values shown in bold are obtained after bias correction following the equation given in \cite{1997ApJ...476L..27W}:
\begin{equation} 
\label{eqn:pdeb}
    P = (P_{\rm obs} - \sigma_{P}^{2}/P_{\rm obs}) \times h(P_{\rm obs} - \sigma_{P}),
\end{equation}
where $h$ is the Heaviside function, $P_{\rm obs}$ is the observed polarization and $\sigma_{P}$ is the 1$\sigma$ error. The final measurements, $p_{\rm db}=0.00 \pm 0.14 \%$ at +1 day rest-frame, and $p_{\rm db}=0.25 \pm 0.18 \%$ at +19 days rest-frame, are consistent with null-polarization. They suggest that the geometry of the photosphere of SN\,2021lwz which is almost spherical, as from our observer point of view, didn't change with time. Similarly, there is no hint that the SN's photosphere interacted with the circumstellar medium (CSM). 

\section{Data modelling} \label{data_modeling}

\subsection{Optical spectrum evolution and chemical composition} \label{spec_model} 

To study the spectroscopic evolution of SN\,2021lwz, we model the available photospheric phase spectra utilizing the code named SYN++ \citep{thomas11}. With this code, it is possible to fit some global parameters that refer to the overall model spectrum, such as the temperature ($T_{\rm phot}$) and the velocity ($v_{\rm phot}$) at the photosphere. The lines of the individual ions can be fitted via local parameters, that code the optical depth ($\tau$), the velocity width of the line forming region ($v_{\rm min}$ and $v_{\rm max}$), the scale height parameter of the optical depth above the photosphere (aux) and the excitation temperature ($T_{\rm exc}$), assuming Local Thermodynamic Equilibrium.

For SN\,2021lwz, we have spectra available at 12 different epochs from -5.4 days to +29.1 days phase, rest-frame, relative to the moment of the maximum light. Table \ref{tab:globparams} presents the date of observation of the modelled high-resolution spectra together with their phases, and the $T_{\rm phot}$ and $v_{\rm phot}$ values of their best-fit models obtained using SYN++. The possibilities, constraints and uncertainties of the usage of SYN++ and its parameters are discussed in details in \citet{ktr22}.

\begin{table}
\caption{SYN++ modelling of SN\,2021lwz.} 
\label{tab:globparams}
\centering
\begin{tabular}{ccccc}
\hline 
Date & MJD & Phase & $T_{\rm phot}$ & $v_{\rm phot}$ \\
     &     & (days)& (K) & (km s$^{-1}$) \\
\hline
\hline
2021-05-13 & 59347 & -3.4 & 20000 & 15000 \\
2021-05-30 & 59364 & 13.2 & 8000 & 7500 \\
2021-06-16 & 59381 & 29.1 & 8000 & 7500 \\
\hline
\end{tabular}
\end{table}

\begin{figure}
\centering
\includegraphics[width=8cm]{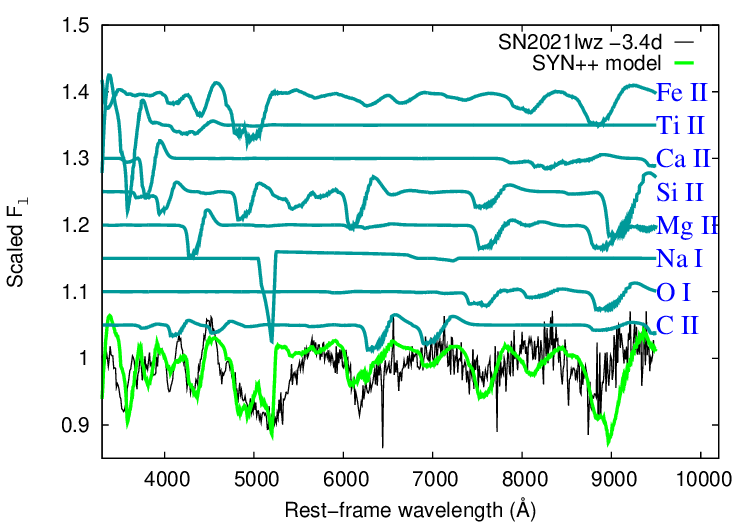}
\includegraphics[width=8cm]{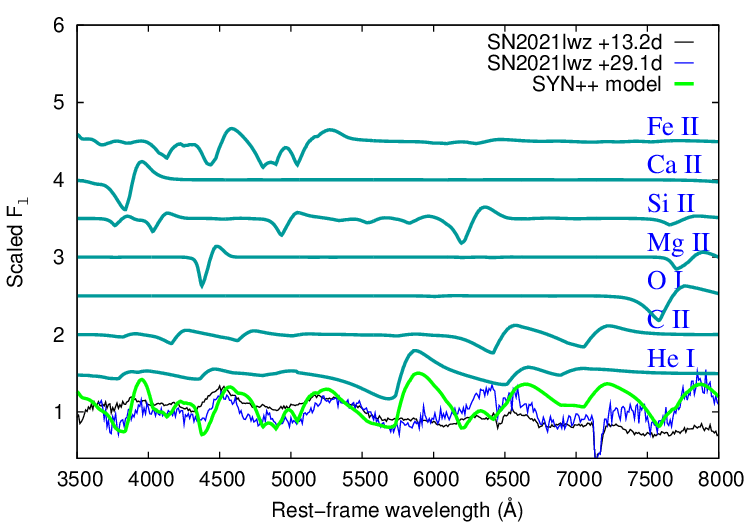}
\caption{Best-fit SYN++ models of the spectra of SN\,2021lwz. Top: the modelling of the -3.4d phase spectrum. Bottom: best-fit model applied to the +13.2d and the +29.1d phase spectrum). All spectra have been corrected for interstellar reddening (E(B-V) = 0.0108) and redshift.
The spectra have been continuum normalized using the Planck curve fitted previously using SYN++ before plotting.
Green colour codes the best-fit model, while vertically shifted, blue lines represent the single ion contributions to the overall model spectra.} 
\label{fig:spmodels}
\end{figure}

Figure \ref{fig:spmodels} shows the spectrum modelling of SN\,2021lwz. In the top panel, 
the modelling of the -3.4d phase spectrum of SN\,2021lwz can be seen. The best-fit $T_{\rm phot}$  and $v_{\rm phot}$ values can be read out from Table \ref{tab:globparams}. The identified elements are \ion{C}{ii}, \ion{O}{i}, \ion{Na}{i}, \ion{Mg}{ii}, \ion{Si}{ii}, 
\ion{Ca}{ii}, \ion{Ti}{ii} and \ion{Fe}{ii}. Regarding this, SN\,2021lwz is more similar to a normal Type Ic supernova rather than a SLSN. The pre-maximum spectrum of $\sim$2/3 of the SLSNe-I usually contains a W-shaped absorption blend between $\sim$4000 and 5000 \AA, while the remaining $\sim$1/3 fraction of the SLSNe-I show similar features to SN~2015bn \citep[see][]{ktr22}. However, the -3.4d phase spectrum of SN\,2021lwz does not show resemblance to any of these SLSN-I subtypes. 

The bottom panel of Figure~\ref{fig:spmodels} shows the modelling of the +13.2d and the +29.1d phase spectra, which are quite similar to each other. By these phases, both the photospheric velocity and temperature decreased significantly ($T_{\rm phot}~=~8000K$ and $v_{\rm phot}~=~7500$K), and besides \ion{Fe}{ii}, \ion{Ca}{ii}, \ion{Si}{ii}, \ion{Mg}{ii}, \ion{O}{i} and \ion{C}{ii}, \ion{He}{i} was included in the best-fit model.

In Table \ref{tab:locparams} in the Appendix, the best-fit local parameter values obtained in SYN++ are collected. The SYN++ modelling analysis suggests SN\,2021lwz is a intermediate Type Ic/Ic-BL transient rather than a SLSN-I. 

To conclude, we point out that any difference between $T_{\rm exc}$ and $T_{\rm phot}$ can be explained with the presence of Non Local Thermodynamic Equilibrium (NLTE) effects. In the case of LTE, $T_{\rm exc}=T_{\rm phot}$ for each ion, however, in supernovae, NLTE effects usually play a role in the spectrum formation. Although SYN++ is not a NLTE code, like TARDIS \citep[][]{2014MNRAS.440..387K}, which can appropriately take into account every difference from LTE, choosing a $T_{\rm exc} \neq T_{\rm phot}$ indicates that there is some divergence from the LTE. If we change the $T_{\rm exc}$ of a given ion in SYN++, the line ratios and strengths become different. We chose the best-fit model accordingly.

\subsection{Bolometric light curve modelling} \label{LCmodeling}

We compare semi-analytic models to the bolometric light curve to determine the most probable power source of the SN. The radioactive source and magnetar source models used by \citet[][Sect.~7]{WestEtal2023} \citep{Arnett1982, kb10, woo10} are employed using the ejecta velocity estimate from Section~\ref{EjectaVelocity}.
Our main conclusion is that the light curve can't be fit with $M_{ni} \le M_{ej}$. The details of this analysis are given in Appendix\,\ref{arnett_appendix}

There are no signs of narrow lines in the spectra indicating interaction, but shock breakout from interaction with extended material \citep{Nakar2010, Nakar2014, Piro2015} has been used to explain the light curve of the similar transient SN\,2018gep \citep{2019ApJ...887..169H, 2021ApJ...915..121P}.  We do not explicitly use this model, but note that similar parameters ($M_{\rm ej} = 8M_\odot$, $E_{\rm ej} = 2$ $\times$ 10$^{52}$ erg, $M_{\rm CSM} = 0.02M_\odot$, and $R_{\rm CSM} =  3$ $\times$ 10$^{14}$ cm \citep{2019ApJ...887..169H}) could likely also reproduce the light curve of SN\,2021lwz.

\subsection{Multi-band light curve modelling}\label{MLCmodeling}

We perform inference on the multi-band photometry using the \textsc{Redback} \citep{Sarin_redback} magnetar-driven supernova model including ejecta acceleration and non-vacuum dipole spin-down \citep{Omand2024}.  We use the \textsc{PyMultiNest} sampler \citep{Buchner2014} implemented in \textsc{bilby} \citep{Ashton2019, Romero-Shaw2020}, and sample in flux density with a Gaussian likelihood. The uncertainty presented is only the statistical uncertainty in the fits, and does not include systematic uncertainty inherent in the simplified one-zone \textsc{Redback} model. 

We sample the explosion time with a uniform prior of up to 100 days before the first observation, an extinction term $A_V$ with a uniform prior between 0 and 2, and a white noise term.  We constrain the total rotational energy of the magnetar to be $E_{\rm rot} \lesssim 10^{53}$ erg.  We fixed the photospheric plateau temperature to a value of 8000 K to keep consistency with the value found in Section~\ref{spec_model}. The physical parameters and priors used are listed in Table \ref{tab:redbackparams}. Uniform priors are denoted by U and log-uniform priors are denoted by L.

The fitted light curve is shown in Figure \ref{fig:lcfit} and the posterior for the parameters in Figure \ref{fig:corner} (see Appendix\,\ref{21lwz_appendix}).  The model fits the peak of the light curve well in all bands but underestimates the flux in the UV bands from around 14 days post-peak. We find an initial luminosity $L_0 \approx 3 \times 10^{45}$ erg s$^{-1}$ and spin-down timescale $t_{\rm SD} \approx$ one day, and an ejecta mass $M_{\rm ej} \approx 0.24 M_\odot$. These values are similar to that of the luminous fast-blue optical transient (LFBOT) AT\,2020xnd \citep[i.e. the ZTF source ZTF20acigmel; see e.g.][]{2021MNRAS.508.5138P}, the ultra-stripped supernova (USSN) SN\,2014ft \citep[see Table 2 in][]{Omand2024}, and iPTF16asu \citep[see Table~\ref{tab:redbackparams}; and][]{2017ApJ...851..107W, 2019MNRAS.489.1110W,2022ApJ...928..114W}, including a similar higher magnetar rotational energy of $\sim 4 \times 10^{50}$ erg, but a lower explosion energy of $\sim 2 \times 10^{49}$ erg.  We find the braking index $n$ to be $\sim$ 4.5 and rule out vacuum dipole ($n = 3$) with $>$ 95\% confidence, similar to SN 2015bn and SN 2007ru \citep{Omand2024}.

\begin{figure}
    \centering
    \includegraphics[width=\linewidth]{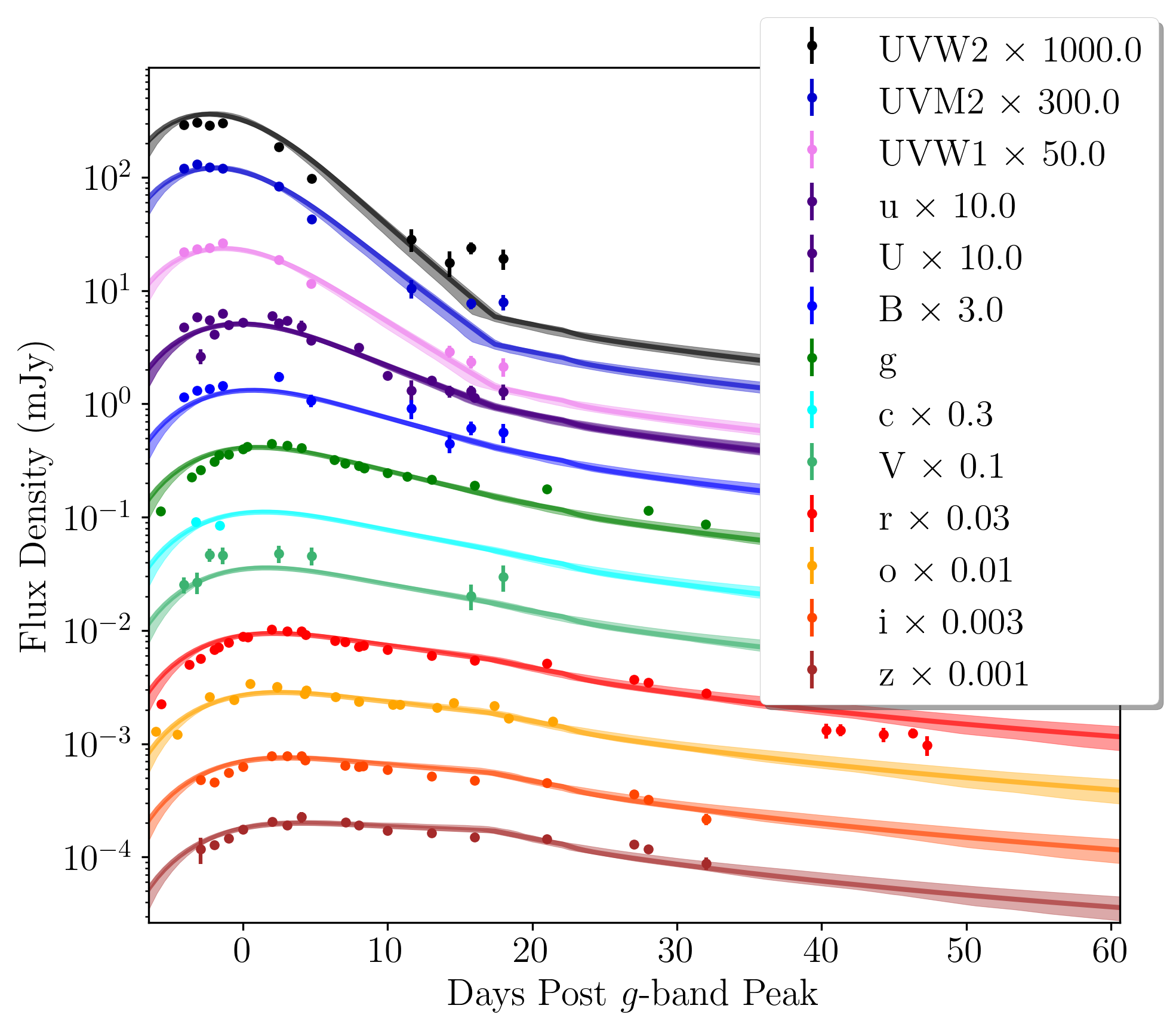}   
\caption{Fitted multi-band light curve for SN\,2021lwz. The solid lines indicate the light curve from the model with the highest likelihood in each band, while the shaded area indicates the 90$\%$ credible interval.}
    \label{fig:lcfit}
\end{figure}

\begin{table*}
	\centering
	\caption{Best fit parameters for the magnetar model obtained on SN\,2021lwz.}
	\label{tab:redbackparams}
	\begin{tabular}{cccccc} 
          Parameter Symbol & Definition & Prior & Best Fit Value & Best Fit Value &  Units\\
           & &  & SN\,2021lwz & iPTF16asu &  Units\\
          \hline
          $\log (L_0)$ & Initial Spin-Down Luminosity & L[40.0,50.0] & $45.42^{+0.28}_{-0.26}$  & 45.45$^{+0.74}_{-0.45}$ & erg s$^{-1}$ \\
          $\log (t_{\rm SD})$ & Spin-Down Timescale & L[2.0,8.0]       & $4.91^{+0.29}_{-0.28}$ & 4.64$^{+0.37}_{-0.61}$ & s \\
          $n$ & Magnetar Braking Index & U[1.5,10]           & $4.56^{+0.80}_{-0.74}$ & 8.41$^{+1.15}_{-1.75}$ & \\
          $M_{\rm ej}$ & Ejecta Mass     & L[0.1,100]          & $0.24^{+0.04}_{-0.04}$ & 0.17$^{+0.03}_{-0.03}$ & $M_\odot$\\          
          $E_{\rm SN}$ &  Supernova Explosion Energy      & L[0.01,2]          & $0.02^{+0.02}_{-0.01}$ & 0.59$^{+0.15}_{-0.07}$ & 10$^{51}$ erg\\
          $\kappa$ & Ejecta Opacity     & U[0.05,0.20]          & $0.06^{+0.01}_{-0.00}$ & 0.06$^{+0.01}_{-0.01}$ & cm$^2$ g$^{-1}$\\
          $\log (\kappa_{\gamma})$ & Ejecta Gamma-Ray Opacity     & L[-4,4]          & $1.65^{+1.54}_{-1.62}$ & 0.27$^{+0.22}_{-0.16}$ & cm$^2$ g$^{-1}$\\
          \hline
	\end{tabular}
\end{table*}

\begin{figure}
    \centering
    \includegraphics[width=0.8\linewidth]{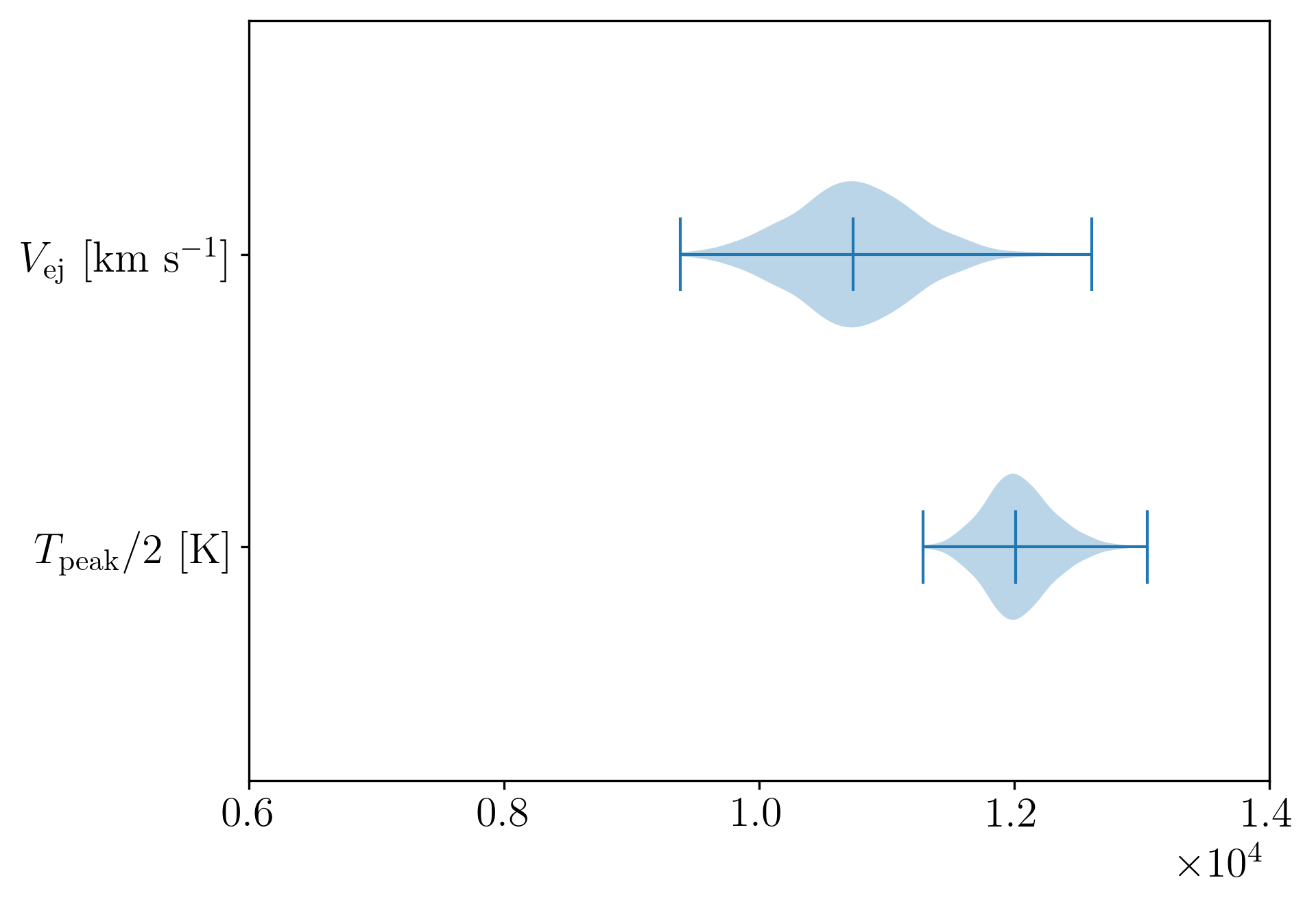}
    \includegraphics[width=0.8\linewidth]{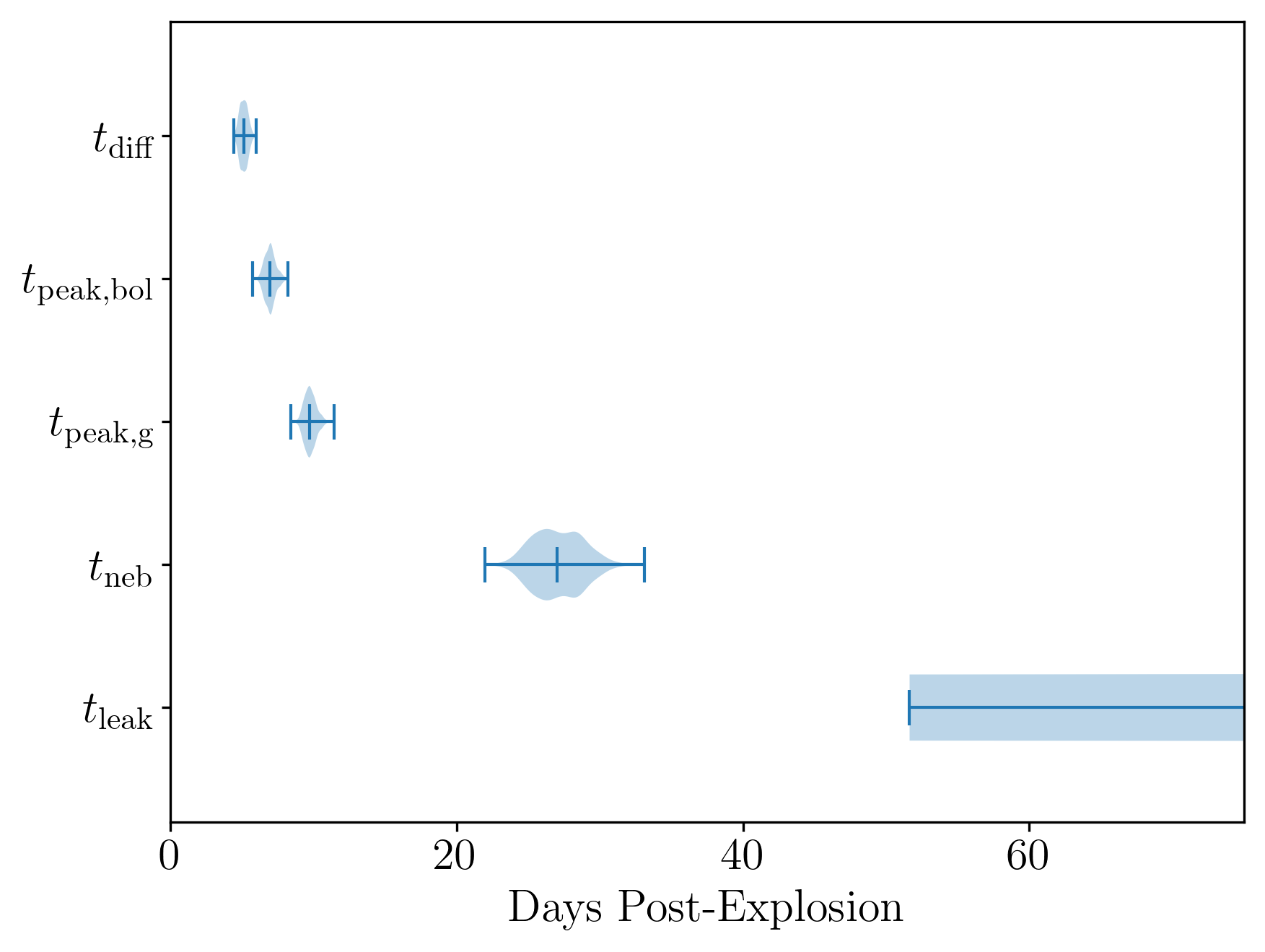}    
    \caption{One-dimensional posteriors of various quantities calculated by the light curve fit.  The median values are indicated by the blue vertical lines within the distribution.  (\textit{Top}) The inferred scale velocity of the ejecta and the temperature of the photospheric at bolometric peak, divided by two to place them on the same scale.  (\textit{Bottom}) The diffusion timescale, gamma-ray leakage timescale, bolometric peak timescale, $g$-band peak timescale, and nebular timescale.)
    }
    \label{fig:violins}
\end{figure}

The distribution of ejecta velocities, peak photospheric temperatures, and several timescales is shown in Figure \ref{fig:violins}.  The final ejecta scale velocity is around 11 000 km s$^{-1}$, which is lower than the inferred photospheric velocity of 15 000 km s$^{-1}$; this implies that most of the material is travelling at much lower velocity than the photosphere at early times, although by peak the photosphere has receded into the bulk of the material.  The photospheric temperature at peak is around 24 000 K, slightly higher than the best-fit model temperature of 20 000 K inferred by the spectral analysis, although not significantly higher. The diffusion time of the supernova,
\begin{align}
    t_{\rm diff} \approx & \sqrt{\frac{3\kappa M_{\rm ej}}{4\pi cv_{\rm ej}}} ,\label{eqn:tdif1} \\
    \approx& \text{ 14.5 days } \left(\frac{\kappa}{0.1 \text{ cm$^2$ g$^{-1}$}}\right)^{0.5} \left(\frac{M_{\rm ej}}{M_\odot}\right)^{0.5}  \left(\frac{v_{\rm ej}}{10 000 \text{ km s$^{-1}$}}\right)^{-0.5} ,
    \label{eqn:tdif}
\end{align}
is very short, only $\sim$ 5 days, due to the low ejecta mass. The bolometric peak timescale is inferred to be $\sim$ 7 days and the $g$-band peak timescale to be $\sim$ 10 days; the discrepancy between these values is consistent with the high photospheric temperature and rapid cooling around peak. From the intrinsic bolometric luminosity we derive $T_{rise, 10\%} = 6.1 \pm 0.3$ days, and $T_{rise, e} = 4.9 \pm 0.3$ days, rest frame. The supernova is expected to enter the nebular phase on the timescale of \citep{Nicholl2017} 
\begin{align}
    t_{\rm neb} \approx & \sqrt{\frac{3\kappa M_{\rm ej}}{4\pi v_{\rm ej}^2}} , \\
    \approx & \text{ 80 days } \left(\frac{\kappa}{0.1 \text{ cm$^2$ g$^{-1}$}}\right)^{0.5} \left(\frac{M_{\rm ej}}{M_\odot}\right)^{0.5}  \left(\frac{v_{\rm ej}}{10 000 \text{ km s$^{-1}$}}\right)^{-1} ,
    \label{eqn:tneb}
\end{align}
which is $\sim$ 27 days post-explosion, or only $\sim$ 17 days post $g$-band peak.  This could imply that the final two observed spectra, at +26d and +29d, are taken during the early nebular phase, even though the spectrum shows no strong evolution between +8 and +29 days.  This could also account for the discrepancy between the model and observed photometric points in the late-phase.  However, it seems more likely that this simplistic estimate of the nebular timescale is not accurate in this circumstance, which could be because of small scale asymmetries or the density profile of the ejecta.  Since Equations \ref{eqn:tdif1}$-$\ref{eqn:tleak} are all derived using a uniform density profile, they could all underestimate the different timescales if the ejecta has a steeper profile \citep[e.g.][]{Suzuki2017}.  Due to the large inferred gamma-ray opacity, the gamma-ray leakage timescale \citep{Wang2015},
\begin{align}
    t_{\rm leak} \approx & \sqrt{\frac{3\kappa_\gamma M_{\rm ej}}{4\pi v_{\rm ej}^2}} , \\
    \approx & \text{ 80 days } \left(\frac{\kappa_\gamma}{0.1 \text{ cm$^2$ g$^{-1}$}}\right)^{0.5} \left(\frac{M_{\rm ej}}{M_\odot}\right)^{0.5}  \left(\frac{v_{\rm ej}}{10 000 \text{ km s$^{-1}$}}\right)^{-1} ,
    \label{eqn:tleak}
\end{align}
is also very long, with a lower limit of $\sim$ 50 days.  This means that a significant amount of non-thermal emission likely does not escape from the supernova over the timescale that we observed it. The inferred lower limit on the gamma-ray opacity is $\sim$ 0.2 cm$^2$ g$^{-1}$, which is higher than the mean value of 0.06$^{+0.07}_{0.05}$ cm$^2$ g$^{-1}$ for SLSNe-I \citep{Gomez2024}, although not a significant outlier. The high gamma-ray opacity here may be inconsistent with a pulsar wind nebula (PWN) with low magnetization \citep{Vurm2021}, which would cause the non-thermal emission to be Comptonized and mostly emitted as hard x-rays and gamma-rays. The likely source of opacity for the PWN is photoelectric absorption, which dominates in the soft x-ray regime (see Figure 4 from \citet{Vurm2021}).  This means that the PWN is likely synchrotron-dominated, and may be a strong radio source at longer timescales \citep[e.g.][]{2018MNRAS.474..573O, Omand2025b}.

These parameters differ significantly from \cite{gomez2022}, and \cite{Gomez2024}, who used the
Modular Open Source Fitter for Transients code \citep[\texttt{MOSFiT};][]{2018ApJS..236....6G}
to fit several luminous supernovae and SLSNe, including SN\,2021lwz, using a pure magnetar and magnetar $+$ nickel model, respectively.  \cite{gomez2022} infers a higher ejecta mass of 0.4 $M_\odot$ and a much lower photospheric velocity of $\sim$ 3000 km s$^{-1}$.  \cite{Gomez2024} infers much higher ejecta and $^{56}$Ni masses of $\sim$ 3 M$\odot$ and $\sim$ 0.8 M$\odot$, respectively, while inferring a much less luminous pulsar of $\sim$ 10$^{40}$ erg s$^{-1}$, meaning the luminosity of the supernova is mostly powered by $^{56}$Ni in that scenario, in contrast to what we find in Section \ref{LCmodeling}.  While the values from \citet{gomez2022} give a similar diffusion time to our inferred parameters, there is a strong tension between their derived photospheric velocity and the value we infer from the spectrum.  They infer a lower value of $2 \times 10^{44}$ erg s$^{-1}$ for the initial pulsar luminosity but a spin-down timescale of $\sim$ 15 days.  The difference in spin-down timescale, as well as their assumption of vacuum-dipole spin down, means the magnetar rotational energy is similar to our inferred parameters.  Any discrepancies likely result from the assumption of constant velocity and decoupling of the magnetar spin-down and ejecta dynamics in the \textsc{mosfit} model, which has been shown via radiation hydrodynamics simulations to be unjustified for high-luminosity magnetars \citep{Suzuki2021}.
 
\subsection{Host SED modelling}\label{host_modeling}

Using the SDSS photometry magnitude (see Table~\ref{tab:host_phot}) and after taking into account the contribution of the Milky-Way reddening, at a redshift $z=0.065$, the absolute magnitude of $M_{g} = -14.47 \pm 0.24 $ mag of the host galaxy of SN\,2021lwz points to a very faint dwarf galaxy.

We model the observed spectral energy distribution (black data points in Figure~\ref{fig:host_sed}) with the software package \program{Prospector} version 1.3 \citep{Leja2017a}.\footnote{\program{Prospector} uses the \program{Flexible Stellar Population Synthesis} (\program{FSPS}) code \citep{Conroy2009a} to generate the underlying physical model and \program{python-fsps} \citep{ForemanMackey2014a} to interface with \program{FSPS} in \program{python}. The \program{FSPS} code also accounts for the contribution from the diffuse gas based on the \program{Cloudy} models from \citet{Byler2017a}. We use the dynamic nested sampling package \program{dynesty} \citep{Speagle2020a} to sample the posterior probability.} We assume a Chabrier IMF \citep{Chabrier2003a} and approximate the star formation history (SFH) by a linearly increasing SFH at early times followed by an exponential decline at late times [functional form $t \times \exp\left(-t/\tau\right)$, where $t$ is the age of the SFH episode and $\tau$ is the $e$-folding timescale]. The model is attenuated with the \citet{Calzetti2000a} model. The priors of the model parameters are set identically to those used by \citet{Schulze2021a}. Figure \ref{fig:host_sed} shows the observed SED (black data points) and its best fit (grey curve). The SED is adequately described by a galaxy template with a mass of $10^{6.66^{+0.35}_{-0.37}}~M_\odot$, and a star-formation rate of $0.04^{+0.03}_{-0.02}~M_\odot\,{\rm yr}^{-1}$. 

\begin{figure}
\begin{center}
\vspace*{2mm}
\centering
\hspace*{0.cm}
\includegraphics[width=8cm,angle=0]{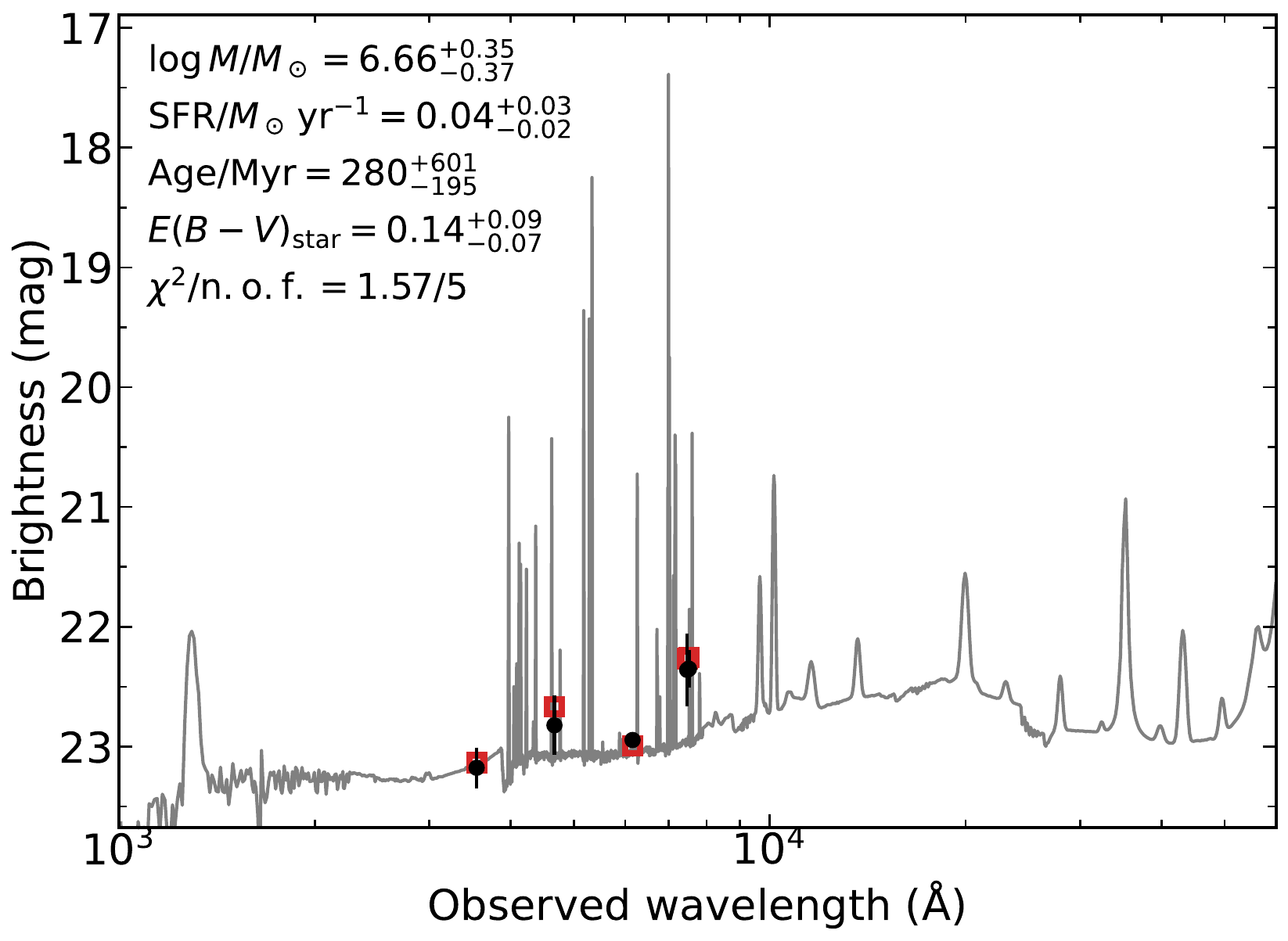}
\vspace*{0.7cm}
\caption{Spectral energy distribution of the host galaxy from 1000 to 60,000~\AA\ (black dots). The solid line displays the best-fitting model of the SED. The red squares represent the model-predicted magnitudes. The fitting parameters are shown in the upper-left corner. The abbreviation `n.o.f.' stands for the number of filters. The five measurements given in Table\ref{tab:host_phot} are considered hence the value of this parameter. Note that the representative wavelengths of the two i-band filters are very similar.
}
\label{fig:host_sed}
\end{center}
\end{figure}

The mass and the star-formation rate are comparable to common star-forming galaxies of that stellar mass \citep[grey band in Figure~\ref{fig:host_mass_sfr};][]{Elbaz2007a}\footnote{The relationship between star-formation and galaxy mass shown in Figure \ref{fig:host_mass_sfr} is based on results from \citet{Brinchmann2004} and is valid at $z\sim0$.}, and they are also similar to those of the host galaxy populations of SNe from the PTF survey \citep{Schulze2018a} albeit at the very low-mass end of the observed distribution. The data relative to USSNe compiled in Table \ref{tab:ussne_list} are given in Appendix\,\ref{ussne_appendix}. 

As a consistency check, we computed a plot similar to the one shown in Figure~\ref{fig:host_sed}, without taking into account the emission lines. We find that the galaxy mass is almost unchanged. On the other hand the SFR departs from a factor 10 from the one discussed in the text and shown in Figure~\ref{fig:host_mass_sfr}. The new reduced chi-square is also higher. All these points make us confident that, despite the low number of data points, the fit shown in Figure~\ref{fig:host_sed} including the emission-lines is better suited than a fit that would not include these lines.

\begin{figure}
\begin{center}
\vspace*{2mm}
\centering
\hspace*{0.cm}
\includegraphics[width=9cm,angle=0]{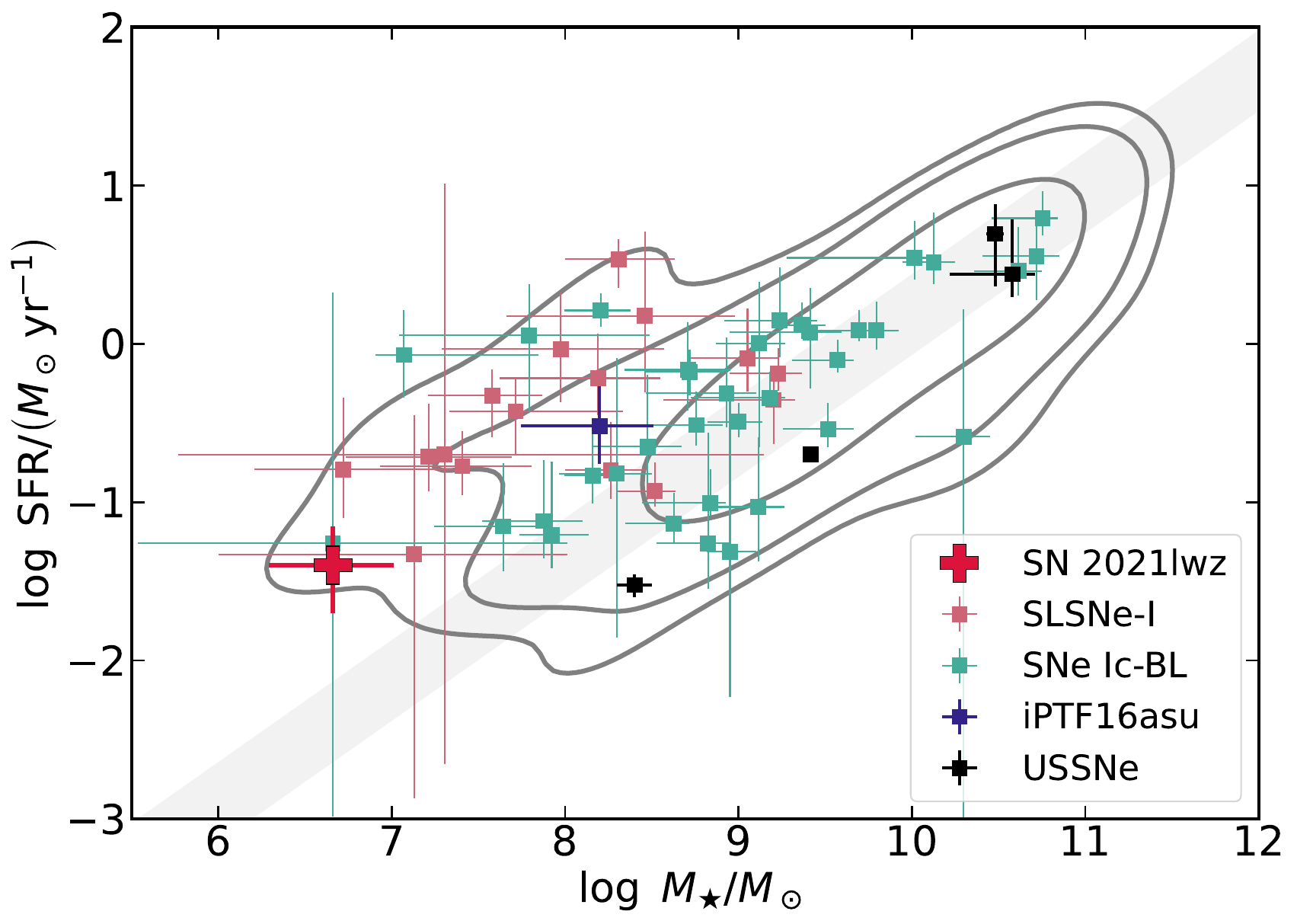}
\vspace*{0.7cm}
\caption{The star-formation rate and stellar mass of SN\,2021lwz's host galaxy of (dark red) in context of the core-collapse supernova sample from Palomar Transient Factory. From the PTF sample, the host properties of iPTF16asu, SNe Ic-BL and SLSNe-I at $z<0.2$ are highlighted. SN\,2021lwz's host lies in the expected parameter space but at the low-mass end of observed mass distribution. Its specific star-formation rate (SFR / mass) is larger than that of the main sequence of star-forming galaxies from SDSS at $z\sim0$ (grey band; \citealt{Elbaz2007a} and references therein) by $\sim$ 1.5 dex, meaning that is a starburst galaxy. The contours indicate the region encircling 68, 90, and 95$\%$ of the sample.
}
\label{fig:host_mass_sfr}
\end{center}
\end{figure}

\section{Discussion} \label{discussion}

Based on our previous modelling and analysis it is not clear what is the nature of SN\,2021lwz. This very fast evolving explosion (t$_{\mathrm{rise}} \sim 7$ days, rest-frame), with a low ejecta mass
($M_{\rm ej} \sim 0.2$ $M_\odot$), is drawn from a very low-brightness ($M_{g} = -14.47 \pm 0.24 $ mag), low mass ($10^{6.66^{+0.35}_{-0.37}}~M_\odot$) dwarf galaxy of SFR ($0.04^{+0.03}_{-0.02}~M_\odot\,{\rm yr}^{-1}$) about ten times larger than the typical star-forming galaxies. 

\begin{table*}
	\centering
	\caption{Data on additional sources used for comparison with SN\,2021lwz. 
 }
	\label{tab:slsn_comparison}
	\begin{tabular}{llllllllll} 
          \hline
          Object &R.A.&Decl.& Type & z&  t$_{\mathrm{rise,rf}}$ $^{(a)}$  & band &  $m$$^{(b)}$ & $M$ $^{(c)}$  & References \\
          Name   &    &     &      &  &   [days]              &        &[mag]&  [mag] & \\
          \hline
          SN\,2018bgv &11:02:30.29&+55:35:55.79 & SLSN-I     & 0.08 &   $ \sim 10$           & g & 17.06 & -20.71  & 1 \\ 
          AT\,2020xnd &22:20:02.03&-02:50:25.27& LFBOT       & 0.2442&  $ < 5.6$             & r & 19.56 & -20.72  & 2 \\ 
          SN\,2014ft  &23:33:27.95&+33:38:46.14& USSN        & 0.1868   & $\sim 5.9^{(d)} $  & r & 19.62 &  -20.06 & 3 \\          
          iPTF\,16asu &12:59:09.28&+13:48:09.19& SL Ic-BL    &  0.187&  $\sim 4 $            & g & 19.14 & -20.53  & 4, 5 \\          
          SN\,2018gep & 16:43:48.22 &+41:02:43.37 & SL Ic-BL & 0.031875 & 5.4$\pm$0.5        & g & 15.93 & -19.84  & 6 \\          
          SN\,1998bw &19:35:03.310 &-52:50:44.81& Ic-BL      & 0.008499 & $\sim$ 17          & V & 13.60 & -19.33  & 7, 8  \\  
          SN\,2021lwz &09:44:47.40&+34:42:44.25& ?           & 0.065&   $\sim $ 7            & g & 17.25 & -20.08  & This work\\   
          \hline            
    \end{tabular}    
\tablefoot{
\tablefoottext{a}{t$_{\mathrm{rise,rf}}$ is the estimated rising time from explosion in the rest-frame.}
\tablefoottext{b}{difference magnitude.}
\tablefoottext{c}{absolute magnitude.}
\tablefoottext{d}{here we refer to the second and largest peak as discussed in \cite{2018Sci...362..201D}.}
} 
\tablebib{
(1)~\citet{2020ApJ...901...61L}; 
(2)~\citet{2021MNRAS.508.5138P}; 
(3)~\citet{2018Sci...362..201D};
(4)~\citet{2017ApJ...851..107W};
(5)~\citet{2022ApJ...928..114W}; 
(6)~\citet{2021ApJ...915..121P}; 
(7)~\citet{1998Natur.395..670G}; 
(8)~\citet{2001ApJ...555..900P}.
}
\end{table*}    

To discuss SN\,2021lwz in a broader context, in Figure~\ref{fig:lcs_comparisons} we show the rest-frame light curves of the different types of transients listed in Table\,\ref{tab:slsn_comparison}; namely SN\,2018bgv (SLSN-I), SN\,1998bw (peculiar Ic-BL), AT\,2020xnd (LFBOT), SN\,2014ft (USSN), iPTF16asu (SL Ic-BL) and SN\,2018gep (SL Ic-BL).
As it is done for SN\,2021lwz, if the photometry is not already corrected for the extinction contribution from our Galaxy, it is done using \cite{2011ApJ...737..103S}. 
As much as possible the filters were chosen by taking into account the redshift of each source so that the rest-frame wavelength ranges covered overlap strongly with each other. 

From the comparison of their rising time parameters given in Table\,\ref{tab:slsn_comparison} all these transients, except SN\,1998bw, reach their maximum light in about 10 or less days.
SN\,2018bgv appears slightly more energetic than SN\,2021lwz. On the other side of the absolute magnitude range, SN\,1998bw is the less energetic transient. SN\,2018bgv and SN\,1998bw whose rest-frame rising times are about twice or more the rising time of SN\,2021lwz, both have smooth light curves decreasing slightly faster than the one of SN\,2021lwz. The light curves of iPTF\,16asu, SN\,2014ft, AT\,2020xnd SN\,2028gep, SN\,2014ft and SN\,2021lwz, all strongly overlap with each other before and up to a few days after maximum light. A few days after peak, the light curves of AT\,2020xnd, SN\,2014ft and SN\,2018gep decrease faster than the light curves of SN\,2021lwz, though. All in all, the ligth curve of iPTF\,16asu is the one looking more similar to the one of SN\,2021lwz. 

\begin{figure}
\begin{center}
\vspace*{2mm}
\centering
\hspace*{0.cm}
\includegraphics[width=9cm,angle=0]{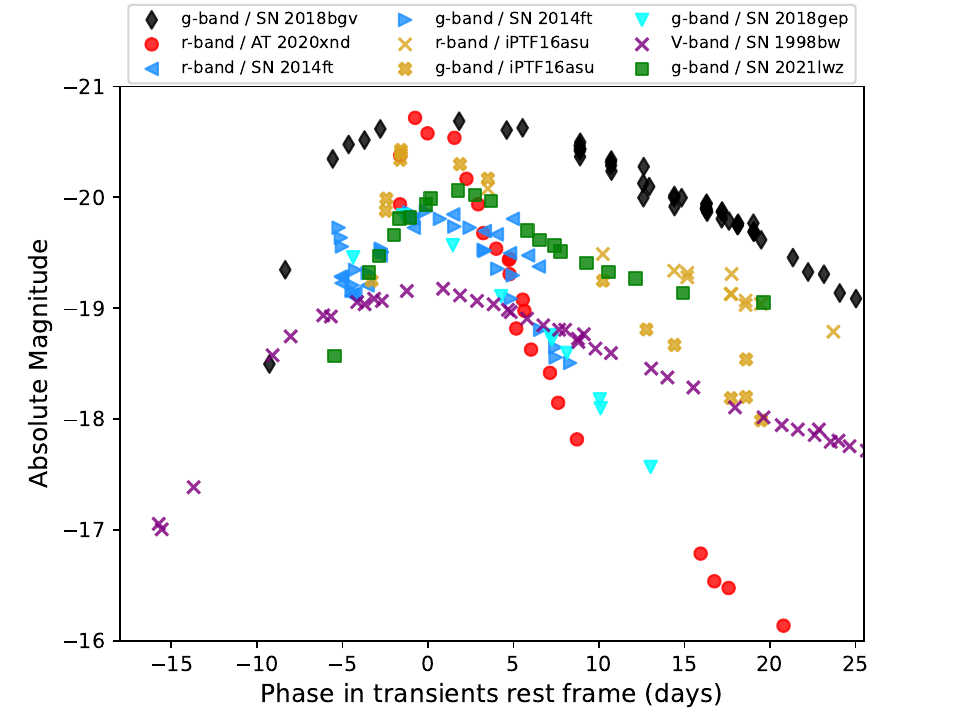}
\vspace*{0.7cm}
\caption{Rest-frame light curves of the fast rising transients (t$_{\mathrm{rise,rf}} \sim < 10 $ days) and of SN\,1998bw (t$_{\mathrm{rise,rf}} \sim 20 $ days) listed in Table\,\ref{tab:slsn_comparison}. The symbols and colours identify the sources, and the filters used are defined in the legend. The filters were chosen to take into account the redshift of each source for a fairer comparison, except for iPTF\,16asu and SN\,2014ft for which g-band data are shown in addition to r-band data to show the rapid increase of these transients. 
The magnitudes are corrected for the Milky-Way extinction, but not for any contribution from the hosts. 
}
\label{fig:lcs_comparisons}
\end{center}
\end{figure}

\subsection{H-poor SLSNe} \label{comp_slsne}

SN\,2018bgv is the fastest evolving ($\sim 10$ days rest-frame) Type-W SLSN-I reported in the literature. It peaks close to $\sim - 20.7 $ mag. It is therefore very interesting for comparisons with SN\,2021lwz. While we didn't model it with our magnetar model, the light curve and a spectral sequence have been studied in detail by \cite{2020ApJ...901...61L}. From black-body fits analysis of the light curve the authors report a photosphere temperature close to 11,000 K and a black-body radius close to 2$\times 10^{15}$ cm, a few days after peak, i.e. a photopshere with a radius about twice larger than, and a temperature about similar to, those of SN\,2021lwz photosphere at similar phases. Close to peak the velocity of the photosphere of SN\,2018bgv appears to be about 1.5 times the velocity estimated for SN\,2021lwz. A fit of the light curve with the magnetar spin-down model using \textsc{mosfit} 
infers an ejecta of mass, $M_{\rm ej} \sim 1.3$ $M_\odot$, i.e. about 5 times higher than the one of SN\,2021lwz as from our modelling estimate, hence suggesting the conditions behind the explosions of SN\,2018bgv and SN\,2021lwz are different. 

As discussed in Section\,\ref{snid_analysis} SN\,2018bgv provides a relatively good match with SN\,2021lwz close to maximum light, but this is obtained at a lower redshift z$\sim 0.055$ than the redshift of SN\,2021lwz. This is mainly because SNID tries to adjust the archetypal W-shape feature attributed to  \ion{O}{ii}, before and close to maximum light, observed in many Type I SLSNe. Indeed, and as discussed in Section\,\ref{spec_model}, for H-poor SLSNe the typical absorption spectral features expected to be observed before maximum light, between 4000 and 5000 \AA\, \citep[see][Figure 1]{ktr23}, in Type 15bn (e.g. \ion{Fe}{iii}) and in Type W (e.g. \ion{O}{ii}), are not detected in the pre-maximum spectra of SN\,2021lwz. Instead,  \ion{C}{ii},  \ion{O}{i}, \ion{Na}{i}, \ion{Mg}{ii}, \ion{Si}{ii}, \ion{Ca}{ii}, \ion{Ti}{ii} and \ion{Fe}{ii} 
are included to reproduce the shape of these spectra, and this despite the fact that from the estimates of the photosphere temperature $T_{\rm phot} \approx 20\,000$\,K and the photosphere velocities $10000 < \nu_{\rm phot} < 15000$ km$/$s before maximum light, one would expect features from a W-Type SLSN-I \citep[see][Figure 7]{ktr22}. We point out, however, that the conditions for these lines to appear are likely limited to a more stringent range of temperature $T_{\rm phot} \sim 14\,000 - 16\,000$\,K once spectral modelling effects, which are not included in the SYN++ modelling, are taken into account \citep[see][]{Saito2024}. However, the study provided by \cite{aamer2025}, obtained on a sample larger than the one used by \cite{ktr23} suggests a continuum of spectral properties, hence that the differences of spectral profiles in type I SLSNe may not correspond to physically different type of events. In this larger sample \ion{Fe}{ii} lines appear to be an important spectral marker formed near the photosphere as it is also found for SN\,2021lw.  

\subsection{LFBOTs} \label{comp_fbot}

The LFBOT AT\,2020xnd and SN\,2021lwz are peaking at about similar absolute magnitudes (see Figure~\ref{fig:lcs_comparisons}), but AT\,2020xnd is a faster evolving event than SN\,2021lwz, mainly because it has a higher initial pulsar luminosity, assuming a magnetar model. 
Two spectra of AT\,2020xnd at phases close to +5 days and +7 days after maximum light are publicly available in the WISeREP repository \footnote{URL to AT\,2020xnd public spectra in WISeREP, {\tt https://www.wiserep.org/object/18199}.}. From the point of view of the photosphere evolution, the two very smooth spectra, which are continuum blue flux dominated do not show any strong absorption features, unlike those seen in the spectra of SN\,2021lwz after maximum light (see Figure~\ref{SpectralEv}). This observational fact, combined with the analysis of their light curves using the magnetar model \citep[for AT\,2020xnd see Table 2 in][]{Omand2024}, suggests that AT\,2020xnd and SN\,2021lwz differ in nature, and do not pertain to the same class of transients. This conclusion is sustained by a visual comparison of some spectra of SN\,2021lwz with some of SN\,2020xnd and SN\,2018cow as shown in Figure\,\ref{fig:spectra_fbots_comp}. The spectral evolution of SN\,2021lwz show absorption features stronger than those observed in the spectra of AT\,2020xnd and AT\,2018cow which, in comparison, look almost featureless.

\begin{figure}
\begin{center}
\vspace*{2mm}
\centering
\hspace*{0.cm}
\includegraphics[width=9cm,angle=0]{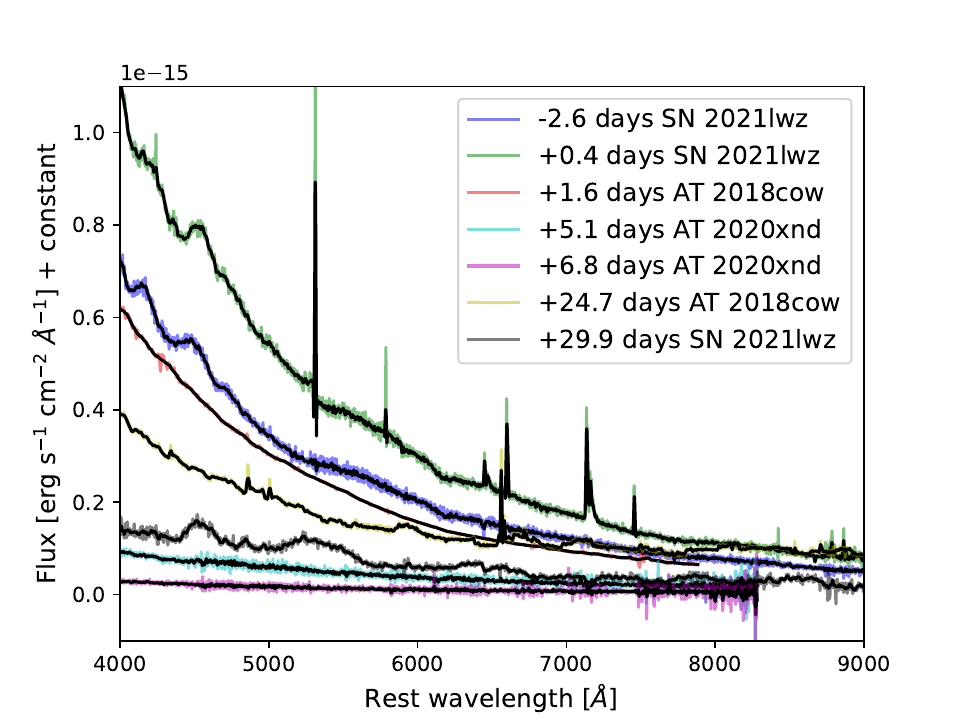}
\vspace*{0.7cm}
\caption{Some spectra of FBOTs AT\,2020xnd and AT\,2018cow are shown for comparisons with some of the spectra obtained on SN\,2021lwz (see Figure~\ref{SpectralEv}). All the spectra have been smoothed using a Savitzky-Golay low pass filter.}
\label{fig:spectra_fbots_comp}
\end{center}
\end{figure}

\subsection{USSN} \label{comp_ussn}

SN\,2014ft has been observed, studied and classified by \cite{2018Sci...362..201D} as a USSN that likely formed a compact neutron star binary. It is recognised as the first ever detection of a USSN \citep[][]{2019ApJ...882...93H}, while previous studies of fast evolving Type Ic-BL SNe like SN\,2005ek \citep[][]{2013ApJ...774...58D} and SN\,2010X \citep[][]{2010ApJ...723L..98K} pointed-out the existence of this class of objects. Posterior to these works, SN\,2019dge (or ZTF18abfcmjw) was added to this class of transients as discussed by \cite{2020ApJ...900...46Y}. More recently SN\,2023zaw has been shown to be a fast evolving USSN of very low ejecta mass ($M_{\rm ej} \sim 0.07$ $M_\odot$) \citep[][]{das2024,moore2025}. 

SN\,2014ft is modelled using the \textsc{Redback} \citep{Sarin_redback} magnetar-driven supernova model in \cite{Omand2024}. SN\,2021lwz and SN\,2014ft have very similar light curves both peaking at absolute magnitudes close to -20.1 mag (Figure~\ref{fig:lcs_comparisons}), and similar fast rising times of order 6--7 days (see Table\,\ref{tab:slsn_comparison}. Assuming the magnetar-driven supernova model is appropriate to describe both explosions, the posterior parameters from the light curve modelling both infer very low ejecta of the same order ($\sim 0.15-0.25 $ $M_\odot$), and supernovae explosion energies of the same order ($\sim 2-6 \times 10^{49}$ erg ). The main difference is that SN\,2021lwz has an initial spin-down luminosity about 100 times higher than the one of SN\,2014ft, and a spin-down timescale about 10 times faster. The braking index of both explosions being of the same order ($\sim 4.5$), this results in SN\,2021lwz having a total rotational energy an order of magnitude higher than SN\,2014ft.

During the first 30 hours after explosion SN\,2014ft showed \ion{He}{ii} $\lambda$4686 \AA\,lines in 3 spectra \citep[see Figure3 A in][]{2018Sci...362..201D}. 
Such early spectra are not available for SN\,2021lwz therefore we don't know if a similar feature was present very early in its photosphere. A similar feature is not detected in the earliest spectra available on SN\,2021lwz obtained about 5 to 7 days after the explosion (see Figure~\ref{SpectralEv}). Hints of \ion{He}{i}, but not of \ion{He}{ii}, are only suggested by the modelling analysis in the post peak spectra of SN\,2021lwz. 
At maximum light the spectra of SN\,2021lwz and SN\,2014ft show common ionization atomic lines from \ion{Ca}{ii}, \ion{Mg}{ii}, \ion{Fe}{ii}, \ion{Ti}{ii}, \ion{Si}{ii}, \ion{C}{ii}, and \ion{O}{i} \citep[see our SYN++ analysis and Figure 5B in][]{2018Sci...362..201D} but absorption lines of \ion{P}{iii} (while not unambiguous) and \ion{Na}{i} observed in the spectrum of SN\,2021lwz are not detected in the spectrum of SN\,2014ft. The distinct temperatures of the photosphere estimated for each supernova close to maximum light, with $T_{\rm phot} \approx 20 000$\,K for SN\,2021lwz (this work), and $T_{\rm phot} \approx 11 000$\,K for SN\,2014ft \citep[Figure 4B in][]{2018Sci...362..201D}, may explain this difference. 
Before and near maximum light SN\,2021lwz has a steeper continuum than SN\,2014ft, showing that it is hotter and likely coming from a more energetic event, an inference consistent with the one provided by the modelling of the light curves. In the near-maximum spectra, around 5000 \AA, SN\,2014ft shows a `basin shaped' feature similarly to SN\,2015bn, while this is missing from the spectra of SN\,2021lwz, here again most probably because of the difference in the temperatures. The post-maximum spectral lines are somewhat more similar, but SN\,2014ft is still cooler than SN\,2021lwz.

Finally, we point out that, contrary to the high specific SFR of the host of SN\,2021lwz, the specific SFR of USSNe shown in Figure \ref{fig:host_mass_sfr} is generally lower than, or about, the average expected value for typical star-forming galaxies (grey band). The stellar masses of the USSNe galaxies are also 2 to 4 orders of magnitudes higher than the mass of SN\,2021lwz's host. The statistics is low with only 4 hosts of USSNe, though.

All the arguments put together it is not clear if SN\,2021lwz has been ultra-stripped or not, the main argument in favour of an ultra-stripped process being that it has a low ejecta mass, as does have SN\,2014ft. 

\subsection{Typical H-poor stripped envelope SNe} \label{comp_sesn}

Hydrogen-poor, stripped envelope, core-collapse supernovae (SESNe), of Types Ib, Ic and Ic-BL, are well studied in the optical \citep[e.g.][]{2011ApJ...741...97D,2019A&A...621A..71T} the NIR \citep[e.g.][]{2022ApJ...925..175S}, and at radio wavelengths \citep[][]{2024ApJ...976...71S}. As has been discussed previously our post-peak optical spectral modelling hints at a marginal detection of \ion{He}{i}, and the possible 1.0830 $\mu$m line. Possible connections and a continuity between typical Type Ib and Ic has been discussed with no clear results about residual detections of  \ion{He}{i} in optical spectra \citep[e.g.][]{2012MNRAS.422...70H,2014ApJ...792L..11P,2018MNRAS.478.4162P,2020A&A...642A.106D,2021ApJ...908..150W}. More recently, \cite{2022ApJ...925..175S} presented the analysis of 75 NIR spectra of 34 SESNe with the aim to discuss NIR strong  \ion{He}{i} lines: $\lambda 1.0830$ $\mu$m and $\lambda 2.0581$ $\mu$m. In all spectra, except a few outliers, most of the absorption lines features observed close to 1\,$\mu$m can be explained by a combination of strong lines (C I $\lambda 1.0693$ $\mu$m,  \ion{He}{i} $\lambda 1.0830$ $\mu$m, and Mg II $\lambda 1.0927 \mu$m), and the authors concluded that there is no distinct difference between He-poor and He-rich groups in this wavelength range. On the other hand, by analysing the 2\,$\mu$m features the authors showed one can clearly discriminate He-rich (optical IIb/Ib) from He-poor (optical Ic/Ic-BL) SESNe and computed SESNe templates using a Principal Component Analysis. The +13 days templates of such He-rich and He-poor SNe are shown in Figure~\ref{SpecNIR} for illustration. A comparison with the +12.6 days NIR spectrum of SN\,2021lwz, which optical spectral modelling points out to a Type Ic-like, disfavours the presence of residual  \ion{He}{i}. Actually, in the vicinity of the  \ion{He}{i} 2\,$\mu$m feature, the spectrum of SN\,2021lwz looks flatter than the He-poor template, as looks the spectrum of the peculiar Type Ic-BL, SN\,1998bw. 

Additional arguments making SN\,2021lwz peculiar with respect to SESNe are the following. It is more luminous at peak than the majority of these objects (see for example our plot in Figure~\ref{fig:mrpeak_vs_trise} for a direct comparisons with Type Ic and Ic-BL absolute magnitudes). It has a low mass ejecta (as inferred from our multi-band modelling), $M_{\rm ej} \sim 0.24^{+0.04}_{-0.04}$ $M_\odot$, which is in contradiction with typical $^{56}$Ni ejecta masses of M$_{\rm{Ni}} \approx 0.2$ and 0.5 $M_\odot$ for SNe Ib/c and SN Ic-BL, respectively \citep[][]{2011ApJ...741...97D}, as well as with the ejecta masses expected for these two Types of SNe, of order M$_{\rm{ej}} \approx 2$ and 5 $M_\odot$, respectively \citep[][]{2011ApJ...741...97D, 2019A&A...621A..71T}. More recently \cite{2023ApJ...955...71R} studied 191 stripped-envelope SNe, including 50 SNe Ib and 96 SNe Ic, of which 26 are SNe Ic-BL. Compared to that SN Ib/Ic/Ic-BL sample, SN\,2021lwz is at least ten times more luminous at peak than the average (2.7 and 4.5 $\times 10^42$ erg s$^{-1}$ for Ib and Ic/Ic-BL, respectively). 
Its ejecta mass is about ten times lower than the average (2.6 $M_\odot$), and also lower than the minimum ejecta mass (0.5 $M_\odot$), making it a low occurrence event in the bottom tail of the distribution.

\subsection{Comparisons with peculiar Type Ic-BL SN\,1998bw}  \label{comp_1998bw}

Before and close to maximum light the photosphere of SN\,2021lwz looks more similar to a normal Type Ic supernova than to any other Type, with  \ion{C}{ii}, \ion{O}{i}, \ion{Na}{i}, \ion{Mg}{ii}, \ion{Si}{ii}, \ion{Ca}{ii}, \ion{Ti}{ii} and \ion{Fe}{ii} lines identified in its earliest spectrum (see Section\,\ref{spec_model}), and the velocity of the spectral features and of the photosphere are close to those of a Type Ic. Slightly broader optical spectral features observed after peak suggest that SN\,2021lwz could have evolved closer to a Type Ic-BL. Actually, before maximum light, the SYN++ modelling shows that the photosphere velocity of SN\,2021lwz is not extraordinarily high ($\sim 15000$ km/s), and the broadness parameter $aux$ (see values displayed in Table~\ref{tab:locparams}) are pointing on a normal Type Ic. After maximum light, the input $aux$ parameter value had to be higher to adjust the formation of several broader spectral features, making SN\,2021lwz may be closer to the limit of a Ic-BL. But still, after peak, the best match is obtained with LSQ12dlf which is a SLSN-I showing velocity lines below the ~30000 km/s seen in classic SNe Ic-BL. The match with the +14 days and +16 days spectra of SN\,1998bw are not too bad above $\lambda \sim 5300 \AA$ but are relatively poor at lower wavelengths. Some attempts to fit the -3.4 days Keck spectrum of SN\,2021lwz with early spectra of SN\,1998bw proved unsuccessful. Some differences between the NIR spectra of both transients can also seen in Figure~\ref{SpecNIR}. All in all, SN\,2021lwz looks more like a SLSN showing the Ic-SL Type after peak.

\subsection{Comparisons with fast evolving luminous Type Ic-BL iPTF16asu}  \label{comp_iptf16asu}

iPTF\,16asu whose origin is not yet well understood has been first discussed by \cite{2017ApJ...851..107W}. It is a luminous, rapidly evolving, stripped-envelope SN peaking at $M_{\rm{g}} \sim -20.5$ mag. With a rest-frame rise time estimated as $\sim$ 4 days it appears as a faster evolving event than SN\,2021lwz. Contrary to SN\,2021lwz which looks like a Type Ic SN before peak, iPTF\,16asu shows a relatively featureless blue continuum, but contrary to SN\,2021lwz, iPTF\,16asu shows Ic-BL features after peak (see some spectra shown in Figure\,\ref{fig:spectra_iptf16asu} for comparison).

In Appendix\,\ref{16asu_appendix} we present the fit of the multi-band light curve of iPTF\,16asu using the same software, model, and priors as for SN\,2021lwz (see Section \ref{MLCmodeling} for details). The light curve fit is shown in Figure \ref{fig:lcfit_asu16} and posterior is shown in Figure \ref{fig:corner_asu16}. From our modelling, the estimated SN explosion energy of iPTF\,16asu is about 30 times stronger than the one of SN\,2021lwz, which could explain the optical spectral differences before peak. 

More in depth light curve modelling of iPTF\,16asu has been discussed by \citet{2019MNRAS.489.1110W}. In Section \ref{LCmodeling} we discuss the difficulty to model the light curve of SN\,2021lwz using the Arnett model, similarly the analysis proposed by these authors underlined the difficulty to model the light curve of iPTF\,16asu with radioactive heating ($^{56}$Ni-only), and also the magnetar $^{56}$Ni models. Our modelling of iPTF\,16asu and SN\,2021lwz includes a fraction of Nickel in the ejecta mass \citep[see Equation 9 in][]{Omand2024}, which given the total ejecta mass estimate $M_{\rm ej} \approx 0.17$ $ M_\odot$ and $M_{\rm ej} \approx 0.24$ $ M_\odot$, of each object, respectively, means the Nickel mass would be excessively low. Energetically speaking, we find that these low ejecta masses are both associated to rotational energies of the same order ($E_{\rm rot}= \frac{n-1}{2} L_{0} t_{\rm{SD}}$). The light curves of both events are very similar and peak at absolute magnitudes close to $\sim$ -20.3 mag. They also have fast rising times sharing a common parameter space ($t_{\rm rise} \lesssim 7 $ days; see Table\,\ref{tab:slsn_comparison} and Figure~\ref{fig:mrpeak_vs_trise}). The g-band data plotted in Figure~\ref{fig:lcs_comparisons} show the steep increase of iPTF\,16asu, while the r-band data show a decreasing light curve very similar to the one of SN\,2021lwz. 

The interpretation of the results obtained by \citet{2019MNRAS.489.1110W}, and additional modelling of the light curves has been proposed by \cite{2022ApJ...928..114W}. The construction of the theoretical bolometric light curve is revised and the authors find that the magnetar plus $^{56}$Ni model can now account for the photometry of iPTF\,16asu with a peak luminosity inferred from the theoretical light curve of $\sim 1.06 \times 10^{44}$\,erg. The best fit value for the ejecta mass is $M_{\rm{ej}} = 0.19$ $M_\odot$ and the best fit parameter of the Nickel mass is $M_{\rm{Ni}} = 0.029$ $M_\odot$. Some results very close to the best fit parameters obtained with our modelling of the multi-light curve of SN\,2021lwz. All these results possibly make SN\,2021lwz part of a class of rare rapidly evolving ($t_{\rm rise, rest frame} < \sim 1$ week), luminous type Ic-BL SNe ($M_{g} \sim -20$ mag or slightly higher), like iPTF\,16asu \citep[see][]{2022ApJ...928..114W}, even though, SN\,2021lwz is an intermediate velocity ($v_{\rm phot} \sim 15000$ km s$^{-1}$) type Ic/Ic-BL object, while iPTF\,16asu is a high velocity type Ic-BL after peak. 

\begin{figure}
\begin{center}
\vspace*{2mm}
\centering
\hspace*{0.cm}
\includegraphics[width=9cm,angle=0]{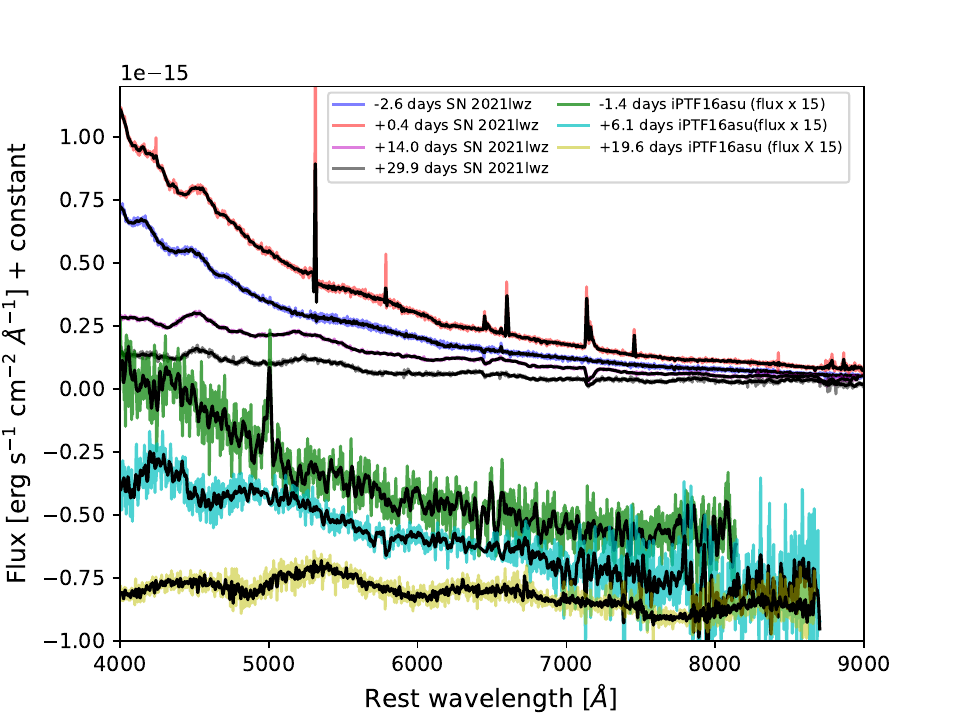}
\vspace*{0.7cm}
\caption{Some spectra of iPTF\,16asu and SN\,2021lwz are shown for comparisons with some of the spectra obtained on SN\,2021lwz (see Figure~\ref{SpectralEv}). All the spectra have been smoothed using a Savitzky-Golay low pass filter.}
\label{fig:spectra_iptf16asu}
\end{center}
\end{figure}

\subsection{Comparisons with fast evolving luminous Type Ic-BL SN\,2018gep}  \label{comp_2018gep}

SN\,2018gep \citep[see][]{2019ApJ...887..169H, 2021ApJ...915...80L, 2021ApJ...915..121P} is another rapidly rising (1.4 $\pm$ 0.1 mag hr$^{-1}$ during the first hours just after explosion), stripped envelope SN, spectroscopically classified as a Ic-BL \citep[see][]{2019ApJ...887..169H}. As a Ic-BL SN its light curve shows similarities with the one of iPTF\,16asu \citep[see][and Figure\,~\ref{fig:lcs_comparisons}]{2021ApJ...915..121P}. With a total fast rise time, $t_{\rm rise, V} = 5.6 \pm 0.5$ days after explosion and a peak absolute magnitude $M_{V} = -19.53 \pm 0.23$ mag it shares a common locus with iPTF\,16asu and SN\,2021lwz on the extreme side of the distribution of SN Ic-BL in this parameter space \citep[see][and our Figure~\ref{fig:mrpeak_vs_trise}]{2021ApJ...915..121P}.
The spectra obtained during the first three days after explosion are blue dominated spectra showing smooth absorption features of highly ionized elements (CIV, O III, CIII) and a very hot photosphere decreasing from $\sim 40,000$ K to $\sim 20,000$ K. Then the W-Type signature attributed to O II and C II absorption lines observed in many Type I SLSNe before and around peak is visible in the spectra of SN\,2018gep obtained during the fourth day after explosion \citep[see Figure 11 in][]{2019ApJ...887..169H}. This marks the main difference with SN\,2021lwz spectra in which this signature is interpreted differently given the assumed redshift of SN\,2021lwz (see Section\ref{spec_model}).

All in all, all these comparisons suggest that SN\,2021lwz has characteristics of, and is probably closely related to, many of these classes, and is therefore most likely some kind of transitional event.

\subsection{Prospects and final remarks}

An interesting test of the magnetar scenario for luminous, fast evolving objects is non-thermal emission. These objects could emit luminous X-ray \citep{Kotera2013, Metzger2014, Murase2015} and radio emission \citep{Murase2015,2018MNRAS.474..573O, Eftekhari2021, Murase2021, Omand2025b}, similar to PTF\,10hgi \citep{Eftekhari2019, Law2019, Mondal2020} or SN\,2012au \citep{Stroh2021} after a few years, but on timescales of a few weeks to a year.  There could also be bright infrared emission on short timescales due to dust heated by the PWN \citep{2019MNRAS.484.5468O}, although this emission would likely fade quickly as the dust cools.  The dominant non-thermal component can be predicted, in principle, from the nebular spectrum of the supernova, since line emission is sensitive to the underlying PWN spectrum \citep{Omand2023}; however, these models are not currently capable of reproducing nebular spectra on timescales of less than a year.

The multi-dimensional structure of these magnetar-driven supernovae could show strong clumping due to Rayleigh-Taylor instabilities between the nebula and ejecta \citep{Blondin2017, Suzuki2017}.  When $E_{\rm rot} > E_{\rm exp}$, such as in SN\,2021lwz, the shock can break through the ejecta \citep{Blondin2017}, leading to a decrease in the ejecta acceleration efficiency.  This process, known as blowout, has been inferred to happen in the Crab Nebula \citep{Omand2025a}.  Clumpy ejecta can cause polarization if the clump size is large enough \citep{Tanaka2017}, but it is currently unknown how to determine the clump size distribution from the inferred engine properties of the supernova.

It has been proposed by \cite{2018PASA...35...32Z} that Type I SLSNe and Type Ic-BL could be produced by jets. If so, they would both show strong polarization from early times. We point out that for SN\,2021lwz our non-detections contribute to rule out this scenario. Additionally, if the energy was injected into the ejecta early, most of it would accelerate the ejecta instead of contributing to the luminosity \citep[][]{Suzuki2021}.    

\section{Conclusions} \label{conclusions}

SN\,2021lwz is a peculiar luminous supernova at a redshift z=0.065 of absolute magnitude $M_{g} \approx$ - 20.1 (AB system) discovered in 
a very low-brightness ($M_{g} = -14.47 \pm 0.24 $ mag), low mass ($10^{6.66^{+0.35}_{-0.37}}~M_\odot$) dwarf galaxy of SFR ($0.04^{+0.03}_{-0.02}~M_\odot\,{\rm yr}^{-1}$) about ten times larger than the typical star-forming galaxies. 

An extensive analysis and modelling of the photometry, spectroscopy and polarimetry data of the SN, followed by comparisons with SNe of well know types and with rarer transients, consistently leads to the following results:

\begin{itemize}
    \item SN\,2021lwz shows a rapidly evolving and luminous optical light curve. The bolometric light curve rises in about $7$ days to a peak luminosity of about $5 \times 10^{43}$ erg/s, at a rate of 0.2 mag day$^{-1}$ close to the peak. From the intrinsic bolometric luminosity we derive $T_{rise, 10\%} = 6.1 \pm 0.3$ days, and $T_{rise, exp} = 4.9 \pm 0.3$ days, rest frame. 
    
    \item SN\,2021lwz has a low mass ejecta $M_{\rm ej} \sim 0.24 $ $M_\odot$ (as from the magnetar model) with estimated velocities of $\sim 15000$ km s$^{-1}$ at $-3.4$ days and of $\sim 7500$ km s$^{-1}$ after peak at $+13..2$ and $+29.1$ days, respectively. Near peak the photosphere has a temperature of $\sim 18000$ K, a radius of  $\sim 10^{15}$ cm and a velocity of $\sim$ 10\,000 km s$^{-1}$, consistent with the ejecta velocity. 

    \item If SN\,2021lwz is a magnetar-driven supernova it is expected to have an initial luminosity $L_0 \approx 3 \times 10^{45}$ erg s$^{-1}$, a spin-down timescale $t_{\rm SD} \approx 1$ day, and an ejecta mass, a final ejecta scale velocity, and a photospheric temperature at peak, of values similar to the ones reported above. 

    \item The magnetar-driven model leads to a very short diffusion time, $t_{\rm diff} \sim$ 5 days, due to the low ejecta mass, to a bolometric peak timescale $\sim$ 7 days and to a $g$-band peak timescale $\sim$ 10 days; the discrepancy between these values (Figure~\ref{fig:violins}) is consistent with the high photospheric temperature and rapid cooling around peak (Figure~\ref{fig:TempEvol}). The SN is expected to have entered the nebular phase on the timescale of $t_{\rm neb} \approx 27$ days $\pm 2$ but the spectroscopy shows the nebular phase should have started after +30 days.  The gamma-ray leakage timescale, $t_{\rm leak}$, is very long. 

    \item The shape of the photosphere of SN\,2021lwz looks close to perfectly spherical and does not show any loss of symmetry between the two phases close to ($\approx$ +1.0 days) and after maximum light ($\approx$ +19. days). The observed symmetry of the system rules out a jet activity to be at the origin of SN\,2021lwz.
      
    \item Spectroscopically, SN\,2021lwz looks like a Type Ic-like SN before maximum light. After maximum light some velocity lines may be a bit larger but the post-peak spectrum looks most like a SLSN and the continuum slope (see Figure\,\ref{SpectralEv}) is much more like a SLSN that a Ic-BL, which is usually much redder from line blanketing.

    \item Given its spectroscopic properties and the low ejecta mass needed to model its light-curve, SN\,2021lwz does not match with many core-collapse H-poor SNe Types.

    \item Using the magnetar model, the low ejecta mass of SN\,2021lwz and its light curve shape possibly make it a USSN but it is not possible to demonstrate it with the data at hand. 

    \item SN\,2021lwz shows similarities with the peculiar Type Ic-BL SN\,1998bw but some differences in their rise time, absolute magnitudes at peak and in their NIR spectra suggest they are two different types of objetcs.

    \item The light curve and multi-band light curve modelling of SN\,2021lwz suggest it shares similarities
    with iPTF\,16asu, and with SN\,2018gep. The three sources show continuum blue dominated spectra before maximum light, even though SN\,2021lwz shows spectral features pointing to a Type Ic-like SNe, while iPTF\,16asu and SN\,2018gep don't. A possible explanation is that iPTF\,16asu and SN\,2018gep are more energetic explosions quickly rising to higher photosphere temperatures than is SN\,2021lwz.
    
\end{itemize}

In conclusion SN\,2021lwz is an uncommon transient showing many similarities with several classes of transients, and with rare transients. It may be an interesting example pointing on how differences in ejecta mass and engine parameters could produce a wide range of engine-driven SESNe. 

With the advent of new high-cadence large-scale deeper surveys like the Vera C. Rubin observatory Large Synoptic Survey Telescope \citep[LSST;][]{2019ApJ...873..111I}, very early detections of bright, low redshift events similar to SN\,2021lwz may be helpful to better understand such phenomena. Similarly, data obtained with the \textit{Euclid} satellite \citep[][]{2025A&A...697A...1E} should be helpful to tackle such problems. The observing strategy is not optimized for transient science but the Quick Data Release 1 has proven to be useful for early and late type detections of some SNe \citep[see][]{2025arXiv250315334D}. 

\begin{acknowledgements} 

The authors would like to thank the anonymous referee for her/his insightfull and constructive comments, Sheng Yang, Anna Ho and \'{O}smar Rodr\'{i}guez for helpful discussions and inputs. We thank Rosa Clavero and David Morate for their kind support on the NOT.
We thank George Khorenzhev (Space Research Institute, Russian Academy of Sciences) for contributing to some observations.

F.P. and C.G.C. acknowledge support from the Spanish Ministerio de Ciencia, Innovación y Universidades (MICINN) under grant numbers PID2022-141915NB-C21.

R.K.T. has been supported by the NKFIH/OTKA FK-134432 grant of the National Research, Development and Innovation Office of Hungary (NKFIH).

C. M. B. O. acknowledges support from the Royal Society (grant Nos. DHF-R1-221175 and DHF-ERE-221005).

N. Sarin acknowledges support from the Kavli foundation. 

T.-W.C. acknowledges the Yushan Fellow Program by the Ministry of Education, Taiwan for the financial support
(MOE-111-YSFMS-0008-001-P1).

T.K. acknowledges support from the Research Council of Finland project 360274.

Based on observations obtained with the Samuel Oschin Telescope 48-inch and the 60-inch Telescope at the Palomar Observatory as part of the Zwicky Transient Facility project. ZTF is supported by the National Science Foundation under Grant No. AST-2034437 and a collaboration including Caltech, IPAC, the Weizmann Institute of Science, the Oskar Klein Center at Stockholm University, the University of Maryland, Deutsches Elektronen-Synchrotron and Humboldt University, the TANGO Consortium of Taiwan, the University of Wisconsin at Milwaukee, Trinity College Dublin, Lawrence Livermore National Laboratories, IN2P3, University of Warwick, Ruhr University Bochum, Cornell University, and Northwestern University. Operations are conducted by COO, IPAC, and UW.

Zwicky Transient Facility access was supported by Northwestern University and the Center for Interdisciplinary Exploration and Research in Astrophysics (CIERA).

The ZTF forced-photometry service was funded under the Heising-Simons Foundation grant No. 12540303 (PI: Graham).

SED Machine is based upon work supported by the National Science Foundation under Grant No. 1106171.

The Gordon and Betty Moore Foundation, through both the Data-Driven Investigator Program and a dedicated grant, provided critical funding for SkyPortal. 

Based on observations made with the Nordic Optical Telescope (NOT), owned in collaboration by the University of Turku and Aarhus University, and operated jointly by Aarhus University, the University of Turku and the University of Oslo, representing Denmark, Finland and Norway, the University of Iceland and Stockholm University at the Observatorio del Roque de los Muchachos, La Palma, Spain, of the Instituto de Astrof\'{i}sica de Canarias. The data presented here were obtained in part with ALFOSC, which is provided by the Instituto de Astrof\'{i}sica de Andalucia (IAA) under a joint agreement with the University of Copenhagen and NOT. Some of the data were obtained during CAT service observation Spanish time. ALFOSC polarimetry imaging data of SN2020ank and calibration data were retrieved from the NOT public archive. 

The Liverpool Telescope is operated on the island of La Palma by Liverpool John Moores University in the Spanish Observatorio del Roque de los Muchachos of the Instituto de Astrof\'{i}sica de Canarias with financial support from the UK Science and Technology Facilities Council.

This work has made use of data from the Asteroid Terrestrial- impact Last Alert System (ATLAS) project. The Asteroid Terrestrial- impact Last Alert System (ATLAS) project is primarily funded to search for near earth asteroids through NASA grants NN12AR55G, 80NSSC18K0284, and 80NSSC18K1575; by-products of the NEO search include images and catalogues from the survey area. This work was partially funded by Kepler/K2 grant J1944/80NSSC19K0112 and HST GO-15889, and STFC grants ST/T000198/1 and ST/S006109/1. The ATLAS science products have been made possible through the contributions of the University of Hawaii Institute for Astronomy, the Queen s University Belfast, the Space Telescope Science Institute, the South African Astronomical Observatory, and The Millennium Institute of Astrophysics (MAS), Chile.

Lasair is supported by the UKRI Science and Technology Facilities Council and is a collaboration between the University of Edinburgh (grant ST/N002512/1) and Queen’s University Belfast (grant ST/N002520/1) within the LSST:UK Science Consortium.

SNID is Copyright (C) 1999-2007 St\'{e}phane Blondin and John L. Tonry, and is available under the GNU General Public License.

This research made use of the Transient Name Server (TNS) which is the official IAU mechanism for reporting new astronomical transients such as supernova candidates, As of January 1, 2016.

This research made use of Legacy Survey Data Release 10 data. The Legacy Surveys consist of three individual and complementary projects: the Dark Energy Camera Legacy Survey (DECaLS; Proposal ID \#2014B-0404; PIs: David Schlegel and Arjun Dey), the Beijing-Arizona Sky Survey (BASS; NOAO Prop. ID \#2015A-0801; PIs: Zhou Xu and Xiaohui Fan), and the Mayall z-band Legacy Survey (MzLS; Prop. ID \#2016A-0453; PI: Arjun Dey). DECaLS, BASS and MzLS together include data obtained, respectively, at the Blanco telescope, Cerro Tololo Inter-American Observatory, NSF’s NOIRLab; the Bok telescope, Steward Observatory, University of Arizona; and the Mayall telescope, Kitt Peak National Observatory, NOIRLab. The Legacy Surveys project is honored to be permitted to conduct astronomical research on Iolkam Du’ag (Kitt Peak), a mountain with particular significance to the Tohono O’odham Nation.

NOIRLab is operated by the Association of Universities for Research in Astronomy (AURA) under a cooperative agreement with the National Science Foundation.

This project used data obtained with the Dark Energy Camera (DECam), which was constructed by the Dark Energy Survey (DES) collaboration. Funding for the DES Projects has been provided by the U.S. Department of Energy, the U.S. National Science Foundation, the Ministry of Science and Education of Spain, the Science and Technology Facilities Council of the United Kingdom, the Higher Education Funding Council for England, the National Center for Supercomputing Applications at the University of Illinois at Urbana-Champaign, the Kavli Institute of Cosmological Physics at the University of Chicago, Center for Cosmology and Astro-Particle Physics at the Ohio State University, the Mitchell Institute for Fundamental Physics and Astronomy at Texas A$\&$M University, Financiadora de Estudos e Projetos, Fundacao Carlos Chagas Filho de Amparo, Financiadora de Estudos e Projetos, Fundacao Carlos Chagas Filho de Amparo a Pesquisa do Estado do Rio de Janeiro, Conselho Nacional de Desenvolvimento Cientifico e Tecnologico and the Ministerio da Ciencia, Tecnologia e Inovacao, the Deutsche Forschungsgemeinschaft and the Collaborating Institutions in the Dark Energy Survey. The Collaborating Institutions are Argonne National Laboratory, the University of California at Santa Cruz, the University of Cambridge, Centro de Investigaciones Energeticas, Medioambientales y Tecnologicas-Madrid, the University of Chicago, University College London, the DES-Brazil Consortium, the University of Edinburgh, the Eidgenossische Technische Hochschule (ETH) Zurich, Fermi National Accelerator Laboratory, the University of Illinois at Urbana-Champaign, the Institut de Ciencies de l’Espai (IEEC/CSIC), the Institut de Fisica d’Altes Energies, Lawrence Berkeley National Laboratory, the Ludwig Maximilians Universitat Munchen and the associated Excellence Cluster Universe, the University of Michigan, NSF’s NOIRLab, the University of Nottingham, the Ohio State University, the University of Pennsylvania, the University of Portsmouth, SLAC National Accelerator Laboratory, Stanford University, the University of Sussex, and Texas A$\&$M University.

The Legacy Surveys imaging of the DESI footprint is supported by the Director, Office of Science, Office of High Energy Physics of the U.S. Department of Energy under Contract No. DE-AC02-05CH1123, by the National Energy Research Scientific Computing Center, a DOE Office of Science User Facility under the same contract; and by the U.S. National Science Foundation, Division of Astronomical Sciences under Contract No. AST-0950945 to NOAO.
\end{acknowledgements} 

\section*{Data Availability}

For science reproducibility purposes, the photometry and the spectra presented in this work are available via WISEReP\footnote{ {\tt
    https://www.wiserep.org/object/18321}.} \citep[][]{2012PASP..124..668Y}.  

\bibliography{biblio}

\begin{appendix}

\section{Host Photometry Table} \label{host_appendix}

The measurements in the different bands of the host galaxy of SN\,2021lwz discussed in Section\ref{data_phot_host}, are given in Table\ref{tab:host_phot}. The magnitudes are not corrected for extinction.

\begin{table}
	\centering
	\caption{Photometry of SN\,2021lwz's host galaxy. 
  }
	\label{tab:host_phot}
	\begin{tabular}{lll} 
          \hline
Survey                & Band & Brightness \\
Telescope/Instrument  &      & (mag)      \\
CFHT/MegaPrime  &$ u $&$ 23.23 \pm 0.16 $ \\
SDSS            &$ g $&$ 22.86 \pm 0.24 $ \\
CFHT/MegaPrime  &$ r $&$ 22.97 \pm 0.05 $ \\
Pan-STARRRS     &$ i $&$ 22.37 \pm 0.15 $ \\
SDSS            &$ i $&$ 22.38 \pm 0.30 $ \\
         \hline
	\end{tabular}
\end{table}

\section{Spectroscopy Table}  \label{spectroscopy_table_appendix}

The log of the spectroscopy obtained on  SN\,2021lwz discussed in Section~\ref{spectra} is given in Table,\ref{tab:spectra}. The rest-frame phase relative to the g-band filter is given using MJD$_{\rm max}=59350$ days.

\begin{table}
	\centering
	\caption{List of spectra obtained on SN\,2021lwz.}
	\label{tab:spectra}
	\begin{tabular}{llc} 
          \hline
          Instrument & Observation Date & Rest-frame phase \\
          & [UT] & (relative to g-band)  \\
          \hline                                                            
P60/SEDM          & 2021-05-11T04:07:47 & -5.4\\
Keck/LRIS$^{(*)}$ & 2021-05-13T05:40:20 & -3.4 \\
P60/SEDM          & 2021-05-15T04:07:59 & -1.6 \\
NOT/ALFOSC        & 2021-05-15T22:02:58 & -0.9 \\
Keck/LRIS         & 2021-05-16T09:40:23 & -0.5 \\
NOT/ALFOSC        & 2021-05-18T21:57:56 & 1.9 \\
P60/SEDM          & 2021-05-24T04:00:59 & 6.8 \\
P60/SEDM          & 2021-05-28T04:04:46 & 10.6 \\
P60/SEDM          & 2021-05-30T05:48:20 & 12.5 \\  
Keck/NIRES        & 2021-05-30T07:15:08 & 12.6 \\
NOT/ALFOSC        & 2021-05-30T21:58:39 & 13.2 \\   
P60/SEDM          & 2021-06-13T04:15:04 & 25.6 \\
NOT/ALFOSC        & 2021-06-16T21:20:28 & 29.1 \\
          \hline
	\end{tabular}
\tablefoot{
\tablefoottext{*}{spectrum publicly available on TNS.}
}    
\end{table}

\section{Photometry Table} \label{photometry_table_appendix}

All the photometry obtained on SN\,2021lwz is given in Table~\ref{tab:sn21lwz_all_phot}.

\begin{table}[h]
    \centering
    \caption{Photometric data of SN\,2021lwz.}
    
    \label{tab:sn21lwz_all_phot}
    \small
    \begin{tabular}{ccccc}
    \hline
    \hline
     MJD & Phase\tablefootmark{\scriptsize a} & Filter & Magnitude & Telescope\\
     & (days) & & (mag) & \\
    \hline
59344.2 & -6.7 & g & 18.76 $\pm$ 0.03 & ZTF \\
59344.2 & -6.7 & r & 19.21 $\pm$ 0.04 & ZTF \\
59345.3 & -5.6 & o & 18.69 $\pm$ 0.05 & ATLAS \\
59346.2 & -4.7 & r & 18.35 $\pm$ 0.02 & ZTF \\
59346.3 & -4.6 & g & 18.01 $\pm$ 0.04 & ZTF \\
59347.0 & -3.9 & sdssg & 17.86 $\pm$ 0.03 & LT \\
59347.0 & -3.9 & sdssr & 18.22 $\pm$ 0.04 & LT \\
59347.0 & -3.9 & sdssi & 18.39 $\pm$ 0.08 & LT \\
59347.0 & -3.9 & sdssu & 17.85 $\pm$ 0.16 & LT \\
59347.0 & -3.9 & sdssz & 18.73 $\pm$ 0.28 & LT \\
59347.4 & -3.5 & UVW1 & 17.41 $\pm$ 0.04 & Swift \\
59347.4 & -3.5 & U & 17.31 $\pm$ 0.05 & Swift \\
59347.4 & -3.5 & B & 17.54 $\pm$ 0.09 & Swift \\
59347.4 & -3.5 & UVW2 & 17.89 $\pm$ 0.04 & Swift \\
\hline
\end{tabular}
\end{table}
\addtocounter{table}{-1}
\begin{table}
    \centering
    \caption{continued.}
    \small
    \begin{tabular}{ccccc}
    \hline
    \hline
     MJD & Phase\tablefootmark{\scriptsize a} & Filter & Magnitude & Telescope\\
     & (days) & & (mag) & \\
    \hline
59347.4 & -3.5 & V & 17.97 $\pm$ 0.18 & Swift \\
59347.4 & -3.5 & UVM2 & 17.55 $\pm$ 0.03 & Swift \\
59347.9 & -3.0 & sdssg & 17.67 $\pm$ 0.03 & LT \\
59347.9 & -3.0 & sdssr & 18.01 $\pm$ 0.03 & LT \\
59347.9 & -3.0 & sdssi & 18.44 $\pm$ 0.03 & LT \\
59347.9 & -3.0 & sdssu & 17.37 $\pm$ 0.04 & LT \\
59347.9 & -3.0 & sdssz & 18.64 $\pm$ 0.07 & LT \\
59348.2 & -2.7 & r & 17.96 $\pm$ 0.02 & ZTF \\
59348.2 & -2.7 & g & 17.53 $\pm$ 0.01 & ZTF \\
59348.3 & -2.6 & c & 17.77 $\pm$ 0.02 & ATLAS \\
59348.3 & -2.6 & UVW1 & 17.34 $\pm$ 0.04 & Swift \\
59348.3 & -2.6 & U & 17.09 $\pm$ 0.06 & Swift \\
59348.3 & -2.6 & B & 17.38 $\pm$ 0.09 & Swift \\
59348.3 & -2.6 & UVW2 & 17.84 $\pm$ 0.04 & Swift \\
59348.3 & -2.6 & V & 17.91 $\pm$ 0.23 & Swift \\
59348.3 & -2.6 & UVM2 & 17.45 $\pm$ 0.04 & Swift \\
59348.9 & -2.0 & sdssg & 17.52 $\pm$ 0.03 & LT \\
59348.9 & -2.0 & sdssr & 17.86 $\pm$ 0.03 & LT \\
59348.9 & -2.0 & sdssi & 18.23 $\pm$ 0.03 & LT \\
59348.9 & -2.0 & sdssu & 17.16 $\pm$ 0.01 & LT \\
59348.9 & -2.0 & sdssz & 18.48 $\pm$ 0.05 & LT \\
59349.3 & -1.6 & UVW1 & 17.32 $\pm$ 0.04 & Swift \\
59349.3 & -1.6 & U & 17.15 $\pm$ 0.06 & Swift \\
59349.3 & -1.6 & B & 17.34 $\pm$ 0.08 & Swift \\
59349.3 & -1.6 & UVW2 & 17.9 $\pm$ 0.04 & Swift \\
59349.3 & -1.6 & V & 17.3 $\pm$ 0.14 & Swift \\
59349.3 & -1.6 & UVM2 & 17.52 $\pm$ 0.04 & Swift \\
59349.3 & -1.6 & o & 17.93 $\pm$ 0.03 & ATLAS \\
59349.9 & -1.0 & sdssg & 17.4 $\pm$ 0.03 & LT \\
59349.9 & -1.0 & sdssr & 17.73 $\pm$ 0.04 & LT \\
59349.9 & -1.0 & sdssi & 18.1 $\pm$ 0.03 & LT \\
59349.9 & -1.0 & sdssu & 17.1 $\pm$ 0.05 & LT \\
59349.9 & -1.0 & sdssz & 18.29 $\pm$ 0.05 & LT \\
59350.2 & -0.7 & g & 17.34 $\pm$ 0.01 & ZTF \\
59350.2 & -0.7 & r & 17.74 $\pm$ 0.02 & ZTF \\
59350.3 & -0.6 & UVW1 & 17.21 $\pm$ 0.05 & Swift \\
59350.3 & -0.6 & U & 17.01 $\pm$ 0.06 & Swift \\
59350.3 & -0.6 & B & 17.29 $\pm$ 0.1 & Swift \\
59350.3 & -0.6 & UVW2 & 17.86 $\pm$ 0.05 & Swift \\
59350.3 & -0.6 & V & 17.31 $\pm$ 0.18 & Swift \\
59350.3 & -0.6 & UVM2 & 17.55 $\pm$ 0.06 & Swift \\
59351.9 & 1.0 & sdssg & 17.27 $\pm$ 0.03 & LT \\
59351.9 & 1.0 & sdssr & 17.56 $\pm$ 0.03 & LT \\
59351.9 & 1.0 & sdssi & 17.86 $\pm$ 0.05 & LT \\
59351.9 & 1.0 & sdssu & 16.96 $\pm$ 0.03 & LT \\
59351.9 & 1.0 & sdssz & 18.12 $\pm$ 0.09 & LT \\
59352.3 & 1.4 & o & 17.65 $\pm$ 0.05 & ATLAS \\
59352.9 & 2.0 & sdssg & 17.31 $\pm$ 0.07 & LT \\
59352.9 & 2.0 & sdssr & 17.61 $\pm$ 0.04 & LT \\
59352.9 & 2.0 & sdssi & 17.86 $\pm$ 0.08 & LT \\
59352.9 & 2.0 & sdssu & 17.06 $\pm$ 0.05 & LT \\
59352.9 & 2.0 & sdssz & 18.2 $\pm$ 0.07 & LT \\
59353.9 & 3.0 & sdssg & 17.36 $\pm$ 0.06 & LT \\
59353.9 & 3.0 & sdssr & 17.6 $\pm$ 0.04 & LT \\
59353.9 & 3.0 & sdssi & 17.86 $\pm$ 0.06 & LT \\
59353.9 & 3.0 & sdssu & 17.2 $\pm$ 0.15 & LT \\
59353.9 & 3.0 & sdssz & 18.02 $\pm$ 0.12 & LT \\
59354.2 & 3.3 & i & 17.95 $\pm$ 0.05 & ZTF \\
59354.2 & 3.3 & r & 17.69 $\pm$ 0.03 & ZTF \\
59354.3 & 3.4 & o & 17.72 $\pm$ 0.03 & ATLAS \\
59354.4 & 3.5 & UVW1 & 17.58 $\pm$ 0.06 & Swift \\
59354.4 & 3.5 & U & 17.21 $\pm$ 0.08 & Swift \\
59354.4 & 3.5 & B & 17.08 $\pm$ 0.1 & Swift \\
59354.4 & 3.5 & UVW2 & 18.38 $\pm$ 0.06 & Swift \\
59354.4 & 3.5 & V & 17.28 $\pm$ 0.19 & Swift \\
\hline
\end{tabular}
\end{table}
\addtocounter{table}{-1}
\begin{table}
    \centering
    \caption{continued.}
    \small
    \begin{tabular}{ccccc}
    \hline
    \hline
     MJD & Phase\tablefootmark{\scriptsize a} & Filter & Magnitude & Telescope\\
     & (days) & & (mag) & \\
    \hline
59354.4 & 3.5 & UVM2 & 17.94 $\pm$ 0.05 & Swift \\
59356.2 & 5.3 & g & 17.63 $\pm$ 0.03 & ZTF \\
59356.2 & 5.3 & r & 17.82 $\pm$ 0.04 & ZTF \\
59356.3 & 5.4 & o & 17.86 $\pm$ 0.04 & ATLAS \\
59356.8 & 5.9 & UVW1 & 18.11 $\pm$ 0.07 & Swift \\
59356.8 & 5.9 & U & 17.6 $\pm$ 0.09 & Swift \\
59356.8 & 5.9 & B & 17.6 $\pm$ 0.14 & Swift \\
59356.8 & 5.9 & UVW2 & 19.08 $\pm$ 0.08 & Swift \\
59356.8 & 5.9 & V & 17.33 $\pm$ 0.2 & Swift \\
59356.8 & 5.9 & UVM2 & 18.67 $\pm$ 0.06 & Swift \\
59357.0 & 6.1 & sdssg & 17.72 $\pm$ 0.04 & LT \\
59357.0 & 6.1 & sdssr & 17.84 $\pm$ 0.04 & LT \\
59357.0 & 6.1 & sdssi & 18.07 $\pm$ 0.04 & LT \\
59357.0 & 6.1 & sdssz & 18.13 $\pm$ 0.07 & LT \\
59357.9 & 7.0 & sdssg & 17.77 $\pm$ 0.07 & LT \\
59357.9 & 7.0 & sdssr & 17.95 $\pm$ 0.03 & LT \\
59357.9 & 7.0 & sdssi & 18.1 $\pm$ 0.03 & LT \\
59357.9 & 7.0 & sdssu & 17.66 $\pm$ 0.04 & LT \\
59357.9 & 7.0 & sdssz & 18.2 $\pm$ 0.05 & LT \\
59358.2 & 7.3 & i & 18.08 $\pm$ 0.04 & ZTF \\
59358.2 & 7.3 & r & 17.93 $\pm$ 0.05 & ZTF \\
59358.2 & 7.3 & g & 17.82 $\pm$ 0.04 & ZTF \\
59359.9 & 9.0 & sdssg & 17.92 $\pm$ 0.09 & LT \\
59359.9 & 9.0 & sdssr & 18.02 $\pm$ 0.04 & LT \\
59359.9 & 9.0 & sdssi & 18.17 $\pm$ 0.03 & LT \\
59359.9 & 9.0 & sdssu & 18.27 $\pm$ 0.01 & LT \\
59359.9 & 9.0 & sdssz & 18.32 $\pm$ 0.06 & LT \\
59360.3 & 9.4 & o & 18.04 $\pm$ 0.06 & ATLAS \\
59361.2 & 10.3 & g & 18.01 $\pm$ 0.05 & ZTF \\
59362.9 & 12.0 & sdssg & 18.07 $\pm$ 0.02 & LT \\
59362.9 & 12.0 & sdssr & 18.15 $\pm$ 0.03 & LT \\
59362.9 & 12.0 & sdssi & 18.32 $\pm$ 0.03 & LT \\
59362.9 & 12.0 & sdssu & 18.38 $\pm$ 0.08 & LT \\
59362.9 & 12.0 & sdssz & 18.37 $\pm$ 0.05 & LT \\
59363.3 & 12.4 & o & 18.1 $\pm$ 0.04 & ATLAS \\
59364.1 & 13.2 & U & 18.7 $\pm$ 0.24 & Swift \\
59364.1 & 13.2 & B & 17.79 $\pm$ 0.21 & Swift \\
59364.1 & 13.2 & UVW2 & 20.42 $\pm$ 0.25 & Swift \\
59364.1 & 13.2 & UVM2 & 20.2 $\pm$ 0.2 & Swift \\
59365.9 & 15.0 & sdssg & 18.2 $\pm$ 0.03 & LT \\
59365.9 & 15.0 & sdssr & 18.26 $\pm$ 0.04 & LT \\
59365.9 & 15.0 & sdssi & 18.4 $\pm$ 0.04 & LT \\
59365.9 & 15.0 & sdssu & 18.76 $\pm$ 0.09 & LT \\
59365.9 & 15.0 & sdssz & 18.46 $\pm$ 0.04 & LT \\
59366.9 & 16.0 & UVW1 & 19.61 $\pm$ 0.13 & Swift \\
59366.9 & 16.0 & U & 18.72 $\pm$ 0.13 & Swift \\
59366.9 & 16.0 & B & 18.56 $\pm$ 0.2 & Swift \\
59366.9 & 16.0 & UVW2 & 20.94 $\pm$ 0.28 & Swift \\
59367.3 & 16.4 & o & 18.06 $\pm$ 0.05 & ATLAS \\
59368.5 & 17.6 & UVW1 & 19.84 $\pm$ 0.14 & Swift \\
59368.5 & 17.6 & U & 18.72 $\pm$ 0.13 & Swift \\
59368.5 & 17.6 & B & 18.22 $\pm$ 0.15 & Swift \\
59368.5 & 17.6 & UVW2 & 20.61 $\pm$ 0.14 & Swift \\
59368.5 & 17.6 & V & 18.21 $\pm$ 0.28 & Swift \\
59368.5 & 17.6 & UVM2 & 20.53 $\pm$ 0.12 & Swift \\
59370.9 & 20.0 & UVW1 & 19.94 $\pm$ 0.2 & Swift \\
59370.9 & 20.0 & U & 18.73 $\pm$ 0.17 & Swift \\
59370.9 & 20.0 & B & 18.31 $\pm$ 0.21 & Swift \\
59370.9 & 20.0 & UVW2 & 20.85 $\pm$ 0.22 & Swift \\
59370.9 & 20.0 & V & 17.79 $\pm$ 0.28 & Swift \\
59370.9 & 20.0 & sdssg & 18.28 $\pm$ 0.03 & LT \\
59370.9 & 20.0 & UVM2 & 20.5 $\pm$ 0.17 & Swift \\
59370.9 & 20.0 & sdssr & 18.32 $\pm$ 0.03 & LT \\
59370.9 & 20.0 & sdssi & 18.45 $\pm$ 0.03 & LT \\
\hline
\end{tabular}
\end{table}
\addtocounter{table}{-1}
\begin{table}
    \centering
    \caption{continued.}
    \small
    \begin{tabular}{ccccc}
    \hline
    \hline
     MJD & Phase\tablefootmark{\scriptsize a} & Filter & Magnitude & Telescope\\
     & (days) & & (mag) & \\
    \hline
59370.9 & 20.0 & sdssz & 18.5 $\pm$ 0.07 & LT \\
59371.3 & 20.4 & o & 18.41 $\pm$ 0.06 & ATLAS \\
59376.9 & 26.0 & sdssr & 18.68 $\pm$ 0.04 & LT \\
59376.9 & 26.0 & sdssi & 18.71 $\pm$ 0.05 & LT \\
59376.9 & 26.0 & sdssz & 18.62 $\pm$ 0.08 & LT \\
59377.9 & 27.0 & sdssg & 18.75 $\pm$ 0.04 & LT \\
59377.9 & 27.0 & sdssr & 18.74 $\pm$ 0.04 & LT \\
59377.9 & 27.0 & sdssi & 18.83 $\pm$ 0.05 & LT \\
59377.9 & 27.0 & sdssz & 18.73 $\pm$ 0.07 & LT \\
59381.9 & 31.0 & sdssg & 19.05 $\pm$ 0.07 & LT \\
59381.9 & 31.0 & sdssr & 18.97 $\pm$ 0.05 & LT \\
59381.9 & 31.0 & sdssi & 19.26 $\pm$ 0.12 & LT \\
59381.9 & 31.0 & sdssz & 19.04 $\pm$ 0.14 & LT \\
59390.2 & 39.3 & r & 19.8 $\pm$ 0.14 & ZTF \\
59391.2 & 40.3 & r & 19.79 $\pm$ 0.14 & ZTF \\
59394.2 & 43.3 & r & 19.89 $\pm$ 0.15 & ZTF \\
59396.2 & 45.3 & r & 19.86 $\pm$ 0.11 & ZTF \\
59397.2 & 46.3 & r & 20.12 $\pm$ 0.21 & ZTF \\
\hline
\end{tabular}
   \tablefoot{The photometry is reported on the AB system and is not corrected for reddening. This table is available in machine readable form. Multiple exposures on any given night are averaged to give the values presented here.\tablefoottext{a}{Rest-frame relative to the g-band maximum (MJD 59350.9).}}
\end{table}

\newpage

\section{Linear Polarimetry material}  \label{polarimetry_table_appendix}

Linear polarimetry of SN\,2021lwz obtained with ALFOSC in one of the V-band is shown in Figure~\ref{fig:eo_images}.
The log of the polarimetry obtained on  SN\,2021lwz using ALFOSC on the NOT (see Section~\ref{data_pol}) is presented in Table,\ref{tab:log_polarimetry}. 

\begin{figure}
\begin{center}
\vspace*{2mm}
\centering
\hspace*{0.cm}
\includegraphics[width=70mm,angle=0]{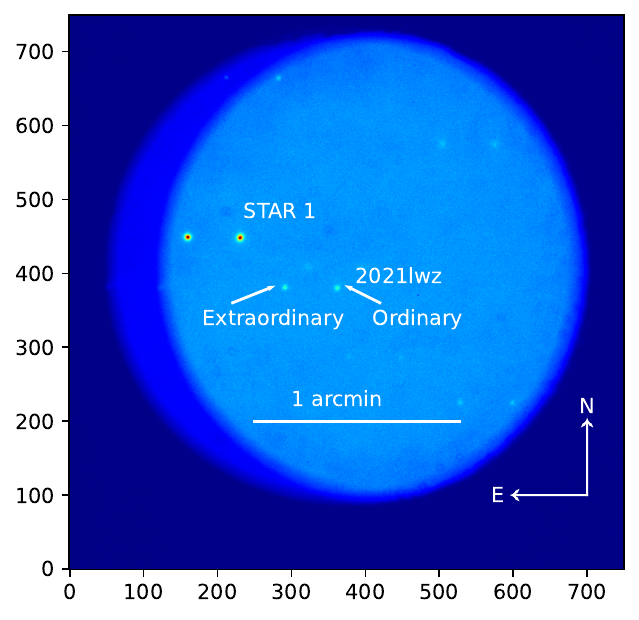}
\vspace*{0.cm}
\caption{Linear polarimetry of SN\,2021lwz with ALFOSC in one of the
  V-band flat-field and bias corrected data frame 
  obtained with the half-wave plate at a position angle of $0.0^{\circ}$. 
  Imaging polarimetry was acquired through half-wave
plates positions angles at $0.0^{\circ}$, $22.5^{\circ}$,
$45.0^{\circ}$ and $67.5^{\circ}$.  The calcite plate splits the light
from the several objects into Ordinary images and
Extraordinary images separated by about $15^{\arcsec}$ from each other. SN\,2021lwz and field star STAR 1, which is used to measure the interstellar polarization component of our Galaxy, are marked in the image.}
\label{fig:eo_images}
\end{center}
\end{figure}

\begin{table*}
	\centering
	\caption{Observations log of the imaging polarimetry observations.}
	\label{tab:log_polarimetry}
	\begin{tabular}{lllcc} 
          \hline
          UT Time & Object & Exp. Time & Filter & Seeing \\
           &  &  [s] & & [$\arcsec$] \\
          \hline
          2021-05-17 21:37:36 &  2021lwz     & $ 1 \times (4 \times 600) $ & V &1.9\\
          2021-05-17 22:53:56 &  HD127769    & $ 2 \times ( 4\times 1) $ & V &1.5\\
          2021-05-17 22:57:30 &  GD 319      & $ 2 \times ( 4 \times 5) $ & V &1.3\\
          \hline
          2021-06-04 21:05:14 &  2021lwz     & $ 1 \times (4 \times 600) $ & V &0.6\\
          2021-06-04 21:26:07 &   GD 319   & $ 2 \times ( 4\times 5) $ & V &0.5\\
          2021-06-04 21:30:53 &  HD127769      & $ 2 \times ( 4 \times 1) $ & V &0.5\\
          \hline
	\end{tabular}
\end{table*}

\begin{table*}
	\centering
	\caption{Polarimetry results on SN\,2021lwz obtained in the V-band Bessel filter.}
	\label{tab:pol_results}
	\begin{tabular}{clcccccccl} 
          \hline
          Date
          &Source
          & $\overline{Q}^{\rm (a)}$
          & $\overline{U}^{\rm(a)}$
          & $P [\%]^{\rm(a)} $
          & $P [\%]^{\rm(b)}$
          & $\theta [^{\circ}]^{\rm(b)}$
          & $P [\%]^{\rm(dc} $
          & $\theta [^{\circ}]^{\rm(c)}$
          & $P [\%]^{\rm(d)} $\\
          \hline
 \hline
2021-05-17 & BD$+$52913  & 0.03 & -0.10 & 0.10$\pm$0.09 & ... & ... & ... & ... & ...  \\
 -- & HD127769 &  0.45 & -1.41 & 1.48$\pm$0.06 & 1.38$\pm$0.11 &  143.8$\pm$2.4 & ... & ... & ... \\
 -- & STAR 1 &  -0.06 & -0.23 & 0.23$\pm$0.15  & 0.16$\pm$0.18 & 117.4$\pm$32.4 & 0.16$\pm$0.18 & 26.2$\pm$32.4 & ...  \\
 -- & 2021lwz $^{\rm(f)}$& 0.01 & 0.14 &  0.15$\pm$0.10  & 0.24$\pm$0.14 & 47.4$\pm$15.9 &  0.24$\pm$0.14 & 136.2$\pm$15.9 &  0.09$\pm$0.14  \\
 -- & ... &  ... & ... &  ... & ... & ... &  ... & ... & 0.0$\pm$0.14 $^{\rm(bc)}$ \\
2021-06-04 & BD$+$52913 &  -0.05 & -0.09 & 0.11$\pm$0.15 & ... & ... & ... & ... & ...  \\
 -- & HD127769 &  0.52 & -2.02 & 2.09$\pm$0.31 & 2.02$\pm$0.34 & 143.3$\pm$4.8 & ... & ... & ... \\
 -- & STAR 1 &  -0.33 & -0.20 & 0.39$\pm$0.21  & 0.30$\pm$0.26 & 100.5$\pm$24.9 & 0.30$\pm$0.26 & 9.9$\pm$24.9  & ...  \\
 -- & 2021lwz $^{\rm(f)}$&  -0.14 & 0.20 &  0.24$\pm$0.11  & 0.30$\pm$0.18 & 53.4$\pm$17.2 &  0.30$\pm$0.18 & 142.8$\pm$17.2 &  0.43$\pm$0.18  \\
 -- & ... &  ... & ... &  ...  & ... & ... &  ... & ... & 0.25$\pm$0.18$^{\rm(bc)}$ \\
        \hline
	\end{tabular}
\tablefoot{
\tablefoottext{a}{Stokes
          parameters, $\overline{Q}$ and $\overline{U}$, directly obtained
          from the ALFOSC data frames Extraordinary and Ordinary images without applying
          any further corrections.}
\tablefoottext{ba}{raw data polarization estimates.}
\tablefoottext{b}{instrumental polarization (IP) corrected.}
\tablefoottext{c}{instrumental
          polarization and zero polarization angle (ZPA) corrected. Assuming HD\,127769 has a polarization
          angle of $52.7^{\circ}$ in V-band the zero
          polarization angle, ZPA, estimates are $91.1 ^{\circ}$ and
          $91.6 ^{\circ}$, at the first and second epochs, respectively.}
\tablefoottext{d}{IP and ZPA
          corrected. $^{\rm(e)}$: IP and ZPA corrected +
          Milky Way Interstellar polarization
          corrected with Stokes parameters obtained on star 1$^{\rm(f)}$.}         
\tablefoottext{bc}{bias corrected.}
}    
\end{table*}

\newpage

\section{Fits obtained with SNID} \label{snid_appendix}

The fits obtained with SNID and discussed in Section\,\ref{snid_analysis} are shown in Figure\,\ref{fig:snid_fits}

\begin{figure*}
	\centering
	\includegraphics[width=90mm, angle=0]{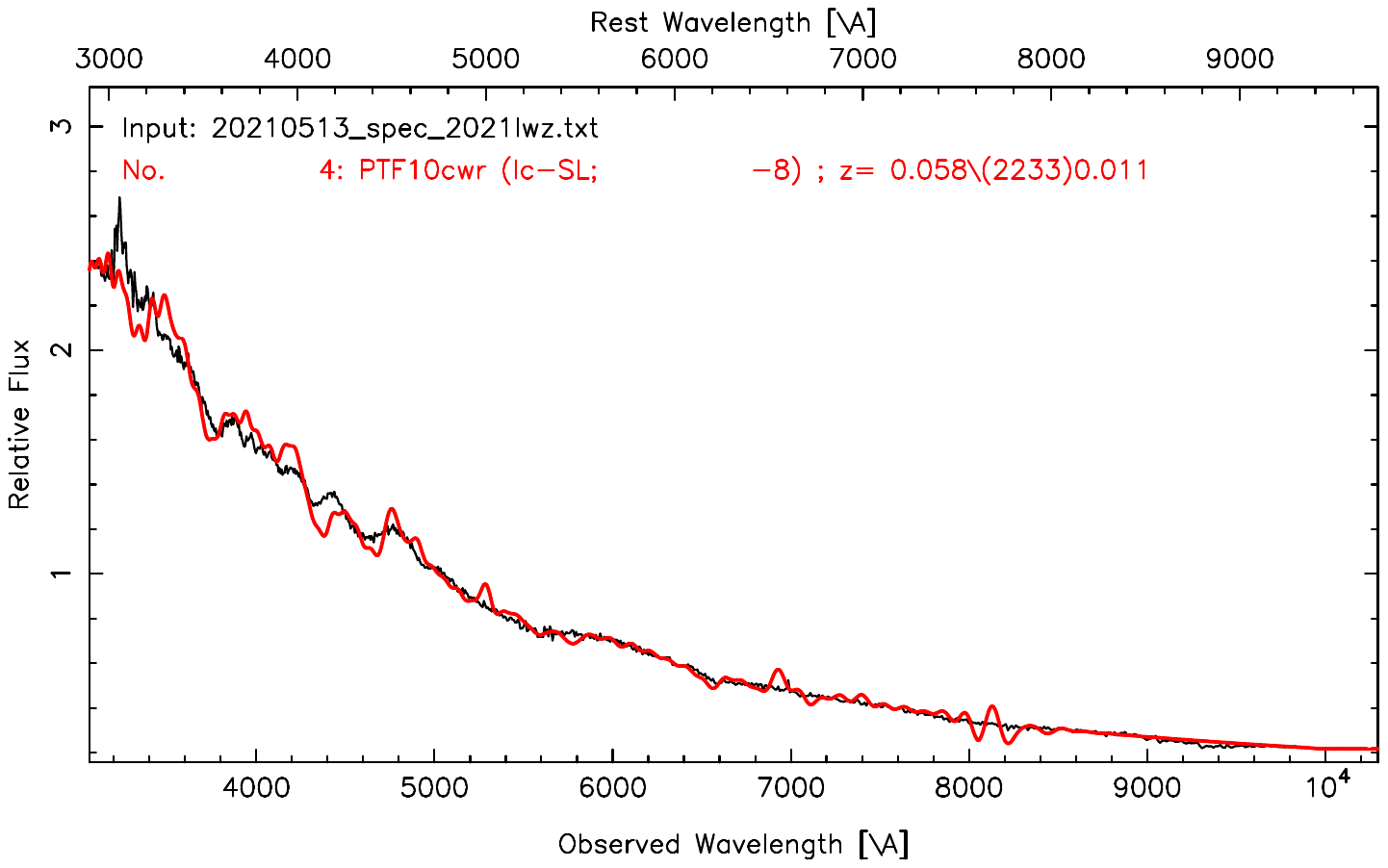}
	\includegraphics[width=90mm, angle=0]{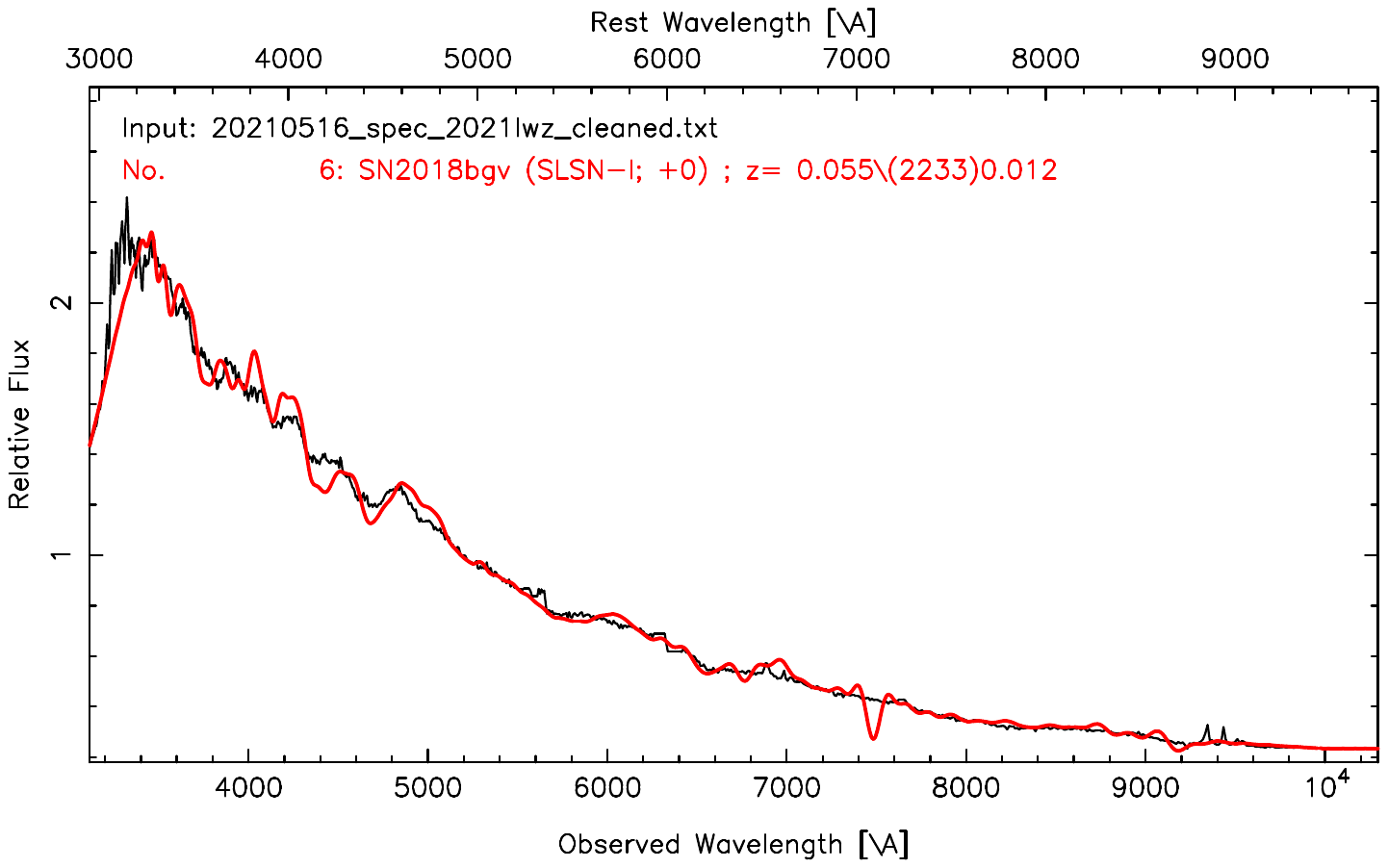}
	\includegraphics[width=90mm, angle=0]{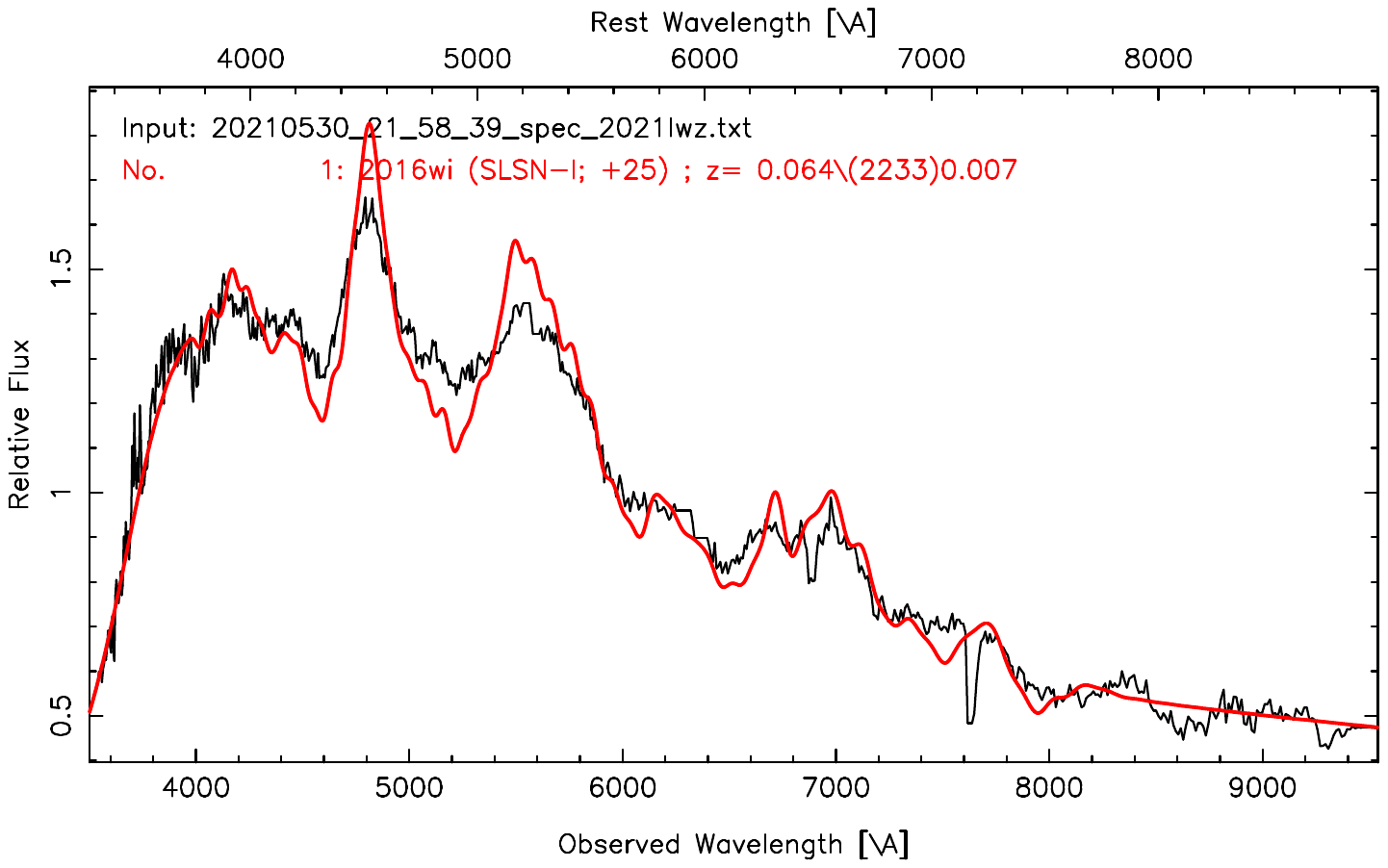}
	\includegraphics[width=90mm, angle=0]{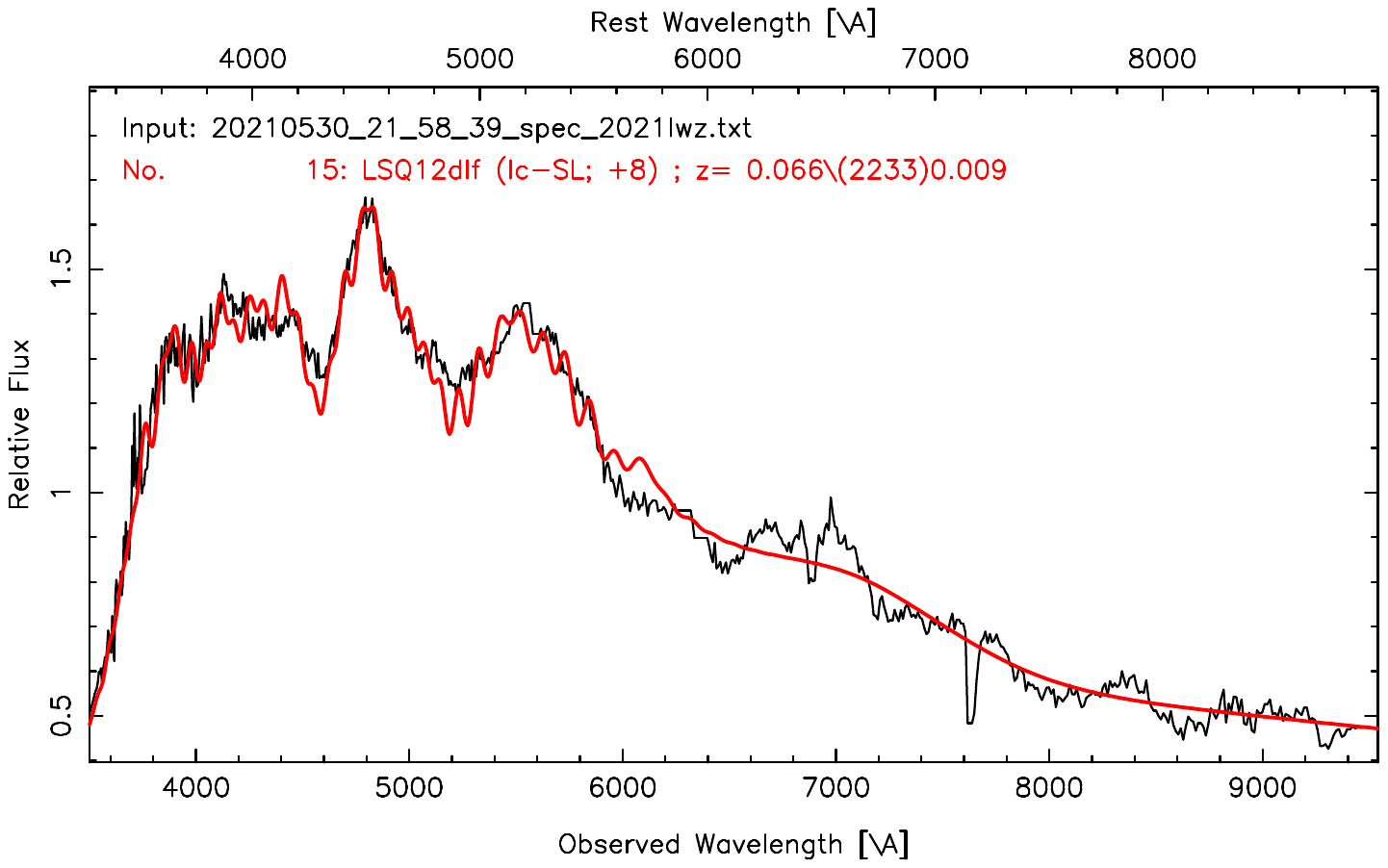}    
	\includegraphics[width=90mm, angle=0]{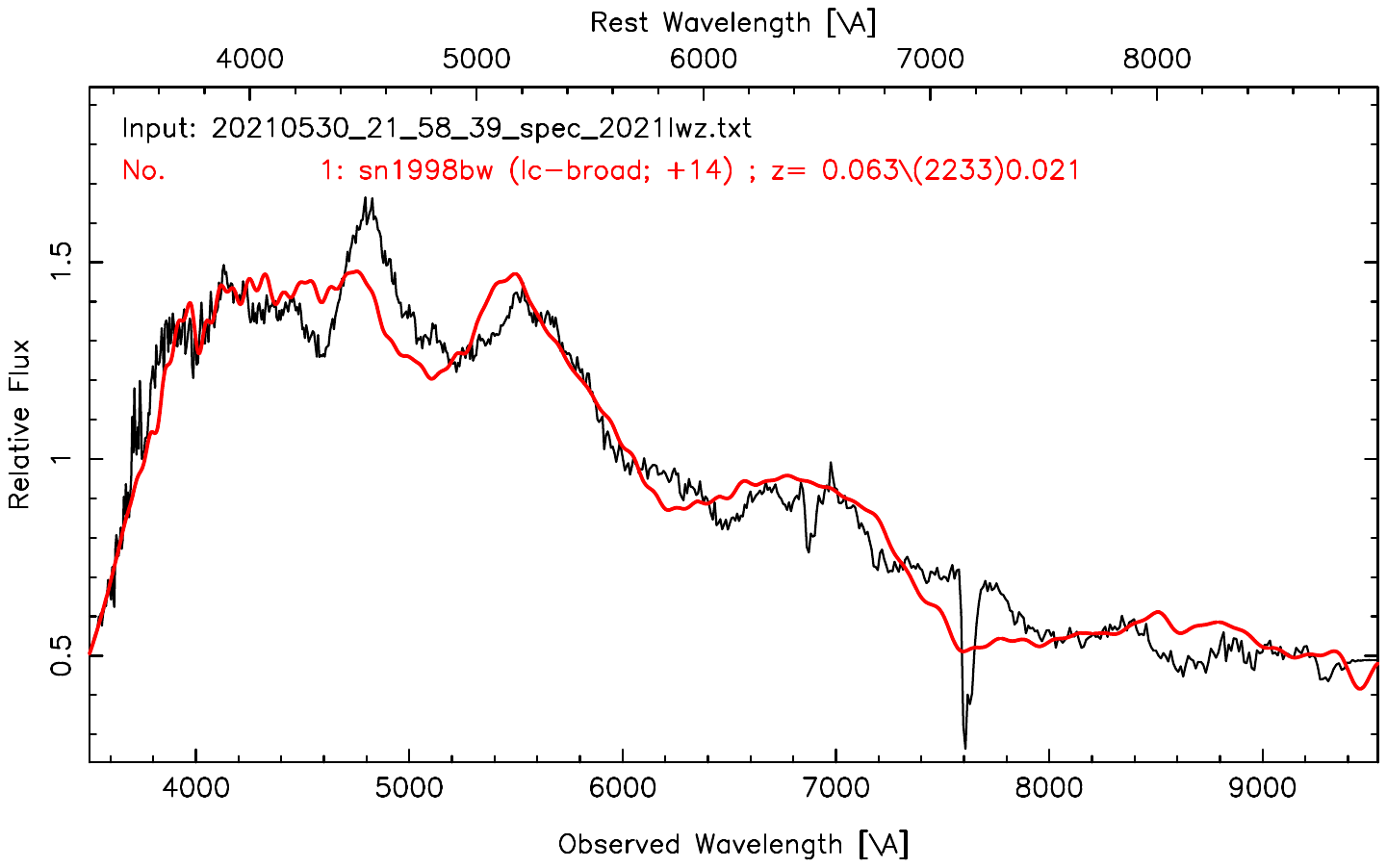}
	\includegraphics[width=90mm, angle=0]{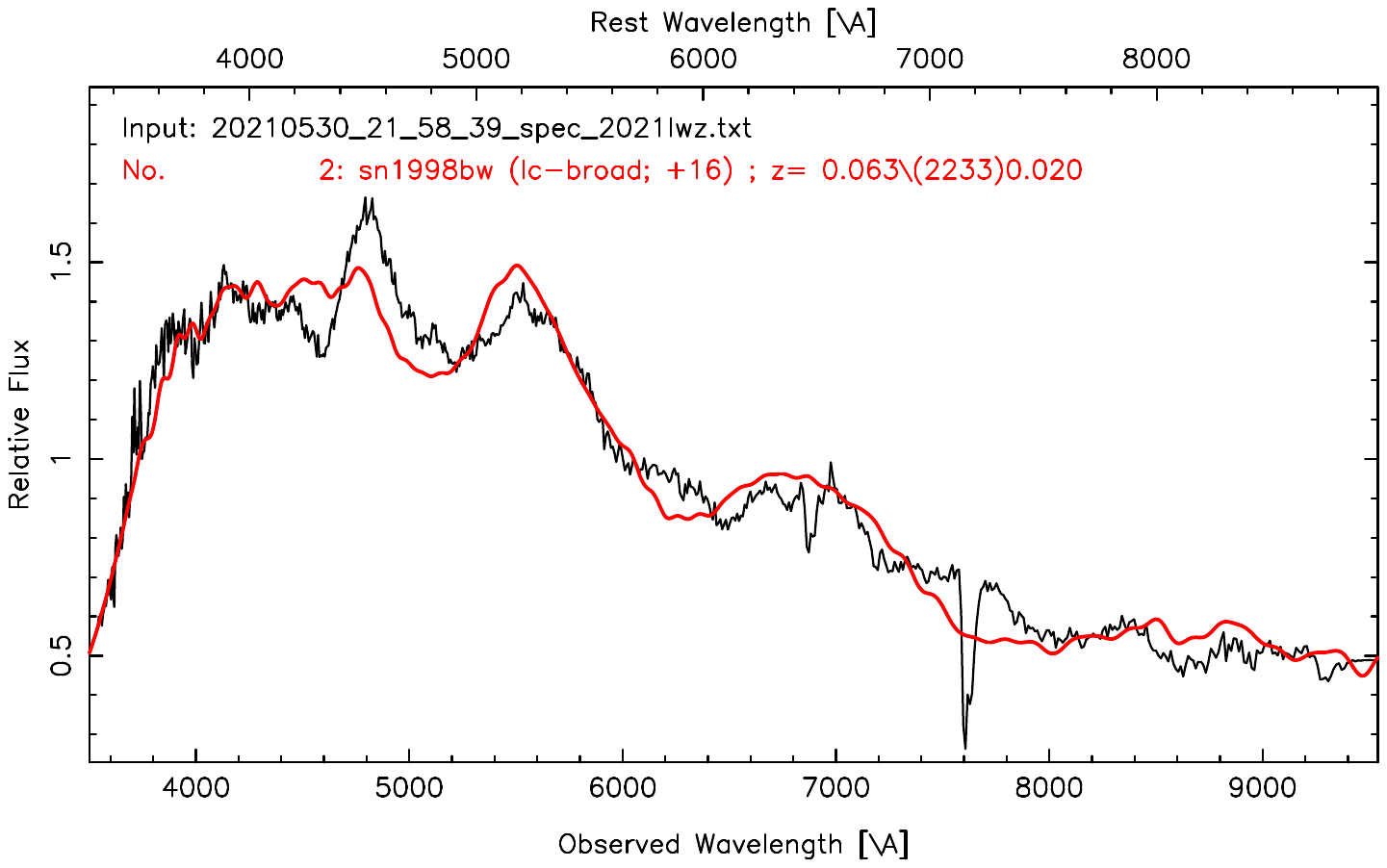}    
    \caption{\label{fig:snid_fits}Fits of some of the SN\,2021lwz spectra using SNID. Top-left: -3.4 days Keck spectrum fit with Type Ic-SL PTF10cwr. Top-right: -0.5 days Keck spectrum fit with fast-rising SLSN-I SN\,2018bgv. Middle: example of a post-peak good fit obtained on the ALFOSC +13.2 days spectrum of SN\,2021lwz with Type H-poor SLSN SN\,2016wi (left), and with Ic-SL SN\,LSQ12dlf. Bottom: fits obtained on the ALFOSC +13.2 days spectrum using SN\,1998bw templates for comparisons.}
\end{figure*}

\section{Spectrum modelling with SYN++} \label{syn++_appendix}

The best-fit global parameter values of the SYN++ modelling of SN\,2021lwz (see Section~\ref{spec_model} )are given in Table,\ref{tab:globparams}. The best-fit local SYN++ parameters of the modelled spectra of SN\,2021lwz are given in this appendix in Table,\ref{tab:locparams}.

\begin{table*}
\caption{Best-fit local SYN++ parameters of the modelled spectra of SN\,2021lwz.}  
\label{tab:locparams}
\centering
\begin{tabular}{lccccccccc}
\hline
\hline
Ions &  \ion{He}{i} &  \ion{C}{ii} & \ion{O}{i} & \ion{Na}{i} & \ion{Mg}{ii} & \ion{Si}{ii}  & \ion{Ca}{ii} & \ion{Ti}{ii} & \ion{Fe}{ii}  \\
\hline
 \multicolumn{10}{l}{-3.4 days phase} \\
\hline
$\log\tau$ & & -1.0 & -1.0 & 0.9 & -0.5 & -0.4  & -0.3 & -1.6 & -0.5 \\
$v_{\rm min}$ & & 15.0 & 15.0 & 15.0 & 15.0 & 15.0  &  15.0 & 15.0 &  15.0 \\
$v_{\rm max}$ & & 50.0 & 50.0 & 50.0 & 50.0 & 50.0  & 50.0 &50.0 & 50.0 \\
aux & & 1.0 & 1.0 & 1.0 & 1.0 & 1.0 & 1.0 & 1.0 & 1.0 \\
$T_{\rm exc}$ & & 20.0 & 20.0 & 5.0 & 20.0 & 20.0  & 20.0 & 7.0 & 12.0 \\
\hline
 \multicolumn{10}{l}{+13.2 days and +29.1 days phase} \\
\hline
$\log\tau$ & -0.3 & -0.8 & -0.3 & & 0.3 & 0.0  & 0.0 & & 0.1  \\
$v_{\rm min}$ &7.5 & 7.5 & 7.5 &  & 7.5 & 7.5   &  7.5  &&  7.5 \\
$v_{\rm max}$ & 50.0& 50.0 & 50.0 &  & 50.0 & 50.0   & 50.0  && 50.0 \\
aux & 10.0 & 5.0 & 3.0 & & 1.0 & 2.0 &  5.0 & & 3.0  \\
$T_{\rm exc}$ & 8.0 & 20.0 & 8.0 & & 8.0 & 20.0  & 15.0 & & 8.0 \\
\hline
\end{tabular}
\tablefoot{
The values of $v_{\rm min}$ and $v_{\rm max}$ are given in 1000 km s$^{-1}$, while  $T_{\rm exc}$ is in given in 1000 K.
}
\end{table*}

\section{Bolometric Light curve Arnett modelling}  \label{arnett_appendix}

In this section we give the details about the bolometric light curve modelling using the Arnett model \citep[see][]{Arnett1982, kb10, woo10, WestEtal2023} which main conclusion (the light curve can't be fit with $M_{ni} \le M_{ej}$) is given in Section\,\ref{LCmodeling}.

Using the \texttt{lmfit.minimize} least squares method we retrieve the following parameter values for the radioactive source model: M$_{\rm ej}=0.3\pm{0.1}$~M$_{\odot}$, M$_{^{56}Ni}=2.4\pm{0.3}$~M$_{\odot}$, v$_{\rm ej}=1\times10^{4}$ km~s$^{-1}$ (see Section~\ref{EjectaVelocity}), E$_{k}\approx0.2\times10^{51}$ erg, t$_{\rm leak}=22\pm{3}$ days and the diffusion timescale was $\approx9$ days. The resulting best fit model is shown in red in Figure~\ref{fig:Models}. Since the total ejecta mass includes the mass of $^{56}$Ni these parameter results must be considered to be unphysical.

By re-parameterising the $^{56}$Ni mass as a fraction of the ejecta mass we then force the fit to be physical. However, as shown by the dashed red line in Figure~\ref{fig:Models}, the fit converges poorly.

We also fit the bolometric light curve with the magnetar model. Parameter values are found to be M$_{\rm ej}=0.3\pm0.07$ M$_{\odot}$, B$_{14}=1.8\pm0.6 \times 10^{14}$ Gauss, P$_{\rm{ms}}=5.8\pm0.7$, v$_{\rm ej}=1\times10^{4}$ km~s$^{-1}$ (see Section~\ref{EjectaVelocity}), t$_{\rm leak}=16\pm3$ days, the diffusion timescale $\approx9$ days, and the spindown timescale $\approx51$ days. The resulting best fit parameter model is plotted in blue together with the bolometric light curve in green in Figure~\ref{fig:Models}.

\begin{figure*}
	\centering
	\includegraphics[width=155mm, angle=0]{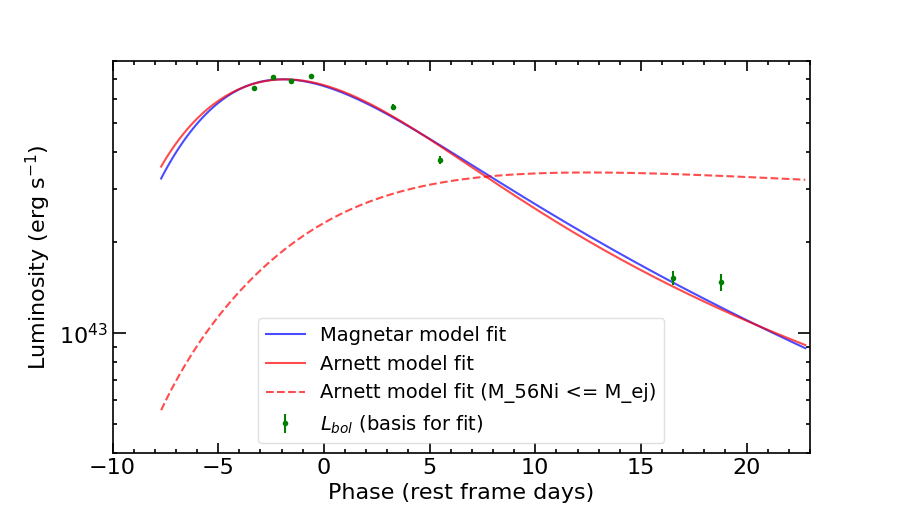}
	\caption{\label{fig:Models} SN\,2021lwz bolometric light curve in green with its best fit Magnetar (in blue) and Arnett radioactive (in red) source models. If the $^{56}$Ni mass is forced to be less than or equal to the ejecta mass then the red dashed line is the best fit one can obtain.}
\end{figure*}

\section{Magnetar-driven model of SN\,2021lwz} \label{21lwz_appendix}

We fit the mulit-band light curve of SN\,2021lwz shown in Figure \ref{fig:lcfit} using the software, model, and priors discussed in Section \ref{MLCmodeling}.  The 
posterior is shown in Figure \ref{fig:corner_asu16}.

\begin{figure*}
    \centering
    \includegraphics[width=0.8\linewidth]{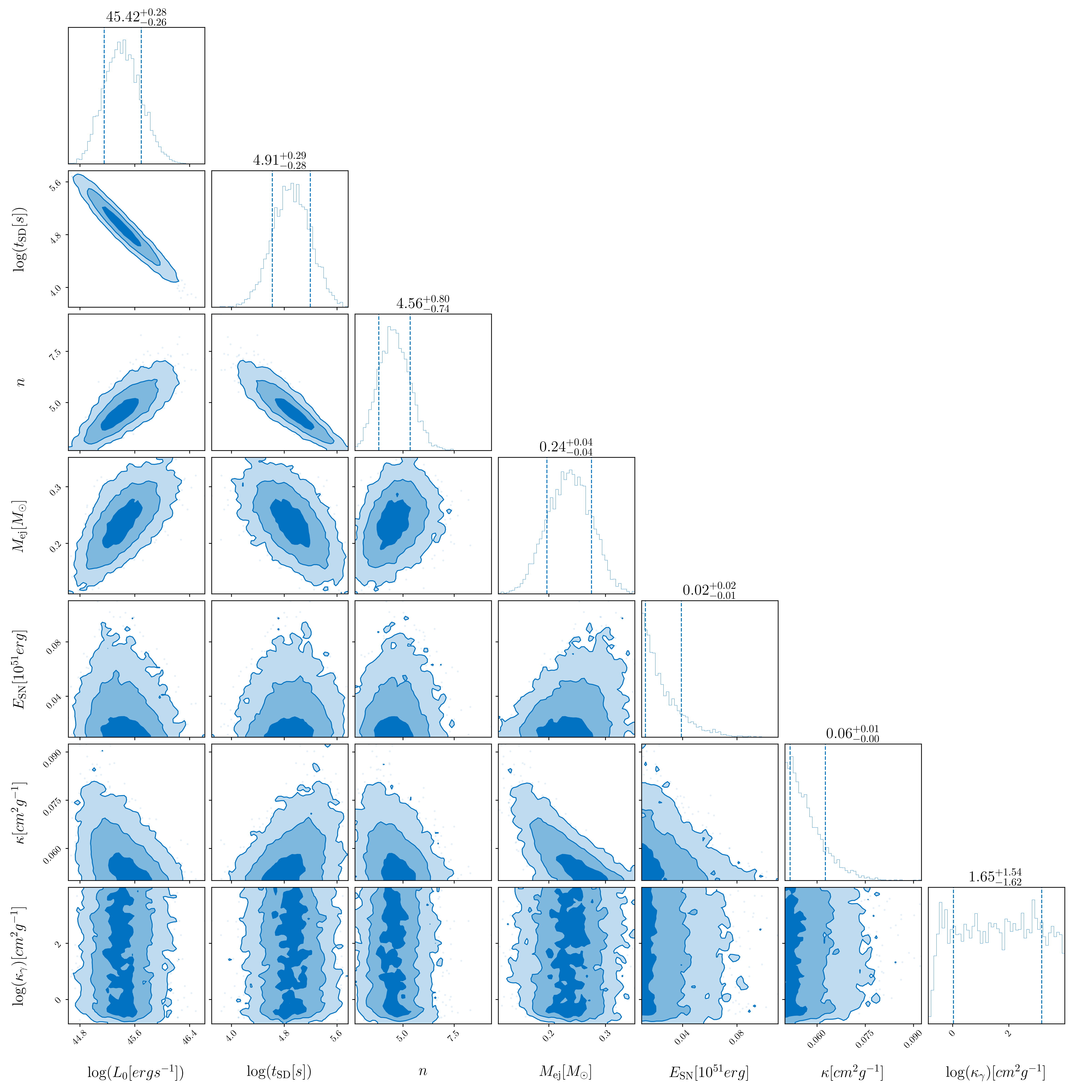}  
    \caption{Posterior distribution of key physical parameters inferred by the light curve fit shown in Figure~\ref{fig:lcfit}.  Median values and 1$\sigma$ uncertainties are given above each parameter.}
    \label{fig:corner}
\end{figure*}

\section{Magnetar-driven model of iPTF16asu} \label{16asu_appendix}

\begin{figure*}
    \centering
    \includegraphics[width=0.5\linewidth]{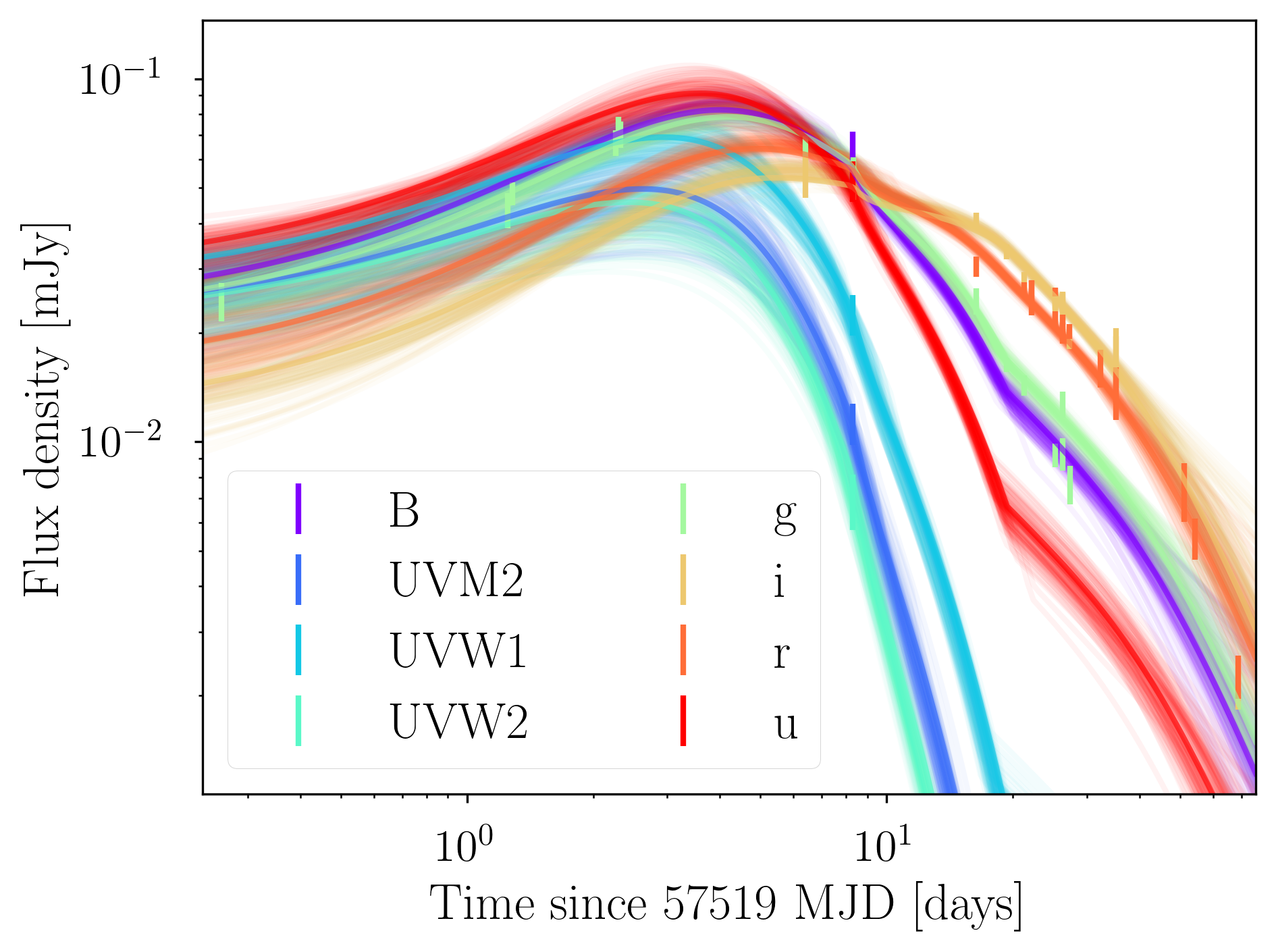}   
    \caption{Fitted multi-band light curve for iPTF16asu. The solid lines indicate the light curve from the model with the highest likelihood in each band, while the shaded area indicates the 90$\%$ credible interval.}
    \label{fig:lcfit_asu16}
\end{figure*}

We fit the mulit-band light curve of iPTF16asu using the same software, model, and priors as for SN\,2021lwz (see Section \ref{MLCmodeling} for details).  The light curve fit is shown in Figure \ref{fig:lcfit_asu16} and posterior is shown in Figure \ref{fig:corner_asu16}.

\begin{figure*}
    \centering
    \includegraphics[width=\linewidth]{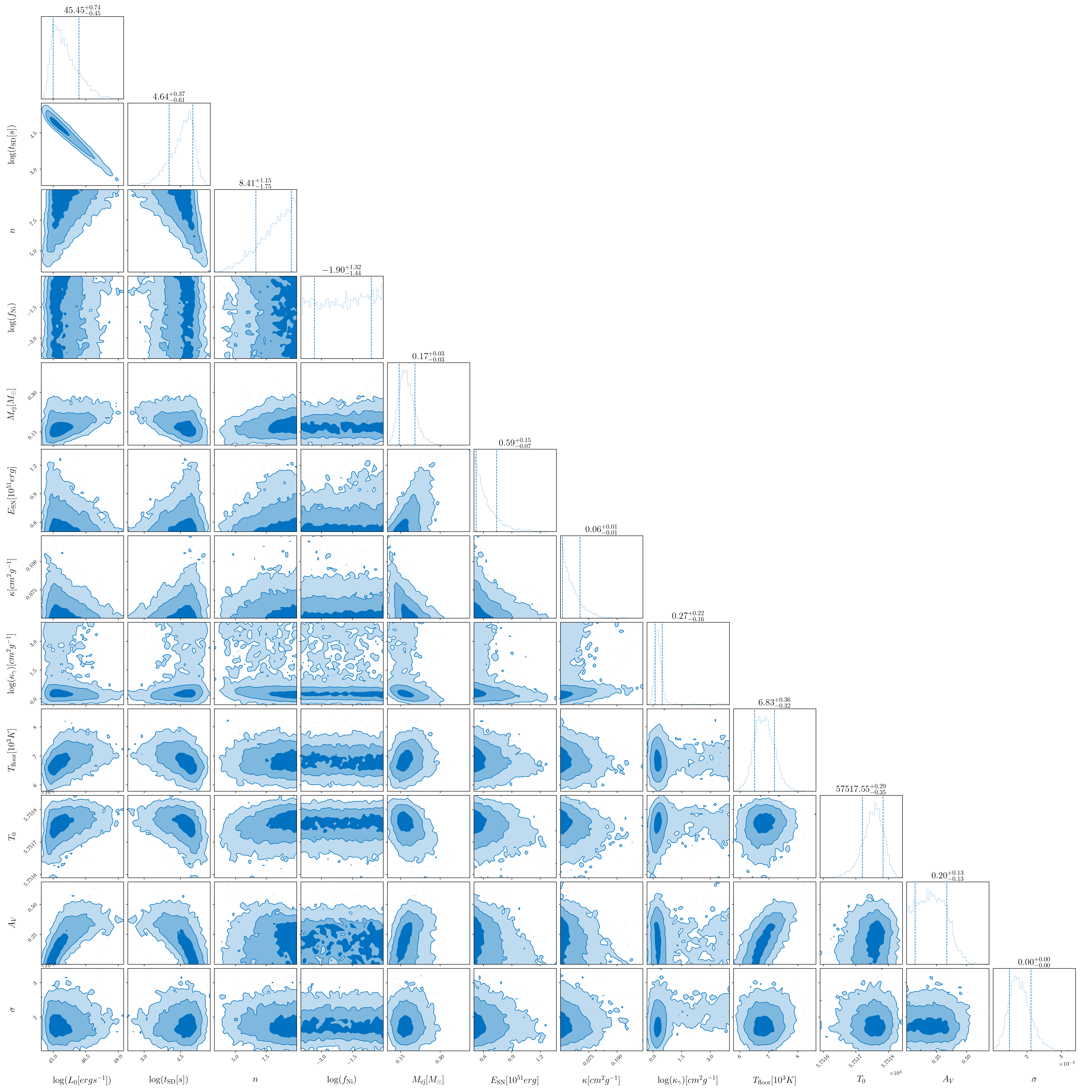}   
    \caption{Posterior distribution of key physical parameters inferred by the light curve fit shown in Figure \ref{fig:lcfit_asu16}.  Median values and 1$\sigma$ uncertainties are given above each parameter.}
    \label{fig:corner_asu16}
\end{figure*}

\section{USSNe Table} \label{ussne_appendix}

The list of USSNe objects displayed in Section~\ref{host_modeling} and their associated properties are given in Table \ref{tab:ussne_list}.

\begin{table*}
\caption{Compilation of the Ultra-Stripped SNe data.}
\label{tab:ussne_list}
\centering
\begin{tabular}{llllclccl}
\hline 
OBJECT	& MASS$_{\rm MED}$ &	MASS$_{\rm SUP}$ &	MASS$_{\rm INF}$ &	SFR$_{\rm MED}$ &	SFR$_{\rm SUP}$ &	SFR$_{\rm INF}$ &	REF. &	TYPE\\
\hline
\hline
SN2021agco &	9.41 &	9.42 &	9.41 &	 -0.70 &	-0.68 &	-0.72 &	1 &	USSN\\ 
SN2019dge &	8.4 &	8.5 &	8.30 &	-1.52 &	-1.46 &	-1.60 &	2 &	USSN\\
SN2019wxt &	10.58 &	10.71 &	10.22 &	0.44 &	0.79 &	0.30 &	3 &	USSN\\
SN2014ft &	10.48 &	10.53 &	10.43 &	0.69 &	0.88 &	0.36 &	4 &	USSN\\
\hline
\end{tabular}
\tablefoot{
These data are used for discussion in Section~\ref{host_modeling}. The masses and SFRs are in log units.
}
\tablebib{
(1)~\citet{2023ApJ...959L..32Y};
(2)~\citet{2020ApJ...900...46Y};
(3)~\citet{2023A&A...675A.201A};
(4)~\citet{2018Sci...362..201D}.
}
\end{table*}

\end{appendix}

\end{document}